%% file: diplom.tex
\documentclass[11pt]{article}

\usepackage[latin1]{inputenc}
\usepackage[breaklinks]{hyperref}

\usepackage[german, english]{babel}

\usepackage{latexsym}

\usepackage{amsmath}
\usepackage{amssymb}

\usepackage{graphicx}
\usepackage{psfig}
\usepackage{color}
\usepackage{epsfig}

\DeclareFontFamily{OT1}{rsfs}{}
\DeclareFontShape{OT1}{rsfs}{m}{n}{ <-7> rsfs5 <7-10> rsfs7 <10->rsfs10}{} 
\DeclareMathAlphabet{\mycal}{OT1}{rsfs}{m}{n}
\newcommand{\scri}{{\mycal I}}

\newcommand{\arccosh}{\mathrm{arccosh}}

\numberwithin{equation}{section}

\newtheorem{theorem}{Theorem}

\begin{document} 

\begin{titlepage}
\begin{center}
\LARGE General Relativistic Static Fluid Solutions \\ 
\LARGE with Cosmological Constant
\end{center}
\mbox{} \\[20mm]
\selectlanguage{german}
\begin{center}
	\Large Diplomarbeit \\[30mm]
	\normalsize von \\
	\large Christian G. Böhmer \\
	\normalsize aus Berlin \\ [35mm]
	eingereicht bei der Mathematisch-Naturwissenschaftlichen Fakultät \\
	der Universität Potsdam \\ [30mm]
	durchgeführt am Max-Planck-Institut für Gravitationsphysik \\
	- Albert-Einstein-Institut - \\
	in Golm \\
	unter Anleitung von Prof. Dr. Bernd G. Schmidt
\end{center} 

\end{titlepage}

\newpage
\selectlanguage{english}
\tableofcontents

\newpage
\input{introduction}

\newpage
\input{tovlambda}

\newpage
\input{secrho}

\newpage
\input{existence-and}

\newpage
\appendix
\input{appendix}

\newpage
\mbox{}
\newpage
\mbox{}
\newpage

\addcontentsline{toc}{section}{References}
\bibliographystyle{plain}
\bibliography{ref}

\newpage
\selectlanguage{german}
\input{deutsch_ent}

\newpage
\section*{Erklärung} 
\addcontentsline{toc}{section}{Erklärung}
Hiermit erkläre ich, da\ss\ ich die vorliegende Arbeit selbstständig
und ohne fremde Hilfe verfa\ss t habe und mich anderer als der 
angegebenen Quellen und Hilfsmittel nicht bedient habe. Alle Stellen,
die wörtlich oder sinngemä\ss\ aus Veröffentlichungen entnommen wurden,
sind als solche gekennzeichnet. 

\mbox{} \\
Golm, 16. Juli 2002 \\[30mm]
Christian Böhmer

\end{document}

%% file: introduction.tex
\section{Introduction}
This diploma thesis analyses static, spherically symmetric perfect 
fluid solutions to Einstein's field equations with cosmological 
constant. New kinds of global solutions are described. 

By a global solution one means an inextendible spacetime 
satisfying the Einstein equations with cosmological constant
with a perfect fluid source. The matter either occupies the whole
space or has finite extend. In the second case a vacuum solution is
joined on as an exterior field.

Global static fluid ball solutions with finite radius at which the 
pressure vanishes are called stellar models.

Recent cosmological observations give strong indications for the
presence of a positive cosmological constant with $\Lambda < 3 \times 10^{-52} \mathrm{m}^{-2}$.
On the other hand Anti-de Sitter spacetimes, having negative
cosmological constant, are important in the low energy limit
of superstring theory.

Therefore it is interesting to analyse solutions to the field equations
with cosmological constant representing for example relativistic stars.

\subsection*{Vanishing cosmological constant $\Lambda=0$}

The first static, spherically symmetric perfect fluid solution with
constant density was already found by Schwarzschild in 1918.
 
In spherical symmetry Tolman~\cite{Tolman} and Oppenheimer 
and Volkoff~\cite{OppenheimerVolkoff} reduced the field equations
to the well known TOV equation.

The boundary of stellar models is defined to be where the pressure vanishes.
At this surface a vacuum solution is joined on as an exterior field. 
In case of vanishing cosmological constant it is the Schwarzschild solution. 
For very simple equations of state Tolman integrated the TOV equation 
and discussed solutions. Although he already included the cosmological constant
in his calculations he did not analyse them. He stated that the
cosmological constant is too small to produce effects. 

Buchdahl~\cite{Buchdahl:1959}  assumed the existence of a global static 
solution, to show that the total mass of a fluid ball is bounded by its radius. 
He showed the strict inequality \mbox{$M < (4/9)R $}, which holds
for fluid balls in which the density does not increase outwards. It implies
that radii of fluid balls are always larger than the black-hole event horizon. 

Geometrical properties of constant density solutions were analysed by
Stephani~\cite{Stephani1,Stephani2}. He showed that they can be embedded
in a five dimensional flat space and that they are conformally flat. The 
cosmological constant can easily be included in his calculations by redefining
some variables. Conformal flatness of constant density solutions is easily
shown with the use of new Buchdahl variables, which will be introduced
in the fourth chapter. It will be seen that the field equations imply the
vanishing of the squared Weyl tensor.

Moreover he remarked that constant density solutions without cosmological 
constant have the spatial geometry of a 3-sphere. With cosmological constant 
the spatial part of the metric may also be Euclidean or hyperbolic, depending
on the choices of constant density and cosmological constant.     

As already said, Buchdahl assumed the existence of a global static 
solution to derive the upper bound on the mass for given radius. 
Rendall and Schmidt~\cite{RendallSchmidt} proved the existence 
of such a global static solution. Baumgarte and Rendall~\cite{BaumgarteRendall} 
later improved the argument with less assumptions on the equation of state and
the pressure. This was further improved by Mars, Mart{\'{\i}}n-Prats 
and Senovilla~\cite{Mars}.

There is a conjecture that asymptotically flat static perfect fluid solutions 
are spherically symmetric. If this is true then this means that spherically
symmetric solutions are the most general static perfect fluid models.

\subsection*{Non-vanishing cosmological constant $\Lambda \neq 0$}

This diploma thesis proves existence of a global solution for cosmological 
constants satisfying $\Lambda < 4\pi \rho_{b}$. $\rho_{b}$ denotes 
the boundary density and is given by the equation of state. 
For a degenerate neutron star one may assume the boundary density
to be the density of iron. In physical units this leads 
to $\Lambda < (4\pi G/c^{2}) \rho_{\mathrm{Fe}} \approx 7 \times 10^{-23} \mathrm{m}^{-2}$. 
This condition is much weaker than the mentioned
cosmological upper bound. 

For larger values the existence cannot be proved with the used arguments.
Finiteness of the radius for a given equation of state is discussed,
a necessary and a sufficient condition are shown.

Another aim is to derive an analogous Buchdahl inequality that 
includes the cosmological term. For positive cosmological constants 
the vacuum region of spacetime may contain a cosmological event horizon. 
With the analogous Buchdahl inequality it will be proved that stellar models 
have an upper bound given by that event horizon. 

Collins~\cite{Collins} stated that for a fixed equation of state and cosmological
constant the choice of central pressure and therefore central density does not
uniquely determine the solution. This is disproved. 

Static perfect fluid solutions with cosmological constant were analysed by 
Kriele~\cite{Kriele} and later by Winter~\cite{Winter}. Both derived the
analogous TOV-$\Lambda$ equation. The first one shows uniqueness
of the solution for given pressure and density distributions, which
already disproved Collins~\cite{Collins}. An analogous type of Buchdahl 
inequality is derived but not discussed in the context 
of upper and lower bounds on radii of stellar objects.
Winter~\cite{Winter} integrates the \mbox{TOV-$\Lambda$} equation from the
boundary inwards to the centre, without proving the existence of
that boundary. This leads to solutions with non-regular centres and
is therefore not suitable for discussing stellar models.

Constant density solutions with cosmological constant were first
analysed by Weyl~\cite{Weyl}. In this remarkable paper these solutions 
are described for different values of the cosmological constant. 
The different possible spatial geometries were already pointed out 
and a possible coordinate singularity was mentioned. 

More than 80 years later Stuchl{\'{\i}}k~\cite{Stuchlik} analysed
these solutions again. He integrated the 
TOV-$\Lambda$ equation for possible values of the cosmological
constant up to the limit \mbox{$\Lambda < 4\pi \rho_{0}$},
where $\rho_{0}$ denotes this constant density.
In these cases constant density solutions describe stellar models. 
The third chapter shows that a coordinate singularity occurs
if $\Lambda \geq 4\pi \rho_{0}$, already mentioned by Weyl~\cite{Weyl}.

If the cosmological constant equals this upper bound the pressure vanishes 
at the mentioned coordinate singularity. In this case it is not 
possible to join the Schwarzschild-de Sitter solution for the 
vacuum. One has to use the Nariai solution \cite{Nariai1,Nariai2} 
to get the metric $C^{1}$ at the boundary. 
For larger cosmological constant the pressure will vanish after 
the coordinate singularity. The volume of group orbits is decreasing 
and there one has to join the Schwarzschild-de Sitter solution 
containing the $r=0$ singularity. Increasing the cosmological 
constant further leads to generalisations of the Einstein static 
universe. These solutions have two regular centres with 
monotonically decreasing or increasing pressure from the
first to the second centre. Certainly the Einstein cosmos itself is a 
solution. Another new kind describes solutions
with a regular centre and increasing divergent pressure. In this case
the spacetime has a geometrical singularity.
These solutions are unphysical and therefore not of great interest.

With the arguments used in this diploma thesis the existence of a 
global solution with given equation of state 
can only be proved for cosmological constants satisfying the 
bound mentioned above, $\Lambda < 4\pi \rho_{b}$. 

Unfortunately, with this restriction solutions of the new kinds do not occur.
Thus one cannot prove existence of the new kinds of solutions for a
prescribed equation of state.

In addition to what was said before this paper shortly investigates 
Newtonian limits of the Schwarzschild-de Sitter solution and of the 
TOV-$\Lambda$ equation. In the first case both possible 
horizons shrink to a point in the limit. In the second, as in the 
case without cosmological term, this leads to the
Euler equation for hydrostatic equilibrium. Nonetheless both 
limits contain a $\Lambda$-corrected gravitational potential.

%% file: tovlambda.tex
\section{Cosmological TOV equation}
\label{costov}
This chapter deals with the Einstein field equations with cosmological constant in the
spherically symmetric and static case. For a generalised Birkhoff theorem 
see \cite{exact,Severa}. 

First a perfect fluid is assumed to be the matter source. 
This directly leads to a $\Lambda$-extended 
Tolman-Oppenheimer-Volkoff equation which will be called  
TOV-$\Lambda$ equation. The  TOV-$\Lambda$ equation together 
with the mean density equation form a system of differential equations.
It can easily be integrated if a constant density is assumed. 

Next a vacuum solution, namely the 
Schwarzschild-de Sitter solution, is derived. It is the unique 
static, spherically symmetric vacuum solution to the field equations with
cosmological constant with group orbits having non-constant volume.
But there is one other static, spherically symmetric vacuum solution, 
the Nariai solution \cite{Nariai1,Nariai2}. The group orbits
of this solution have constant volume. The Nariai solution is needed 
to join interior and exterior solution in a special case but
will not be derived explicitly.

Finally in this introductory chapter Newtonian limits are derived. 
Limits of the Schwarzschild-de Sitter solution 
and the TOV-$\Lambda$ equation are shown. The second leads 
to the fundamental equation of Newtonian astrophysics 
with cosmological constant.

\subsection{The Cosmological Constant} 
Originally the Einstein field equations read
\begin{align}
	G_{\mu \nu} = \kappa T_{\mu \nu},
	\label{gmunu}
\end{align}
where $\kappa$ is the coupling constant. 
Later Einstein \cite{Einstein} introduced the cosmological constant 
mainly to get a static cosmological solution. These field equations are
\begin{align}
	G_{\mu \nu} + \Lambda g_{\mu \nu}  = \kappa T_{\mu \nu}.
	\label{eqn_gmunu}
\end{align}
The effect of $\Lambda$ can be seen as a special type of an energy-momentum tensor. 
It acts as an unusual fluid 
\begin{align*}
	 T_{\mu \nu}^{\Lambda} = -\frac{\Lambda}{\kappa} g_{\mu \nu} = 
	(\rho^{\Lambda}+\frac{P^{\Lambda}}{c^{2}})u_{\mu}u_{\nu}+P^{\Lambda}g_{\mu \nu},
\end{align*}
with $P^{\Lambda}=-\Lambda / \kappa$ and equation of state
$P^{\Lambda}=-\rho^{\Lambda} c^{2}$.

\subsection{Remarks on the Newtonian limit}
The metric for a static, spherically symmetric spacetime can be written
\begin{align}
	ds^{2}=-c^{2}e^{\nu (r)}dt^{2}+e^{a(r)}dr^2+r^{2}(d\theta ^{2} 
	+\sin^{2} \negmedspace \theta \ d\phi^{2}).
\label{eqn_sss.metric}
\end{align}
Appendix~B of \cite{HawkingEllis} shows that $r^{2}$ in front of the sphere metric
is no loss of generality for a perfect fluid. This is true for vanishing cosmological
constant. With cosmological term there is one vacuum solution with group orbits
of constant volume, the mentioned Nariai solution \cite{Nariai1,Nariai2}.

Metric (\ref{eqn_sss.metric}) contains the constant $c$ representing 
the speed of light. Define $\lambda=1/c^{2}$, roughly 
speaking the limit $\lambda \rightarrow 0$ corresponds to the
Newtonian theory. But equations could be non-regular
in the limit $\lambda =0$. Therefore, 
following~\cite{Ehlers:1989,Ehlers:1991,Ehlers:1997}\nocite{Ferrarese:1991}, 
the static, spherically symmetric metric has to be written
\begin{align}
	ds^{2}=-\frac{1}{\lambda} e^{\lambda \nu (r)}dt^{2}+e^{a(r)}dr^2+r^{2}d\Omega^{2},
\label{eqn_sss.metric_limit}
\end{align} 
with field equations
\begin{align}
	G_{\mu \nu} + \lambda \Lambda g_{\mu \nu}  = \kappa T_{\mu \nu},
	\label{eqn_gmunu_limit}
\end{align}
where $d\Omega^{2} = d\theta ^{2} +\sin^{2} \negmedspace \theta \ d\phi^{2}$.
The new term $\lambda \nu (r)$ in (\ref{eqn_sss.metric_limit}) ensures regularity 
of the energy-momentum conservation in the limit if matter is present. 
The $\lambda \Lambda$ term in the field equations (\ref{eqn_gmunu_limit}) makes 
sure that the gravitational potential is regular in the limit $\lambda =0$.

\subsection{TOV-$\Lambda$ equation}
\label{subsec_tovl}
Assume a perfect fluid to be the source of the gravitational field. The derivation of  
the analogous Tolman-Oppenheimer-Volkoff \cite{OppenheimerVolkoff,Tolman} equation is shown.
Consider metric (\ref{eqn_sss.metric_limit})
\begin{align*}
	ds^{2}=-\frac{1}{\lambda} e^{\lambda \nu (r)}dt^{2}+e^{a(r)}dr^2+r^{2}d\Omega^{2}.
\end{align*} 
Units where $G=1$ will be used, thus $\kappa=8\pi \lambda^{2}$. $\lambda=1$ corresponds to
Einstein's theory of gravitation in geometrised units. 

The field equations (\ref{eqn_gmunu_limit}) for a perfect fluid can be taken from appendix~A, 
(\ref{G00})-(\ref{G33}) with (\ref{tmunu}). These are three independent equations, 
which imply energy-momentum conservation (\ref{emc}). Thus one may either use three 
independent field equations or one uses two field equations and the energy-momentum 
conservation equation. The aim of this section is to derive a system of differential 
equations for pressure and mean density, defined later. Hence it is more
effective to do the second. So consider the first two field equations and the conservation equation. 
This gives
\begin{align}
\frac{1}{\lambda r^{2}}e^{\lambda \nu(r)}\frac{d}{dr}\left(r-re^{-a(r)}\right) - \Lambda e^{\lambda \nu(r)} &= 8\pi \rho(r)e^{\lambda \nu(r)} 
\label{a} \\
\frac{1}{r^{2}}\left(1+r\lambda \nu'(r)-e^{a(r)}\right) +\lambda\Lambda e^{a(r)} &= 8\pi \lambda^{2} P(r)e^{a(r)}
\label{b} \\
-\frac{\nu'(r)}{2}(\lambda P(r)+\rho) &= P'(r).
\label{c} 
\end{align}
These are three independent ordinary differential equations. But there are four unknown functions. 
Mathematically but not physically one of the four functions can be chosen freely. 
From a physical point of view there are two possibilities. Either a matter distribution $\rho = \rho(r)$
or an equation of state $\rho = \rho(P)$ is prescribed. 
The most physical case is to prescribe an equation of state.
Equation (\ref{a}) can easily be integrated. By putting the constant of integration
equal to zero because of regularity at the centre one gets
\begin{align}
	e^{-a(r)}=1-\lambda 8\pi \frac{1}{r}\int_{0}^{r}s^{2}\rho(s)ds-\lambda \frac{\Lambda}{r}\int_{0}^{r}s^{2}ds.
	\label{eqn_e-ar}
\end{align}
Now the definition of `mass up to $r$' is used
\begin{align}
	m(r)=4\pi\int_{0}^{r}s^2\rho(s)ds.
	\label{mass}
\end{align}
The second integral of (\ref{eqn_e-ar}) can be evaluated to finally give
\begin{align}
	e^{-a(r)}=1-\lambda \frac{2m(r)}{r}-\lambda \frac{\Lambda}{3}r^{2}.
\end{align}
In addition define the mean density up to $r$ by
\begin{align}
	w(r)=\frac{m(r)}{r^{3}},
	\label{md}
\end{align}
then the above metric component takes the form
\begin{align}
	e^{-a(r)}=1-\lambda 2w(r)r^{2}-\lambda \frac{\Lambda}{3}r^{2}.
	\label{ea}
\end{align}
Therefore one may write metric~(\ref{eqn_sss.metric_limit}) as
\begin{align}
	ds^{2}=-\frac{1}{\lambda}e^{\lambda \nu (r)}dt^{2}+\frac{dr^2}{1-\lambda 2 w(r)r^{2}-\lambda \frac{\Lambda}{3}r^{2}}
	+r^{2}d\Omega^{2}.
\label{eqn_gen.metric}
\end{align}
The function $\nu'(r)$ can be eliminated from (\ref{b}) and (\ref{c}). 
The second field equation (\ref{b}) yields to 
\begin{align}
	\frac{\nu'(r)}{2}=r\frac{\lambda 4\pi P(r)+w(r)-\frac{\Lambda}{3}}{1-\lambda 2w(r)r^{2}-\lambda \frac{\Lambda}{3}r^{2}}.
	\label{nu'b}
\end{align}
This with (\ref{c}) implies the TOV-$\Lambda$ equation
\cite{Stuchlik,Winter}
\begin{align}
	P'(r)=-r\frac{\left( \lambda 4\pi P(r) + w(r) -\frac{\Lambda }{3} \right) 
		       \left( \lambda P(r)+ \rho(r) \right)}
		      {1-\lambda 2 w(r)r^{2}-\lambda \frac{\Lambda}{3}r^{2} }.
	\label{tov}
\end{align}
Putting $\Lambda =0$ in (\ref{tov}) one finds the Tolman-Oppenheimer-Volkoff equation, without cosmological term, 
short TOV equation, see \cite{OppenheimerVolkoff,Tolman}.

If an equation of state $\rho = \rho(P)$ is given, the conservation equation (\ref{c})~can be integrated to give
\begin{align}
	\nu(r)=- \int_{P_{c}}^{P(r)}\frac{2 d P}{\lambda P+\rho(P)},
	\label{int}
\end{align}
where $P_{c}$ denotes the central pressure. Using the definition of $m(r)$ then (\ref{md}) and (\ref{tov})
form an integro-differential system for $\rho(r)$ and $P(r)$.
Differentiating (\ref{md}) with respect to $r$ implies
\begin{align}
	w'(r)=\frac{1}{r}\left(4\pi \rho(P(r)) - 3w(r) \right).
	\label{mdd}
\end{align}
Therefore given $\rho=\rho(P)$ equations (\ref{tov}) and (\ref{mdd}) are forming a system of
differential equations in $P(r)$ and $w(r)$. The solution for $w(r)$ together with (\ref{ea}) gives the function
$a(r)$, whereas the solution of $P(r)$ in (\ref{int}) defines $\nu(r)$.

There is a alternative way of finding the TOV-$\Lambda$ equation if the TOV equation is given. 
Looking back at the Einstein field equations with cosmological constant (\ref{a}), (\ref{b})
it is found that they may be derived by a simple substitution. Putting
\begin{align*}
	\rho_{\mathrm{eff}} & = \rho + \frac{\Lambda}{8\pi }, \\
	P_{\mathrm{eff}} & = P - \frac{\Lambda}{\lambda 8\pi },
\end{align*}
in the field equations without cosmological term (\ref{gmunu}), gives the $\Lambda$ term properly.
Therefore the effective energy-momentum tensor is defined by
\begin{align}
	T_{\mu \nu}^{\mathrm{eff}} := T_{\mu \nu} - \frac{\Lambda}{\kappa}g_{\mu \nu}.
	\label{eqn_teff}	
\end{align}
In addition it should be remarked how $w_{\mathrm{eff}}$ is given,
\begin{align*}
	w_{\mathrm{eff}} = w + \frac{\Lambda}{6 }.
\end{align*}
The possibility of writing the equations with effective values is of interest later.

\subsection{Schwarzschild-anti-de Sitter and Schwarzschild-de Sitter solution}
\label{subsec_sadssds}
A vacuum solution ($T_{\mu \nu}=0$) to the Einstein field equations with cosmological constant is
derived. For a vanishing cosmological constant the Schwarzschild solution follows,
for vanishing mass the metric gives the de Sitter cosmology. 
Again one uses metric (\ref{eqn_sss.metric_limit})
\begin{align*}
	ds^{2}=-\frac{1}{\lambda}e^{\lambda \nu(r)}dt^{2}+e^{a(r)}dr^2+r^{2}d\Omega^{2}.
\end{align*}
One has to find a solution to the system of equations
\begin{align*}
	G_{\mu \nu}+\lambda \Lambda g_{\mu \nu}=0.
\end{align*}
In the static, spherically symmetric case this system reduces to three equations.
But there are only two unknown functions. The third equation is not independent.
Thus it suffices to consider the first two field equations (\ref{G00}),(\ref{G11}), given in appendix A.
\begin{align}
  	\frac{1}{\lambda r^{2}}e^{\lambda \nu (r)}\frac{d}{dr}\left( r-re^{-a(r)}\right)-\Lambda e^{\lambda \nu(r)} & =0
	\label{eqn_first} \\	
	\frac{1}{r^{2}}\left(1+r\lambda \nu'(r)-e^{a(r)}\right)+\lambda \Lambda e^{a(r)} & =0.
	\label{eqn_second}
\end{align}
The first equation gives
\begin{align*}
	\frac{d}{dr}\left( r-re^{-a(r)}\right)-\lambda \Lambda r^2 = 0
\end{align*}
and can be integrated to
\begin{align*}
	e^{-a(r)}=1-\lambda \frac{2M}{r}-\lambda \frac{\Lambda}{3} r^{2},
\end{align*}
where $2M$ is a constant of integration. The additional $\lambda$ 
in front of the constant of integration is to get interior (\ref{eqn_gen.metric}) and exterior 
metric continuous at a boundary.

Again using equation (\ref{eqn_first}), calculating the derivative and multiplying with $-e^{a(r)}$ one finds
\begin{align*}
	-\left(e^{a(r)}+r a'(r)-1\right)+\lambda \Lambda r^{2}e^{a(r)}=0.
\end{align*}
The second field equation (\ref{eqn_second}) can be put in the following form
\begin{align*}
	\left(e^{a(r)}-r \lambda \nu'(r)-1\right)-\lambda \Lambda r^{2}e^{a(r)}=0.
\end{align*}
Adding up both equations gives $a'(r)+\lambda\nu'(r)=0$. This can now be expressed with logarithms to give
\begin{align*}
	\frac{d}{dr}\ln\left(e^{a(r)}e^{\lambda \nu(r)}\right)=0.
\end{align*}
One integrates this equation and may set the constant of integration equal to one because the 
time coordinate can be rescaled. This leads to
\begin{align*}
	e^{a(r)}e^{\lambda \nu(r)}=1,
\end{align*}
and implies the static, spherically symmetric vacuum 
solution to the field equations with cosmological constant, see ~\cite{exact}.
It is the unique solution which group orbits do not have constant volume.  

With the above the following metric is obtained
\begin{align}
	ds^{2}=-\frac{1}{\lambda}\left(1-\lambda \frac{2M}{r}-\lambda \frac{\Lambda}{3}r^{2}\right)dt^{2}+
	\frac{dr^{2}}{1-\lambda \frac{2M}{r}-\lambda \frac{\Lambda}{3}r^{2}} +r^{2}d\Omega^{2}.
	\label{sds}
\end{align}
Metric (\ref{sds}) is usually called Schwarzschild-de Sitter metric, although it was first
published by Kottler (1918). It is well defined for radii satisfying $g^{rr}(r)>0$.

Figure~\ref{fig: Penrose} shows Penrose-Carter diagram
for the Schwarzschild-de Sitter space with $\lambda=1$. 
The constants $M$ and $\Lambda$ are chosen
such that $9\Lambda M^{2} <1$, which means that there are two horizons, 
see \cite{Geyer,GibbonsHawking}. If $9\Lambda M^{2} =1$ then there
is only one horizon and if $9\Lambda M^{2} >1$ the spacetime
does not contain a horizon.

Figure~\ref{fig: Penrose_anti1} shows Penrose-Carter diagram
for Schwarzschild-anti-de Sitter space with $\lambda=1$. 
For positive mass there is always the black-hole event horizon.

Setting $\Lambda =0$ in (\ref{sds}) gives the usual Schwarzschild metric. 
For a vanishing mass the de Sitter metric is obtained. 

\subsection{Newtonian limits}
The Newtonian limits of the Schwarz\-schild-anti-de Sitter and Schwarz\-schild-de 
Sitter metric and of the TOV-$\Lambda$ equation are derived. The frame 
theory approach by Ehlers \cite{Ehlers:1989,Ehlers:1991,Ehlers:1997} will be used to define a 
one-parameter family of general relativistic spacetime models with Newtonian limit.
Including the cosmological term gives a Newtonian limit with a $\Lambda$-corrected 
gravitational potential.

\subsubsection{Limit of Schwarzschild-anti-de Sitter and Schwarz\-schild-de Sitter models}
A parametrisation of the Schwarzschild-de Sitter metric is given by (\ref{sds})
\begin{align*}
	ds^{2}=-\frac{1}{\lambda}\left(1-\lambda \frac{2M}{r}-\lambda \frac{\Lambda}{3}r^{2}\right)dt^{2}+
	\frac{dr^{2}}{1-\lambda \frac{2M}{r}-\lambda \frac{\Lambda}{3}r^{2}}+r^{2}d\Omega^{2}. 
\end{align*}
In the limit $\lambda=0$ the family 
$\mathcal{M}(\lambda)=\left\{-\lambda g_{\alpha \beta}(\lambda),g^{\alpha \beta}(\lambda),
\Gamma_{\alpha \beta}^{\gamma}(\lambda)\right\}$ 
converges to a field of a mass point at the origin. The black-hole event horizon and the cosmological 
event horizon both shrink to a point.
The gravitational field is given by
\begin{align}
	\Gamma_{tt}^{r}[\lambda=0] = \frac{M}{r^{2}}-\frac{\Lambda}{3}r,
	\label{newton}
\end{align}
all others vanish. 
This corresponds to Newton's equation with cosmological term
\begin{align*}
	\nabla \phi(r)=\frac{M}{r^{2}}-\frac{\Lambda}{3} r.
\end{align*}

\subsubsection{Limit of the TOV-$\Lambda$ equation}
The parametrised metric (\ref{eqn_gen.metric}) is given by
\begin{align*}
	ds^{2}=-\frac{1}{\lambda}e^{\lambda \nu(r)}dt^{2}+\frac{dr^2}{1-\lambda 2w(r)r^{2}-\lambda \frac{\Lambda}{3}r^{2}}
	+r^{2}d\Omega^{2}.
\end{align*}
The parametrisation of the TOV-$\Lambda$ equation (\ref{tov}) is
\begin{align*}
	P'(r)=-r\frac{\left( 4\pi \lambda P(r)+w(r)-\frac{\Lambda}{3} \right) 
		       \left(\lambda P(r)+\rho(r) \right)}
		      {1-2\lambda w(r)r^{2}-\lambda \frac{\Lambda}{3}r^{2}}.
\end{align*}
This differential equation depends on $\lambda$, it is regular in
the limit $\lambda =0$. Solutions lead to a family of models 
with Newtonian limit. 

The parametrised TOV-$\Lambda$ equation leads to 
the ``fundamental equation of Newtonian astrophysics'' \cite{Weinberg} 
with cosmological term 
\begin{align*}
	-r^{2} P'(r)[\lambda=0]=\rho(r)\left(m(r)-\frac{\Lambda}{3} r^{3}\right).
\end{align*}
The gravitational field reads
\begin{align}
	\Gamma_{tt}^{r}[\lambda=0] = \frac{\nu'(r)}{2}[\lambda=0],
	\label{G_ttr}
\end{align}
all others vanish. Using (\ref{nu'b}) in (\ref{G_ttr}) reproduces (\ref{newton}). But with the
energy-momentum conservation (\ref{c}) it gives
\begin{align}
	\Gamma_{tt}^{r}[\lambda=0]=-\frac{P'(r)}{\rho(r)},
	\label{euler}
\end{align}
which is the Euler equation for hydrostatic equilibrium in spherical symmetry, usually written as
\begin{align*}
	\rho(r) \vec{f} = \nabla P(r).
\end{align*}

%% file: secrho.tex
\section{Solutions with constant density}
\label{secrho}

For practical reasons the notation is changed to geometrised units where $c^{2}=1/\lambda=1$.

Assume a positive constant density distribution $\rho =\rho_{0}=\text{const.}$ 
Then $w=\frac{4\pi}{3} \rho_{0}$ gives (\ref{tov}) in the form
\begin{align}
        P'(r)=-r\frac{\left( 4\pi P(r) + \frac{4\pi}{3}\rho_{0} -\frac{\Lambda}{3} \right)
                       \left( P(r) + \rho_{0} \right)}
                      {1-\left( \frac{8\pi}{3}\rho_{0} + \frac{\Lambda}{3} \right)r^{2}} .
\label{tov_const}
\end{align}

In the following all solutions to the above differential equation are derived, see 
\cite{Stephani1,Stuchlik,Tolman}. In \cite{Stephani1,Tolman} the equation was integrated but not 
discussed for the different possible values of $\Lambda$. In \cite{Stuchlik} all cases with 
$\Lambda < 4\pi \rho_{0}$ were discussed using dimensionless variables. This complicated the
physical interpretation of the results and was an unnecessary restriction to the cosmological 
constant. 

The central pressure $P_{c}=P(r=0)$ is always assumed to be positive and finite.
Using that the density is constant in (\ref{int}) metric (\ref{eqn_gen.metric}) 
reads 
\begin{align}
        ds^{2}=-\left(\frac{P_{c}+\rho_{0}}{P(r)+\rho_{0}}\right)^{2}dt^{2}
        +\frac{dr^2}{1-\left( \frac{8\pi}{3}\rho_{0} + \frac{\Lambda}{3} \right)r^{2}}
        +r^{2}d\Omega^{2}.
\label{eqn_metric}
\end{align}
The metric is well defined for radial coordinates $r \in \left[0,\hat{r}\right)$
if $\Lambda > -8\pi \rho_{0}$, where $\hat{r}$ denotes the zero of $g^{rr}$.
If $\Lambda \leq -8\pi \rho_{0}$ the metric is well defined for all $r$.

Solutions of differential equation (\ref{tov_const}) are uniquely determined
by the three parameters $\rho_{0}$, $P_{c}$ and $\Lambda$. Therefore
one has a 3-parameter family of solutions.

The two cases of negative and positive cosmological constant are
discussed separately. In the first case there is no cosmological event horizon
whereas in the second there is. Analogous Buchdahl \cite{Buchdahl:1959} 
inequalities (\ref{eqn_buchdahl}) will be derived for the different cases.

Figures of the pressure are shown. The constant density $\rho_{0}$ and the central
pressure $P_{c}$ are equal in all figures, $\rho_{0}=P_{c}=1$. Only the cosmological 
constant varies. This follows the approach of the chapter and helps to compare 
the different functions.

The end of the chapter summarises the different solutions. An overview is given.

\subsection{Spatial geometry of solutions}
\label{subsec_spatial}
Before starting to solve the TOV-$\Lambda$ equation one should have a closer 
look at metric (\ref{eqn_metric}). 

The aim of this section is to describe the spatial geometry of metric (\ref{eqn_metric})
because it depends on the choice of the constant density $\rho_{0}$ and the
cosmological constant $\Lambda$. For vanishing cosmological constant this
was done in~\cite{Stephani2}. 

For spherically symmetric spacetimes there exists an invariantly 
defined mass function \cite{Zannias}, the quasilocal mass.

For metric~(\ref{eqn_metric}) quasilocal mass \cite{Zannias} is given by
\begin{align}
	2 m_{\mathrm{q}}(r) = \frac{8 \pi}{3} \rho_{0} r^{3} + \frac{\Lambda}{3} r^{3}.
	\label{eqn_qlm}
\end{align} 
If $R$ denotes the boundary of a stellar object then
\begin{align*}
	2 m_{\mathrm{q}}(R) = 2M + \frac{\Lambda}{3} R^{3},
\end{align*}
where $M$ denotes the total mass of the object. Let 
\begin{align*}
        k = \frac{8\pi}{3}\rho_{0} + \frac{\Lambda}{3},
\end{align*}
then the the spatial part of metric (\ref{eqn_metric}) reads
\begin{align}
        d\sigma^{2}=\frac{dr^2}{1-k r^{2}}+r^{2}d\Omega^{2},
        \label{eqn_sigma}
\end{align}
and the quasilocal mass (\ref{eqn_qlm}) is given by
\begin{align}
	2 m_{\mathrm{q}}(r) = k r^{3},
\end{align}

For all positive values of $k$ spatial metric~(\ref{eqn_sigma}) describes one half of 
a 3-sphere of radius $1/\sqrt{k}$. The quasilocal mass $m_{\mathrm{q}}(r)$ is 
positive. Introducing a new coordinate by
\begin{align}
        r = \frac{1}{\sqrt{k}} \sin \alpha,
        \label{eqn_sine}
\end{align}
gives
\begin{align*}
        d\sigma^{2}=\frac{1}{k}(d\alpha^{2}+\sin^{2} \negmedspace \alpha \ d\Omega^{2}).
\end{align*}
This shows that (\ref{eqn_metric}) has a coordinate singularity. 
The metric with radial coordinate $r$ only describes one half of the 3-sphere. 
Geometrically the coordinate singularity is easily explained. The volume of 
group orbits \mbox{$r=\mbox{const.}$} has an extremum at the equator $\hat{r}=1/\sqrt{k}$.
Thus this coordinate cannot be used to describe the second half of the 3-sphere.

Differential equation (\ref{tov_const}) is singular at $\hat{r}=1/\sqrt{k}$. With the
new coordinate $\alpha$ the differential equation is regular at the
the corresponding $\alpha(\hat{r})=\pi/2$. 

If $k=0$ then $m_{\mathrm{q}}(r)=0$ and the spatial metric is purely Euclidean in spherical coordinates. In Cartesian
coordinates this simply reads
\begin{align*}
        d\sigma^{2}=d x^{2} + d y^{2} + d z^{2}.
\end{align*}

Finally, negative values of $k$ in the spacial metric correspond to three-dimensional
hyperboloids with negative quasilocal mass. Then using
\begin{align}
        r = \frac{1}{\sqrt{-k}} \sinh \alpha,
        \label{eqn_sinehyp}
\end{align}
gives
\begin{align*}
        d\sigma^{2}=\frac{1}{(-k)}(d\alpha^{2} + \sinh^{2} \negmedspace \alpha \ d\Omega^{2}).
\end{align*}

The coordinate singularity becomes important if $\Lambda$ exceeds an upper limit.
The restriction to $\Lambda < 4\pi \rho_{0}$ in \cite{Stuchlik} is due to this fact.
Up to this limit the radial coordinate $r$ and the new coordinate $\alpha$ are used side by side.
Although the radial coordinate is less elegant the physical picture is clearer.

In appendix~B it will be shown that constant density solutions are 
conformally flat \cite{Stephani1}. Conformally flat constant density 
solutions were analysed in \cite{exact} and were called generalised solutions 
of the interior Schwarzschild solution.

If the density is decreasing or increasing outwards solutions are
not conformally flat.

It will also be seen that models with a real geometric singularity exists. They are
physically not relevant because the pressure is divergent at the singularity. 

\subsection{Solutions with negative cosmological constant}
In the Schwarzschild-anti-de Sitter spacetime there exists a black-hole
event horizon. It will be shown that radii of stellar objects are always larger than
this radius corresponding to the black-hole event horizon.

\subsubsection{Stellar models with spatially hyperbolic geometry} 
\subsubsection*{$\Lambda < -8 \pi \rho_{0}$}
If  $\Lambda < -8 \pi \rho_{0}$ then $k$ and $m_{\mathrm{q}}(r)$ are negative. With (\ref{eqn_sinehyp}) the 
denominator of (\ref{tov_const}) can be written as $(\cosh^{2} \negmedspace \alpha)$
and the differential equation does not have a singularity. 

The volume of group orbits has no extrema, thus metric~(\ref{eqn_metric})
has no coordinate singularities and is well defined for all $r$.

Write the differential equation (\ref{tov_const}) as
\begin{align}
        d \left[ \ln \frac{3P +\rho_{0} -\frac{\Lambda}{4\pi}}{P+\rho_{0}} \right] =
        d \left[ \ln \cosh \alpha \right].
\end{align}
Integration gives
\begin{align}
        \frac{3P(\alpha) +\rho_{0} -\frac{\Lambda}{4\pi}}{P(\alpha)+\rho_{0}} =
                C \cosh \alpha .
        \label{pr_implicit}
\end{align}
$C$ is the constant of integration, evaluated by defining \mbox{$P(\alpha =0)=P_{c}$} 
to be the central pressure.
At the centre it is found that
\begin{align}
        C = \frac{3P_{c} + \rho_{0}  -\frac{\Lambda}{4\pi}}{P_{c}+\rho_{0}}.
\label{constantC}
\end{align}
Throughout this calculation it is assumed that $8 \pi \rho_{0} < -\Lambda$. 
With $P_{c}>0$ this implies that $C>3$ and $(1-\Lambda/4 \pi \rho_{0})>3 $, 
which can easily be verified.
The constant $C$ will be used instead of its explicit expression given by (\ref{constantC}). 
Then the pressure reads
\begin{align}
        P(\alpha)=\rho_{0} \frac{ \left(1-\frac{\Lambda}{4 \pi \rho_{0}}\right) -C \cosh \alpha }{C \cosh \alpha-3}.
\label{pr<}
\end{align}
The function $P(\alpha)$ is well defined and monotonically decreasing for all $\alpha$.

The pressure converges to $-\rho_{0}$ as
$\alpha$ or as the radius $r$ tends to infinity. Thus a radius $R$, where the pressure vanishes, always exists.
It is given by
\begin{align}
        R^{2}=\frac{3}{|8 \pi \rho_{0}+\Lambda |}
        \left\{\frac{1}{C^{2}}\left(1-\frac{\Lambda}{4 \pi \rho_{0}}\right)^{2} -1\right\}.
\label{eqn_prpc<}
\end{align}
Writing it in $\alpha$, where $\alpha_{b}$ corresponds to $R$
\begin{align}
        \cosh \alpha_{b} = \frac{1}{C} \left(1-\frac{\Lambda}{4 \pi \rho_{0}}\right),
\label{eqn_coshalpha<}
\end{align}
shows that the new coordinate $\alpha$ simplifies the expressions . Therefore the 
radial coordinate will only be used to give a physical picture or if it simplifies the 
calculations. 

From equation (\ref{eqn_coshalpha<}) one can deduce the inverse function. Then one has
the central pressure as a function of $\alpha_{b}$, $P_{c}=P_{c}(\alpha_{b})$. To do this the 
explicit expression for $C$ from (\ref{constantC}) is needed. One finds
\begin{align}
        P_{c}=\rho_{0} \frac{\left(1-\frac{\Lambda}{4 \pi \rho_{0}}\right) \left(\cosh \alpha_{b} -1 \right)}
                {\left(1-\frac{\Lambda}{4 \pi \rho_{0}}\right) - 3\cosh \alpha_{b}}.
\label{eqn_pcpr<}
\end{align}

The central pressure given by (\ref{eqn_pcpr<}) should be finite. Therefore one obtains
an analogue of the Buchdahl inequality. It reads
\begin{align}
        \cosh \alpha_{b} < \frac{1}{9}\left(1-\frac{\Lambda}{4 \pi \rho_{0}}\right).
\label{eqn_buchdahl<}
\end{align}
Thus there exists an upper bound for $\alpha_{b}$ for given $\Lambda$ and $\rho_{0}$.
Use that 
\begin{align*}
	\sinh(\arccosh (\alpha)) = \sqrt{\alpha^{2}-1},
\end{align*}
then the corresponding radius $R$ is given by
\begin{align}
        R^{2} < \frac{\frac{1}{3} \left(4-\frac{\Lambda}{4 \pi \rho_{0}} \right)}{4 \pi \rho_{0}}.
\label{eqn_R<}
\end{align}
Since the cosmological constant is negative the right-hand side 
of (\ref{eqn_R<}) is well defined. Using the definition of 
mass $M=(4\pi/3)\rho_{0} R^{3}$ one can rewrite (\ref{eqn_R<})
in terms of $M$, $R$ and $\Lambda$. This leads to
\begin{align}
	3M < \frac{2}{3}R + R\sqrt{\frac{4}{9}-\frac{\Lambda}{3} R^{2}}.
	\label{eqn_3m<_hyp}
\end{align}
This is the wanted analogue of the Buchdahl inequality (\ref{eqn_buchdahl}).

The explicit derivation of the last equation is shown in section~\ref{exis}.\ref{theorem_buch},
where the generalised Buchdahl inequality is derived.

To emphasise that the cosmological constant is negative write~(\ref{eqn_3m<_hyp})
\begin{align*}
	3M < \frac{2}{3}R + R\sqrt{\frac{4}{9}+\frac{|\Lambda|}{3} R^{2}}.
\end{align*}

At $r=R$, where the pressure vanishes, the Schwarzschild-anti-de Sitter 
solution (\ref{sds}) is joined. Joining interior and exterior solution is needed
in this and many of the following cases. Therefore the next section shows this
procedure explicitly.

\begin{figure}
\noindent
\centering\epsfig{figure=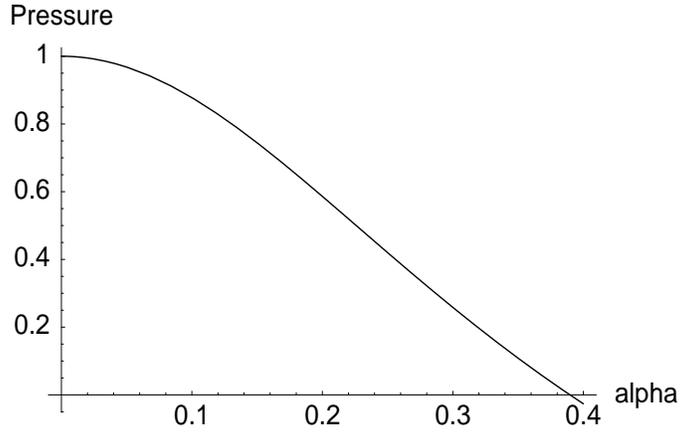,width=9cm,height=6cm} 
\caption{Pressure function $P(\alpha)$ with $\rho_{0}=1$, $P_{c}=1$ and
	$\Lambda=-10\pi$}
\label{fig:pressure_hyp}
\end{figure}

\begin{figure}
\noindent
\begin{minipage}[h]{.46\linewidth}
\centering\epsfig{figure=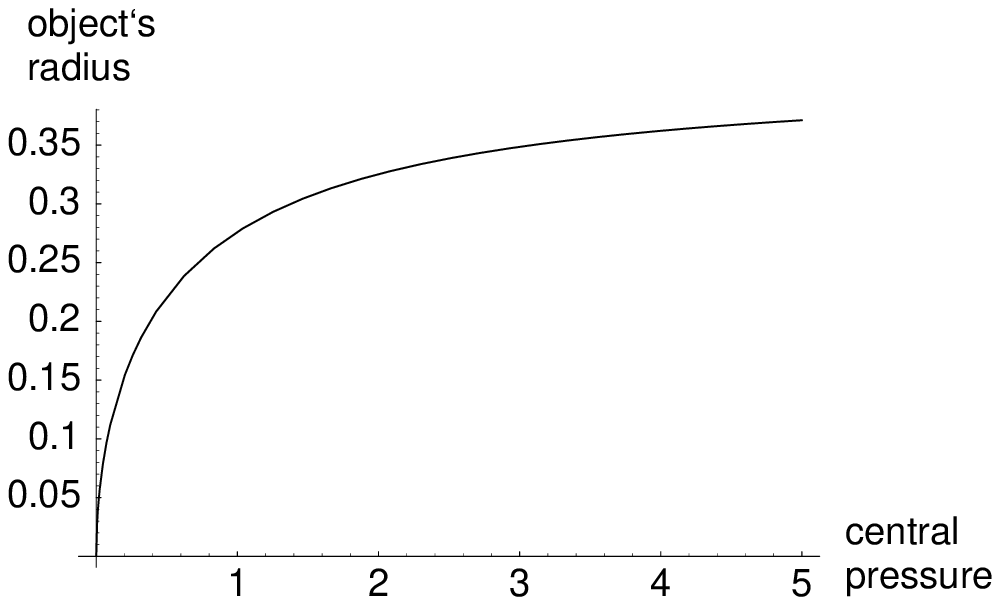,width=\linewidth} 
\caption{Radius as a function of central pressure $R=R(P_{c})$}
\label{fig:r_pc_hyp}
\end{minipage}\hfill
\begin{minipage}[h]{.46\linewidth}
\centering\epsfig{figure=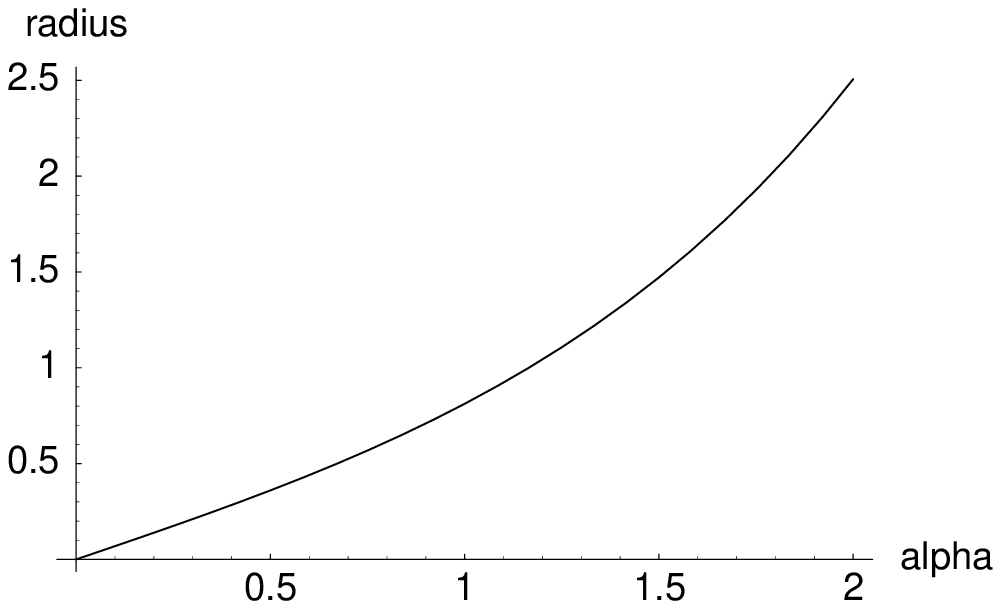,width=\linewidth} 
\caption{Radial coordinate \mbox{$r=r(\alpha)$}}
\label{fig:radius_hyp}
\end{minipage}
\end{figure}

Figure~\ref{fig:pressure_hyp} shows the typical behaviour of function $P(\alpha)$ (\ref{pr<}).
Figures~\ref{fig:r_pc_hyp} and~\ref{fig:radius_hyp} show the radius of the
object as a function of the central pressure (\ref{eqn_prpc<}) and the radius as a function of
the new variable $\alpha$ (\ref{eqn_sinehyp}), respectively. Constant
density and central pressure are both set to one, $\rho_{0}=1$, $P_{c}=1$.  

\subsubsection{Joining interior and exterior solution} 
At the $P=0$ surface the Schwarzschild-anti-de Sitter solution 
(\ref{sds}) is joined. Since this is needed for all solutions
describing stellar objects the procedure of joining interior and
exterior solution is shown explicitly.

Gauss coordinates to the $r=\mbox{const.}$ hypersurfaces in the interior
are given by $\chi=\alpha/\sqrt{k}$. Then the interior metric (\ref{eqn_metric}) reads
\begin{align*}
	ds^{2}=-\left(\frac{P_{c}+\rho_{0}}{P(\chi)+\rho_{0}}\right)^{2}dt^{2}
        	+d\chi^2
        	+\left(\frac{\sinh \sqrt{-k} \chi}{\sqrt{-k}} \right)^{2}d\Omega^{2}.	
\end{align*}
The exterior metric, or the vacuum part of the solution, is given by (\ref{sds}).
Gauss coordinates to the $r=\mbox{const.}$ hypersurfaces are defined by
\begin{align*}
	d\chi = \frac{dr}{\sqrt{1-\frac{2 M}{r} -\frac{\Lambda}{3}r^{2} }}.
\end{align*}
The functions $\chi=\chi(r)$ and $r=r(\chi)$ are not given explicitly because they
involve elliptic functions. However, one does not need the explicit form to show that the
metric is $C^{1}$ at $\chi_{b}$, where the pressure vanishes. In Gauss coordinates
the exterior metric becomes
 \begin{align}
	ds^{2}=-\left(1-\frac{2 M}{r(\chi)} -\frac{\Lambda}{3}r(\chi)^{2} \right) dt^{2}
        	+d\chi^2 + r(\chi)^{2} d\Omega^{2}.	
\end{align}
Both metrics are joined at $P(\chi=\chi_{b})=0$. Continuity of the metric
at $\chi_{b}$ implies that
\begin{align}
	\frac{\sinh \sqrt{-k} \chi_{b}}{\sqrt{-k}} = r(\chi_{b}).	
	\label{eqn_sinh}
\end{align}
To get the $g_{tt}$-component continuous one has to rescale the time
in the interior. With
\begin{align*}
	t \rightarrow \frac{\left(1-\frac{2 M}{r(\chi_{b})} -\frac{\Lambda}{3}r(\chi_{b})^{2} \right)}
	{\left( \frac{P_{c} + \rho_{0}}{\rho_{0}} \right)^{2}} t,
\end{align*}
both metrics are joined continuously at $\chi_{b}$.
Continuity of the first derivatives imply 
\begin{align}
	\cosh \sqrt{-k} \chi_{b} = \frac{dr}{d\chi}(\chi_{b})=\sqrt{1-\frac{2 M}{r(\chi_{b})} -\frac{\Lambda}{3}r(\chi_{b})}.
	\label{eqn_cosh}
\end{align}
If one shows that the derivatives of the $g_{tt}$-components are continuous at $\chi_{b}$ one is done.
The interior part implies
\begin{align*}
	\frac{d g_{tt}^\mathrm{int}}{d \chi}(\chi_{b}) = 
	2 \left(1-\frac{2 M}{r(\chi_{b})} -\frac{\Lambda}{3}r(\chi_{b}) \right) \frac{P'(\chi_{b})}{\rho_{0}},
\end{align*}
where $P'(\chi_{b})$ can be derived from (\ref{tov_const}) to give
\begin{align*}
	\frac{d P}{d \chi}(\chi_{b}) = -\frac{1}{\sqrt{-k}} \frac{\sinh \sqrt{-k} \chi_{b}}{\cosh \sqrt{-k} \chi_{b}}
	\left(\frac{4\pi}{3} \rho_{0} - \frac{\Lambda}{3} \right) \rho_{0}.
\end{align*}
With (\ref{eqn_cosh}) both equations combine to 
\begin{align}
	\frac{d g_{tt}^\mathrm{int}}{d \chi}(\chi_{b}) = -2 \frac{\sinh \sqrt{-k} \chi_{b}}{\sqrt{-k}}
	\cosh \sqrt{-k} \chi_{b}\left(\frac{4\pi}{3} \rho_{0} - \frac{\Lambda}{3} \right).	
	\label{derivative_int}
\end{align}
For the exterior metric one finds
\begin{align*}
	\frac{d g_{tt}^\mathrm{ext}}{d \chi}(\chi_{b}) = -2r(\chi_{b})
	\left( \frac{M}{r(\chi_{b})^{3}} -\frac{\Lambda}{3} \right) \sqrt{1-\frac{2 M}{r(\chi_{b})} -\frac{\Lambda}{3}r(\chi_{b})}.
\end{align*}
Use (\ref{eqn_sinh}) for the first term and (\ref{eqn_cosh}) for the last term.
Furthermore the mass is given by
\begin{align*}
	M = \frac{4\pi}{3} \rho_{0} \left(\frac{\sinh \sqrt{-k} \chi_{b}}{\sqrt{-k}}\right)^{3} 
	= \frac{4\pi}{3} \rho_{0} r(\chi_{b})^{3}, 
\end{align*}
see (\ref{mass}). Then this derivative evaluated at $\chi_{b}$ reads
\begin{align}
	\frac{d g_{tt}^\mathrm{ext}}{d \chi}(\chi_{b}) = -2 \frac{\sinh \sqrt{-k} \chi_{b}}{\sqrt{-k}}
	\cosh \sqrt{-k} \chi_{b}\left(\frac{4\pi}{3} \rho_{0} - \frac{\Lambda}{3} \right),	
\end{align}
which equals equation (\ref{derivative_int}). Thus the metric is $C^{1}$ at $\chi_{b}$.

Since the density is not continuous at the boundary the Ricci tensor is not, either. 
Therefore the metric is at most $C^{1}$. This cannot be improved.

\subsubsection{Stellar models with spatially Euclidean geometry}
\subsubsection*{$\Lambda =-8 \pi \rho_{0}$}
Assume that cosmological constant and constant density are chosen such that
$8\pi \rho_{0} = -\Lambda$.  Then $k=0$, $m_{\mathrm{q}}(r)=0$ and the denominator of 
(\ref{tov_const}) becomes one.
The differential equation simplifies to
\begin{align}
        \frac{d P}{dr}=-4\pi r\left( P+\rho_{0} \right)^{2},
\label{tov_special}
\end{align}
and the $t=\text{const.}$ hypersurfaces of (\ref{eqn_metric}) are purely Euclidean.
As in the former case metric (\ref{eqn_metric}) is well defined for all $r$.

Equation (\ref{tov_special}) can be integrated. If the constant of integration is fixed at the centre by 
$P(r=0)=P_{c}$ one obtains
\begin{align}
        P(r)=\frac{1}{2\pi r^{2}+\frac{1}{P_{c}+ \rho_{0}}} - \rho_{0}.
\label{pr_special}
\end{align}
The denominator of the pressure distribution cannot vanish because
central pressure and density are assumed to be positive. 
Therefore (\ref{pr_special}) has no singularities.

As the radius tends to infinity the fraction tends to zero and
thus the pressure converges to $-\rho_{0}$.

This implies that there always exits a radius $R$ where $P(r=R)=0$.
Therefore all solutions to (\ref{tov_special}) in (\ref{eqn_metric}) 
describe stellar objects. Their radius is given by 
\begin{align}
	R^{2}=\frac{1}{2\pi} \left( \frac{1}{\rho_{0}} - \frac{1}{P_{c}+\rho_{0}} \right).
	\label{eqn_prpc=}
\end{align}

One finds that the radius $R$ is bounded by $1/\sqrt{2\pi \rho_{0}}$.
Inserting the definition of the density yields to
\begin{align}
	R^{2} < \frac{1}{2 \pi \rho_{0}} = \frac{1}{2 \pi} \frac{4 \pi R^{3}}{3M},
\label{eqn_R<1/2pilambda}
\end{align}
which implies 
\begin{align}
	3 M < 2 R. 
\label{eqn_M<23R}
\end{align}
Thus (\ref{eqn_M<23R}) is the analogous Buchdahl inequality to
equation~(\ref{eqn_buchdahl}) and equals (\ref{eqn_3m<_hyp}) with
$\Lambda=-8\pi \rho_{0}=-6M/R^{3}$.

At $r=R$, where the pressure vanishes, the Schwarzschild-anti-de Sitter
solution (\ref{sds}) is joined uniquely by the same procedure as before.
Since the density is not continuous at the boundary the Ricci tensor is not, either. 
Thus the metric is again at most $C^{1}$.

\begin{figure}[!ht]
\noindent
\begin{minipage}[h]{.46\linewidth}
\centering\epsfig{figure=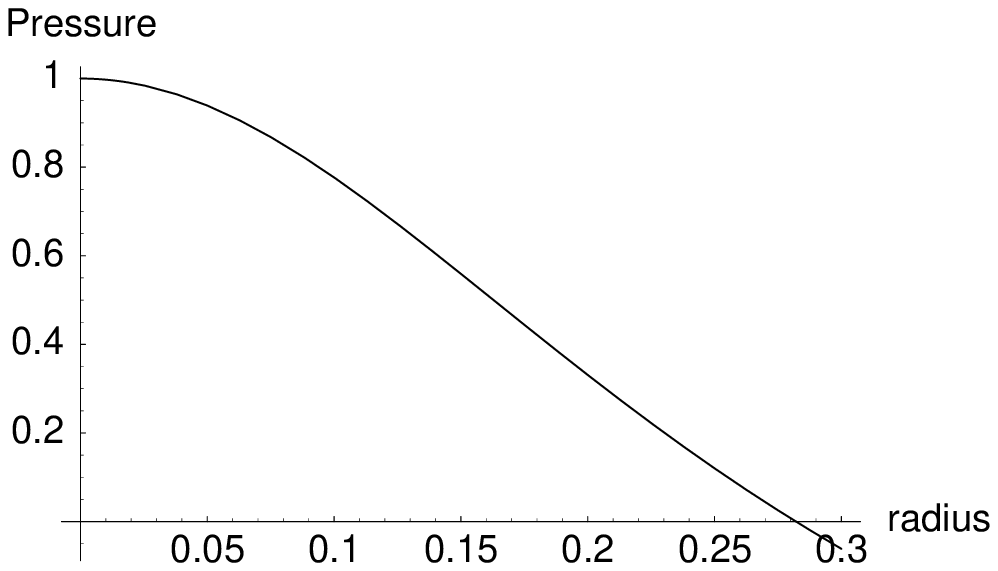,width=\linewidth} 
\caption{Pressure function $P(r)$ with $\rho_{0}=1$, $P_{c}=1$ and
	$\Lambda=-8\pi$}
\label{fig:pressure_euc}
\end{minipage}\hfill
\begin{minipage}[h]{.46\linewidth}
\centering\epsfig{figure=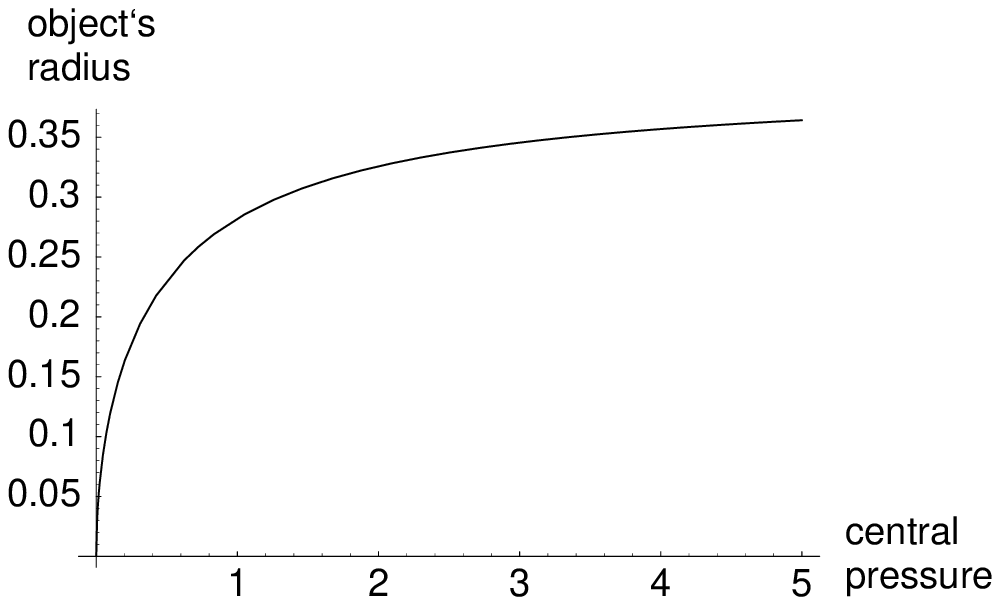,width=\linewidth} 
\caption{Radius as a function of central pressure $R=R(P_{c})$}
\label{fig:r_pc_euc}
\end{minipage}
\end{figure}

Figure~\ref{fig:pressure_euc} shows pressure (\ref{pr_special})
with spatially Euclidean geometry. The radius $R$ where the pressure vanishes
as a function of central pressure $P_{c}$ (\ref{eqn_prpc=}) is show in figure~\ref{fig:r_pc_euc}.
As before $\rho_{0}=1$, $P_{c}=1$.

\newpage

\subsubsection{Stellar models with spatially spherical geometry}
\label{lambda>-8pirho}
\subsubsection*{$\Lambda > -8 \pi \rho_{0}$}
If $\Lambda > -8 \pi \rho_{0}$ then $k$ and $m_{\mathrm{q}}(r)$ are positive.
Equation (\ref{tov_const}) has a singularity at
\begin{align}
        r = \hat{r} = \frac{1}{\sqrt{\frac{8\pi}{3}\rho_{0}+\frac{\Lambda}{3}}}.
\label{eqn_def_hatr}
\end{align}
In $\alpha$ the singularity $\hat{r}$ corresponds to $\alpha(\hat{r}) = \pi /2$.

As already mentioned in section~\ref{subsec_spatial} the spatial part of
metric~(\ref{eqn_metric}) now describes part of a 3-sphere of radius $1/\sqrt{k}$. 
The metric is well defined for radii less than $\hat{r}$.

Integration of (\ref{tov_const}) gives
\begin{align}
        P(\alpha)=\rho_{0}\frac{\left(\frac{\Lambda}{4 \pi \rho_{0}} -1\right)+C\cos \alpha}{3-C\cos \alpha}.
\label{eqn_pr<>}
\end{align}
One finds 
\begin{align}
        P(\pi /2)=\frac{\rho_{0}}{3} \left(\frac{\Lambda}{4 \pi \rho_{0}}-1\right),
\label{eqn_phatr}
\end{align}
which is less than zero if $\Lambda < 4 \pi \rho_{0}$,
which is the restriction in \cite{Stuchlik}.  
Since this is the considered case the singularity 
of the pressure gradient is not important yet.

$C$ is a constant of integration and can be expressed by the central pressure $P_{c}$.
Again one obtains
\begin{align}
        C=\frac{3P_{c} + \rho_{0}  -\frac{\Lambda}{4\pi}}{P_{c}+\rho_{0}}.      
\label{constantC2}
\end{align}
It is the same constant of integration as in (\ref{constantC}). But possible value
for $C$ are different. The considered values of the cosmological constant imply
that $1<C<3$. The pressure is decreasing for all $\alpha$.

It was already stated that  $P(\pi /2) < 0$. 
Thus there exists an $\alpha_{b}$ such that $P(\alpha_{b})=0$.
$\alpha_{b}$ can be derived from equation (\ref{eqn_pr<>}) and one obtains
the analogue of (\ref{eqn_prpc<})
\begin{align}
       \cos \alpha_{b} = \frac{1}{C} \left( 1-\frac{\Lambda}{4 \pi \rho_{0}} \right).
\label{eqn_prpc<>}
\end{align}

It remains to derive the analogue of the Buchdahl inequality for this case. 
One uses (\ref{eqn_prpc<>}) to find $P_{c}=P_{c}(\alpha_{b})$
\begin{align}
        P_{c}=\rho_{0} \frac{\left(1-\frac{\Lambda}{4 \pi \rho_{0}}\right) \left(1-\cos \alpha_{b} \right)}
                {3\cos \alpha_{b} - \left(1-\frac{\Lambda}{4 \pi \rho_{0}}\right)},  
\label{eqn_pcpr<>}
\end{align}
which is similar to (\ref{eqn_prpc<}).
Finiteness of the central pressure in (\ref{eqn_pcpr<>}) gives 
\begin{align}
        \cos \alpha_{b} > \frac{1}{3}\left(1-\frac{\Lambda}{4 \pi \rho_{0}}\right).
        \label{eqn_Buch}
\end{align}
With 
\begin{align*}
	\sin(\arccos(\alpha)) = \sqrt{1-\alpha^{2}}, 
\end{align*}
this yields to the wanted analogue
\begin{align}
        R^{2} < \frac{\frac{1}{3} \left(4-\frac{\Lambda}{4 \pi \rho_{0}} \right)}{4 \pi \rho_{0}},
\label{eqn_buchdahl<>}
\end{align}
which equals (\ref{eqn_R<}). The greater sign reversed because $(\arccos\alpha)$ is decreasing.
Thus it can also be rewritten to give
\begin{align}
	3M < \frac{2}{3}R + R\sqrt{\frac{4}{9}-\frac{\Lambda}{3} R^{2}},
	\label{eqn_3m<_lneg}
\end{align}
which equals~(\ref{eqn_3m<_hyp}). To emphasise that the cosmological
constant is negative one may write $+|\Lambda|$ rather than $-\Lambda$.

At $\alpha=\alpha_{b}$ or $r=R$ the Schwarzschild-anti-de Sitter solution 
is joined and the metric is $C^{1}$. Without further assumptions this cannot be improved.

\begin{figure}
\noindent
\centering\epsfig{figure=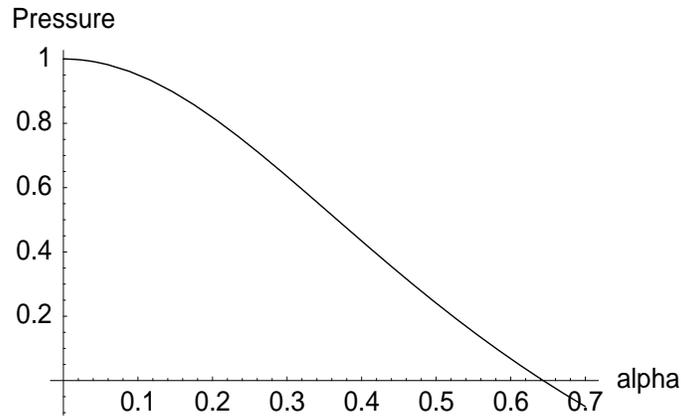,width=9cm,height=6cm} 
\caption{Pressure function $P(\alpha)$ with $\rho_{0}=1$, $P_{c}=1$ and
	$\Lambda=-4\pi$}
\label{fig:press_lneg}
\end{figure}

\begin{figure}
\noindent
\begin{minipage}[h]{.46\linewidth}
\centering\epsfig{figure=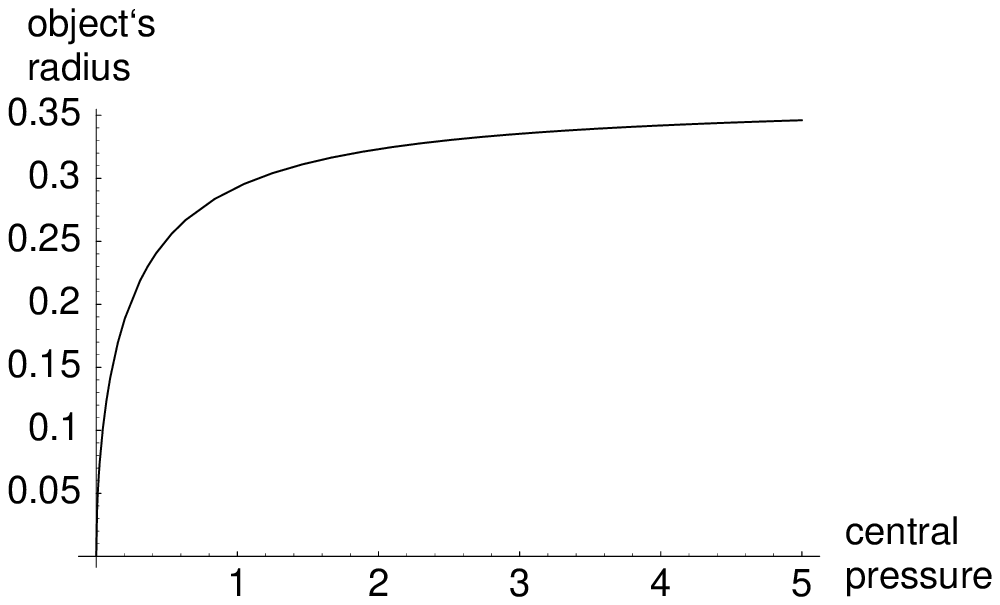,width=\linewidth} 
\caption{Radius as a function of central pressure $R=R(P_{c})$}
\label{fig:r_pc_lneg}
\end{minipage}\hfill
\begin{minipage}[h]{.46\linewidth}
\centering\epsfig{figure=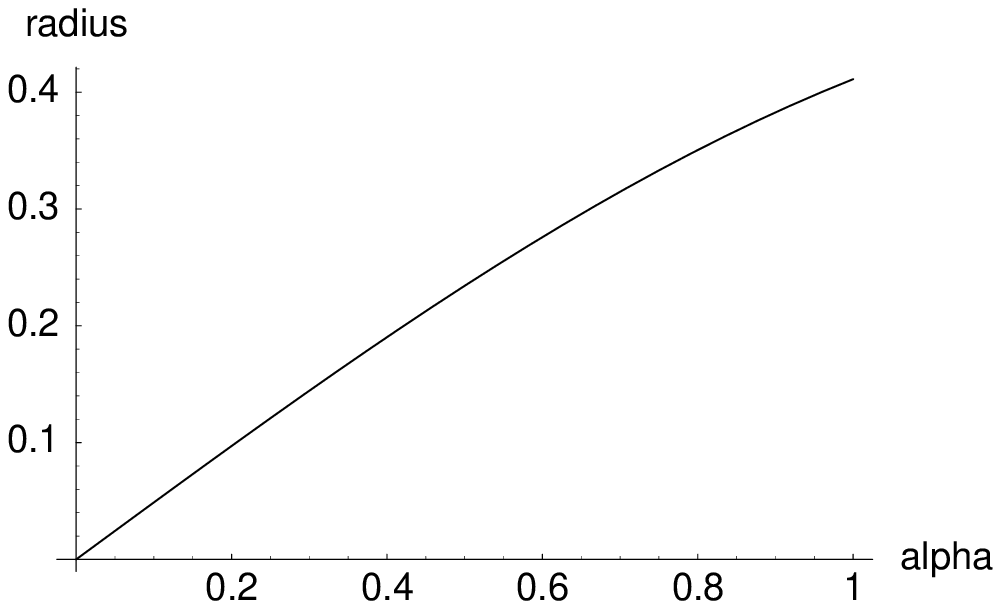,width=\linewidth} 
\caption{Radial coordinate \mbox{$r=r(\alpha)$}}
\label{fig:radius_lneg}
\end{minipage}
\end{figure}

The pressure (\ref{eqn_pr<>}) is shown in figure~\ref{fig:press_lneg}.
Figures~\ref{fig:r_pc_lneg} and~\ref{fig:radius_lneg} show the radius of the
object as a function of the central pressure (\ref{eqn_prpc<>}) and the radius as a function of
the new variable $\alpha$ (\ref{eqn_sine}), respectively. 

\subsection{Solutions with vanishing cosmological constant}
\begin{figure}[!h]
\noindent
\centering\epsfig{figure=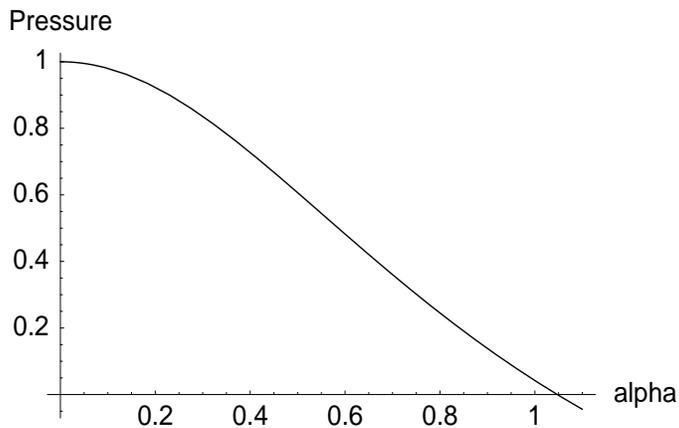,width=9cm,height=6cm} 
\caption{Pressure function $P(\alpha)$ with $\rho_{0}=1$, $P_{c}=1$ and
	$\Lambda=0$}
\label{fig:press_lzero}
\end{figure}

\begin{figure}
\noindent
\begin{minipage}[h]{.46\linewidth}
\centering\epsfig{figure=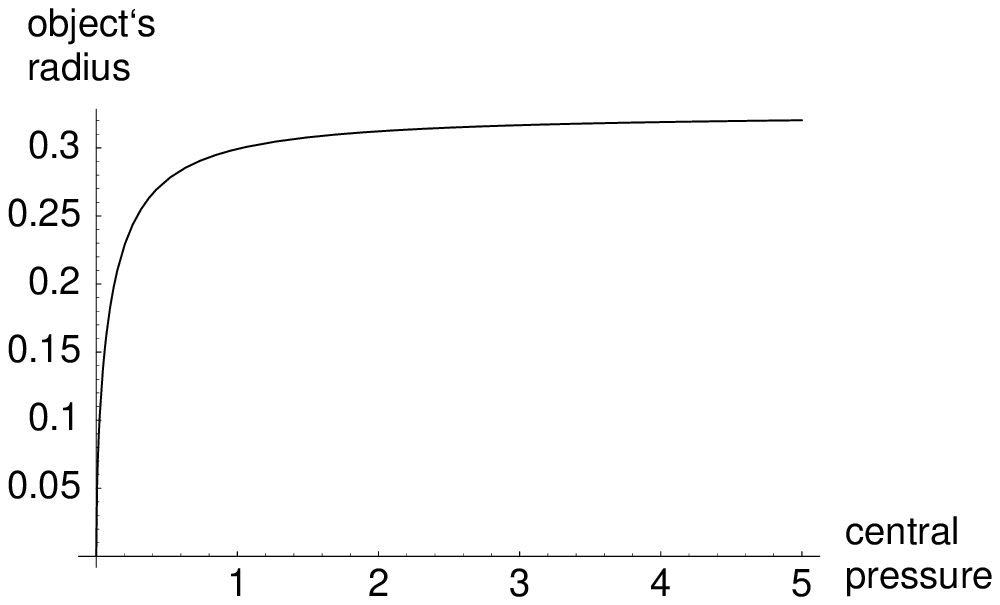,width=\linewidth} 
\caption{Radius as a function of central pressure $R=R(P_{c})$}
\label{fig:r_pc_lzero}
\end{minipage}\hfill
\begin{minipage}[h]{.46\linewidth}
\centering\epsfig{figure=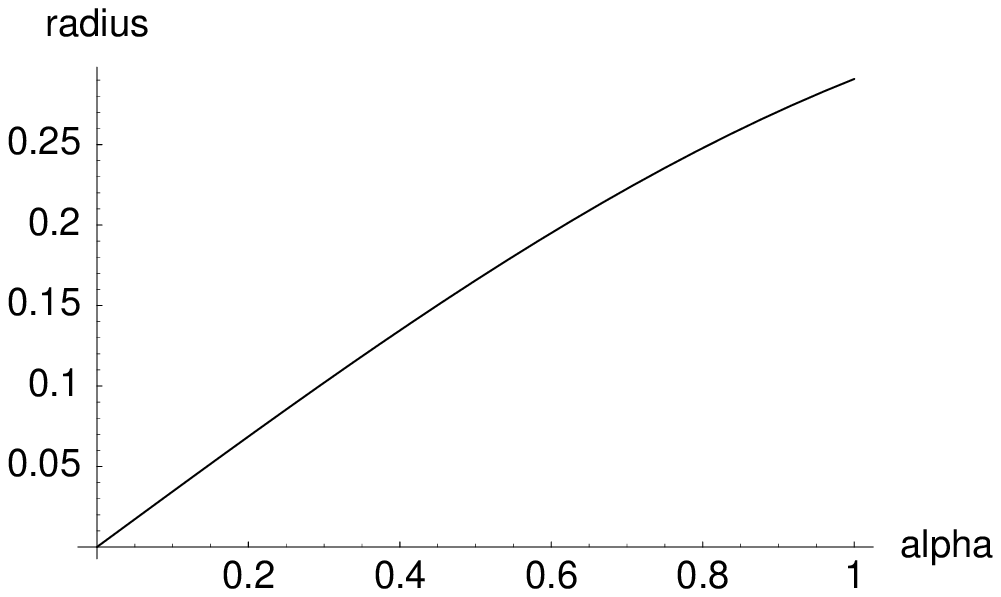,width=\linewidth} 
\caption{Radial coordinate \mbox{$r=r(\alpha)$}}
\label{fig:radius_lzero}
\end{minipage}
\end{figure}

Assume a vanishing cosmological constant. Then one can use all equations of the
former case with $\Lambda=0$. Only one of these relations will be shown, namely the
Buchdahl inequality \cite{Buchdahl:1959}. It is derived from (\ref{eqn_buchdahl<>})
\begin{align}
	R^{2} < \frac{1}{3\pi \rho_{0}},
\end{align}
using $M=(4\pi/3)\rho_{0} R^{3}$ leads to
\begin{align}
	M < \frac{4}{9} R.
	\label{eqn_buchdahl}
\end{align} 
Because of the analogous inequalities of the former cases write 
\begin{align}
	3M < \frac{4}{3} R.
	\label{eqn_buchdahl3m}
\end{align} 
For vanishing cosmological constant the pressure is shown in figure~\ref{fig:press_lzero}.
Figures~\ref{fig:r_pc_lzero} and~\ref{fig:radius_lzero} show the radius of the
object as a function of the central pressure and the radius as a function of
the new variable $\alpha$ (\ref{eqn_sine}), respectively.
The plotted pressure and central pressure are given by  (\ref{eqn_pr<>})
and (\ref{eqn_prpc<>}) with $\Lambda=0$.
 
\newpage

\subsection{Solutions with positive cosmological constant}
In the Schwarzschild-de Sitter spacetime there may exist a black-hole event horizon and there
may also exist a cosmological event horizon. It depends on $M$ and $\Lambda$ in the
Schwarzschild-de Sitter metric~(\ref{sds}) which of the cases occur.
The three possible cases were mentioned in the end of section~\ref{subsec_sadssds}.

\subsubsection{Stellar models with spatially spherical geometry}
\subsubsection*{$0<\Lambda < 4 \pi \rho_{0}$}
Integration of the TOV-$\Lambda$ equation (\ref{tov_const}) leads to
\begin{align}
        P(\alpha) &=\rho_{0}\frac{\left(\frac{\Lambda}{4 \pi \rho_{0}} -1\right)+C\cos \alpha}{3-C\cos \alpha}.
        \label{eqn_pr<>2} 
\end{align}
Because of (\ref{eqn_phatr}) the boundary $P(R)=0$ exists.  
As in section~\ref{lambda>-8pirho} one finds 
\begin{align}
        R^{2} &< \frac{\frac{1}{3} \left(4-\frac{\Lambda}{4 \pi \rho_{0}} \right)}{4 \pi \rho_{0}},
        \label{eqn_buchdahl<>2}
\end{align}
and written in terms of $M$, $R$ and $\Lambda$ again gives
\begin{align}
	3M < \frac{2}{3}R + R\sqrt{\frac{4}{9}-\frac{\Lambda}{3} R^{2}}.
	\label{eqn_3m<_lpos}
\end{align}
Since the cosmological constant is positive the square root term is well
defined if 
\begin{align}
	R \leq \sqrt{\frac{4}{3}} \frac{1}{\sqrt{\Lambda}}.
\end{align}
In this section $0<\Lambda < 4 \pi \rho_{0}$ is assumed. Using the definition
of mass this can be rewritten to give
\begin{align*}
	\Lambda < 4\pi \rho_{0} = \frac{3M}{R^{3}}.
\end{align*}
This leads to
\begin{align}
	9\Lambda M^{2} < \left( \frac{3M}{R} \right)^{3},
	\label{discri}
\end{align}
where the right-hand side of~(\ref{discri}) has an upper bound 
given by~(\ref{eqn_3m<_lpos}). If in addition $R < 1/\sqrt{\Lambda}$
then the right-hand side of~(\ref{discri}) is bounded by one and
the exterior Schwarzschild-de Sitter spacetime has two
horizons, see section~\ref{subsec_sadssds}. Without the
additional assumption the exterior spacetime may also
have one or may also have no horizon. 

The Schwarzschild-de Sitter solution is joined at $r=R$, where \mbox{$P(R)=0$}.
Defining $M=(4\pi/3)\rho_{0}R^{3}$ and rescaling the time in the interior the metric will be continuous at $R$. 
Again Gauss coordinates relative to the hypersurface $P(R)=0$ can be used to get the metric $C^{1}$ at $R$.
Since the density is not continuous at the boundary this cannot be improved. 

One can construct solutions without singularities.
The group orbits are increasing up to $R$. 
The boundary of the stellar object $r=R$ can be put in region $I$ of 
Penrose-Carter figure~\ref{fig: Penrose} where the time-like 
killing vector is future directed. This leads to figure~\ref{fig: Penrose1a},
the spacetime still contains an infinite sequence of singularities $r=0$
and space-like infinities $r=\infty$.
It is possible to put a second object with boundary $r=R$ in region $IV$ of Penrose-Carter
figure~\ref{fig: Penrose1a} where the time-like killing vector is past directed.
This leads to figure~\ref{fig: Penrose2} and shows that there 
are no singularities in the spacetime. This spacetime is not globally
static because of the remaining dynamical parts of the the
Schwarzschild-de Sitter spacetime, regions $II_{C}$ and $III_{C}$.
Penrose-Carter diagrams are in appendix~C.

\begin{figure}
\noindent
\centering\epsfig{figure=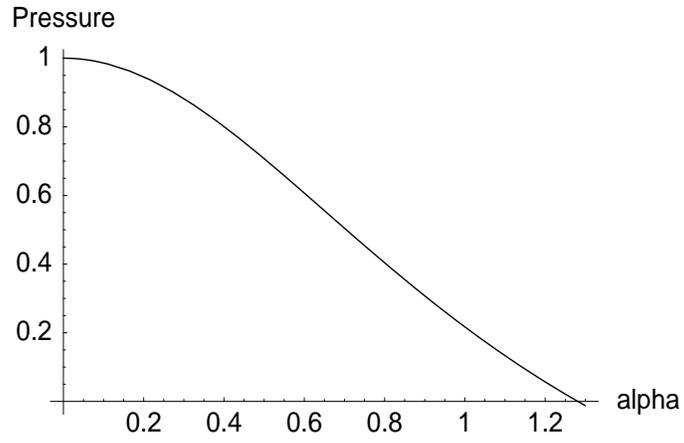,width=9cm,height=6cm} 
\caption{Pressure function $P(\alpha)$ with $\rho_{0}=1$, $P_{c}=1$ and
	$\Lambda=2\pi$}
\label{fig:press_lpos}
\end{figure}

\begin{figure}
\noindent
\begin{minipage}[h]{.46\linewidth}
\centering\epsfig{figure=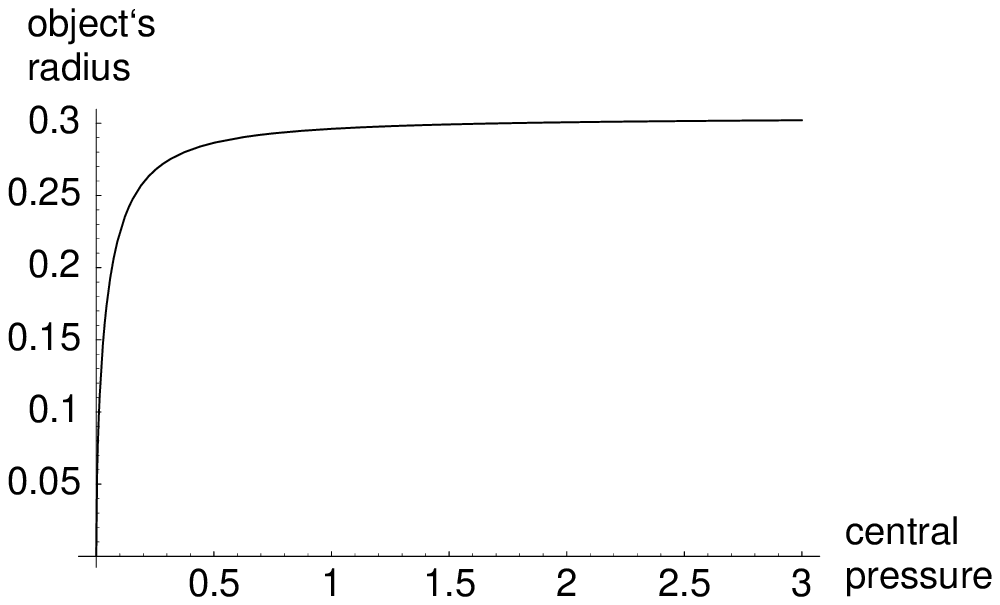,width=\linewidth} 
\caption{Radius as a function of central pressure $R=R(P_{c})$}
\label{fig:r_pc_lpos}
\end{minipage}\hfill
\begin{minipage}[h]{.46\linewidth}
\centering\epsfig{figure=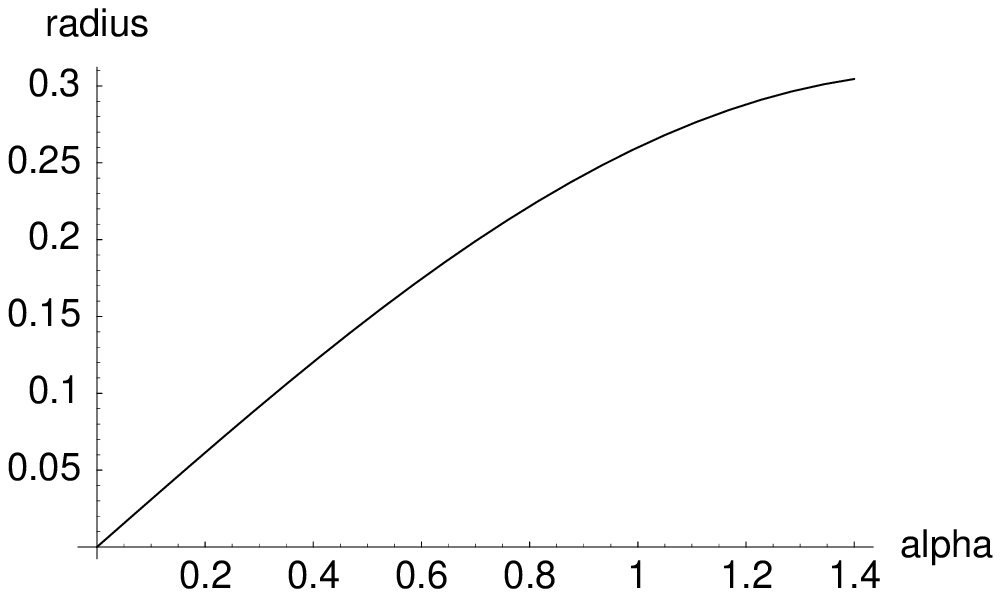,width=\linewidth} 
\caption{Radial coordinate \mbox{$r=r(\alpha)$}}
\label{fig:radius_lpos}
\end{minipage}
\end{figure}

Similarly to the last figures pressure (\ref{eqn_pr<>2}), central pressure (\ref{eqn_prpc<>})
and the new variable $\alpha$ (\ref{eqn_sine}) are shown in 
figure~\ref{fig:press_lpos}, figure~\ref{fig:r_pc_lpos} and 
figure~\ref{fig:radius_lpos}, respectively.

\newpage

\subsubsection{Solutions with exterior Nariai metric}
\label{subsec_ext_nariai}

\subsubsection*{$\Lambda = 4 \pi \rho_{0}$}
In this special case integration of (\ref{tov_const}) gives
\begin{align}
        P(\alpha) = \rho_{0} \frac{C\cos \alpha}{3-C\cos \alpha}.
\label{eqn_pr_4pi}
\end{align}
The pressure (\ref{eqn_pr_4pi}) vanishes at $\alpha = \alpha_{b} = \pi /2$, 
the coordinate singularity in the radial coordinate $r$. 

One would like to join the Schwarzschild-de Sitter metric. 
But with $\Lambda = 4 \pi \rho_{0}$ and $M=(4\pi/3)\rho_{0}R^{3}$ it reads
\begin{align}
	ds^{2}=-\left(1-\frac{3M}{r}\right) dt^{2} + \frac{dr^2}{1-\frac{3M}{r}} + r^{2}d\Omega^{2}.
	\label{sds_3m}
\end{align}
The corresponding radius to $\alpha_{b}$ is $R=3M$. Therefore one would join both solutions 
at the horizon $R=3M$. Since $\Lambda = 4 \pi \rho_{0}$ the radius $R$ is also given
by the cosmological constant $R=1/\sqrt{\Lambda}$. Thus the black-hole event horizon $3M$
and the cosmological event horizon $1/\sqrt{\Lambda}$ are both located at $R$.

The interior metric reads
\begin{align}
          ds^{2}= -\left(1-\frac{P_{c}}{P_{c} + \rho_{0}}\cos \alpha\right)^{2} \left(\frac{P_{c}+\rho_{0}}{\rho_{0}}\right)^{2} dt^{2} 
          + \frac{1}{\Lambda}\left[d\alpha^{2}  + d\Omega^{2}\right].
\label{eqn_interior_4pi}
\end{align}
The volume of group orbits of metric~(\ref{sds_3m}) is increasing whereas
the group orbits of~(\ref{eqn_interior_4pi}) have constant volume. Therefore
it is not possible to join the vacuum solution~(\ref{sds_3m}) on as an 
exterior field to get the metric $C^{1}$ at the $P=0$ surface.

But there is the other spherically symmetric vacuum solution to 
the Einstein field equations with cosmological constant, the 
Nariai solution \cite{Nariai1,Nariai2}, mentioned in chapter~\ref{costov}. 
Its metric is 
\begin{align}
        ds^{2}=\frac{1}{\Lambda}
        \left[-(A \cos \log r + B \sin \log r)^{2} dt^{2} + \frac{1}{r^{2}} dr^{2}  + d\Omega^{2} \right],
\label{eqn_nariai_r}
\end{align}
where $A$ and $B$ are arbitrary constants. With $r=e^{\alpha}$ this becomes
\begin{align}
        ds^{2}=\frac{1}{\Lambda} 
\left[-(A \cos \alpha + B \sin \alpha)^{2} dt^{2} +  d\alpha^{2}  + d\Omega^{2} \right].
\label{eqn_nariai_a}
\end{align}
Metrics (\ref{eqn_interior_4pi}) and (\ref{eqn_nariai_a}) can be joined by fixing the constants $A$ and $B$. With
\begin{align*}
         A = - \frac{P_{c}}{\rho_{0}} \sqrt{\Lambda}, \mbox{\ } B = \frac{P_{c} + \rho_{0}}{\rho_{0}} \sqrt{\Lambda},
\end{align*}
the metric is $C^{1}$ at $\alpha=\pi /2$. As before, since the density is not continuous, the metric
is at most $C^{1}$.

The three figures show $P(\alpha)$, $R=R(P_{c})$ and $r=r(\alpha)$, 
respectively.

\begin{figure}[h]
\noindent
\centering\epsfig{figure=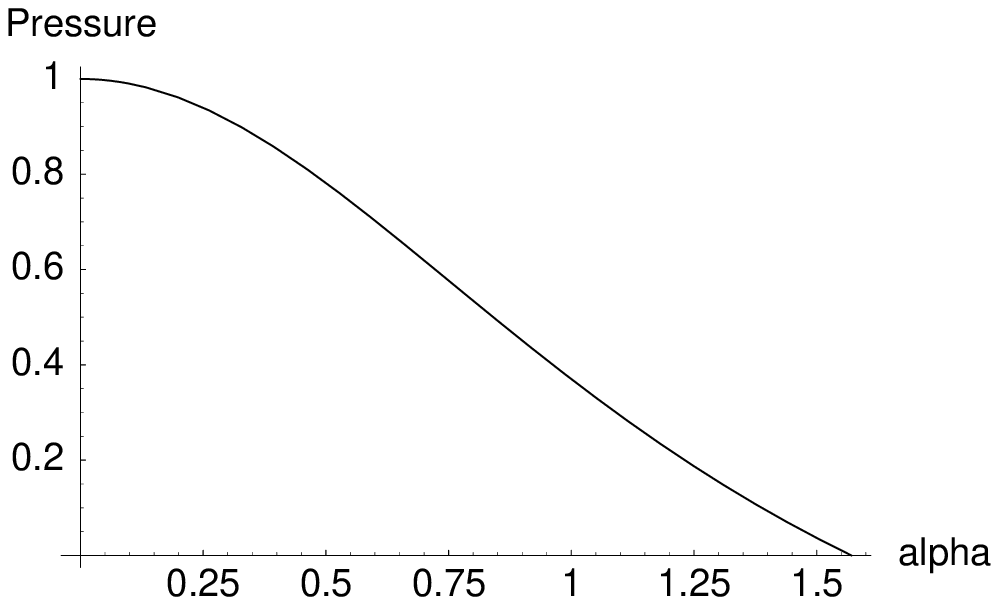,width=9cm,height=6cm} 
\caption{Pressure function $P(\alpha)$ with $\rho_{0}=1$, $P_{c}=1$ and
	$\Lambda=4\pi$}
\label{fig:press_nariai}
\end{figure}

\begin{figure}[h]
\noindent
\begin{minipage}[h]{.46\linewidth}
\centering\epsfig{figure=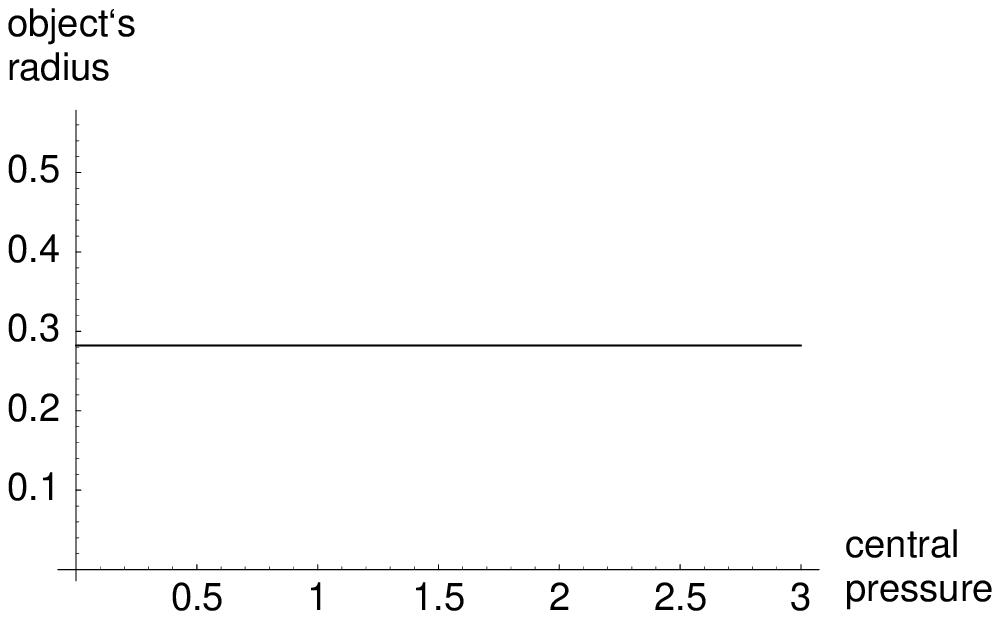,width=\linewidth} 
\caption{Radius as a function of central pressure $R=R(P_{c})$}
\label{fig:r_pc_nariai}
\end{minipage}\hfill
\begin{minipage}[h]{.46\linewidth}
\centering\epsfig{figure=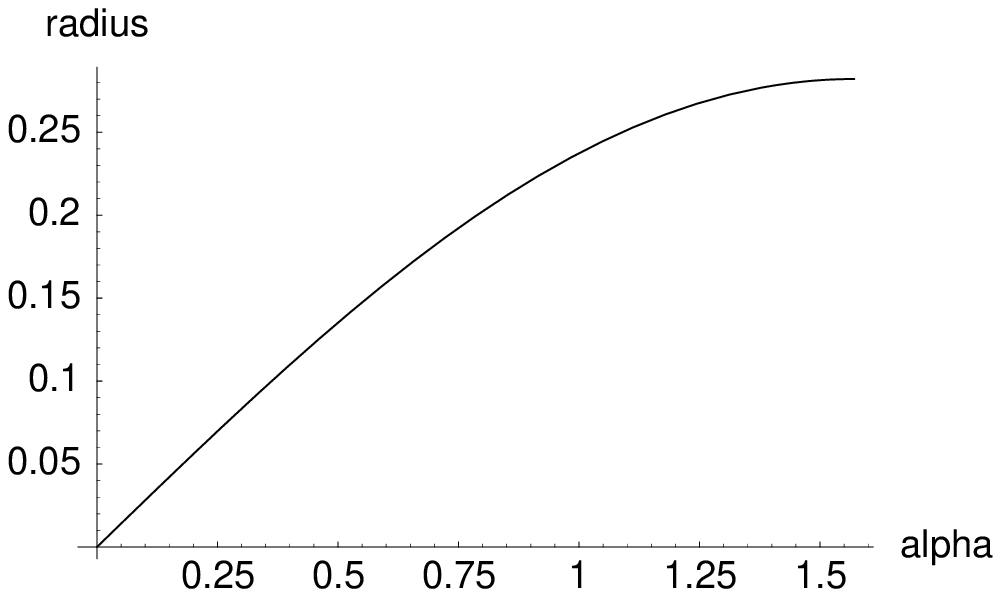,width=\linewidth} 
\caption{Radial coordinate \mbox{$r=r(\alpha)$}}
\label{fig:radius_nariai}
\end{minipage}
\end{figure}

\newpage

\subsubsection{Solutions with decreasing group orbits at the boundary}
\label{subsec_ext_sing}

\subsubsection*{$4 \pi \rho_{0} < \Lambda < \Lambda_{0}$}
Integration of (\ref{tov_const}) gives the pressure
\begin{align}
        P(\alpha) &=\rho_{0}\frac{\left(\frac{\Lambda}{4 \pi \rho_{0}} -1\right)+C\cos \alpha}{3-C\cos \alpha}.
	\label{eqn_press_lvanish}
\end{align}
Assume that the pressure vanishes before the second centre $\alpha=\pi$ of
the 3-sphere is reached. The condition $P(\alpha=\pi)<0$ leads to an upper bound for the cosmological
constant. This bound is given by
\begin{align}
        \Lambda_{0} := 4 \pi \rho_{0} \left( \frac{4 \, P_{c}/\rho_{0} +2} {P_{c}/\rho_{0}+2} \right).
\end{align}
Then $4 \pi \rho_{0} < \Lambda < \Lambda_{0}$ implies the following:

The pressure is decreasing near the centre and vanishes for some $\alpha_{b}$, where 
\mbox{$\pi /2 < \alpha_{b} < \pi$}. Equations (\ref{eqn_sigma}) and
(\ref{eqn_sine}) imply that the volume of group orbits is decreasing if $\alpha > \pi /2$. 

At $\alpha_{b}$ one uniquely joins the Schwarzschild-de Sitter solution by $M=(4\pi/3)\rho_{0}R^{3}$. With
Gauss coordinates relative to the $P(\alpha_{b})=0$ hypersurface the metric will be $C^{1}$. But there is a crucial
difference to the former case with exterior Schwarzschild-de Sitter solution. 
Because of the decreasing group orbits at the boundary there is still the singularity $r=0$ in the
vacuum spacetime. Penrose-Carter diagram~\ref{fig: Penrose3} shows this interesting solution.  

\begin{figure}
\noindent
\centering\epsfig{figure=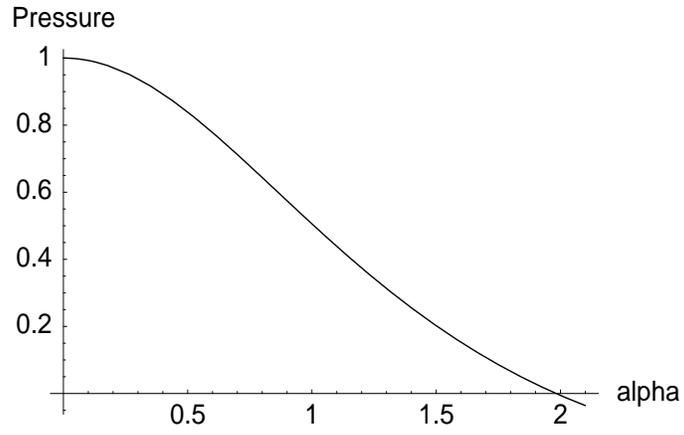,width=9cm,height=6cm} 
\caption{Pressure function $P(\alpha)$ with $\rho_{0}=1$, $P_{c}=1$ and
	$\Lambda=6\pi$}
\label{fig:press_lvanish}
\end{figure}

\begin{figure}
\noindent
\begin{minipage}[h]{.46\linewidth}
\centering\epsfig{figure=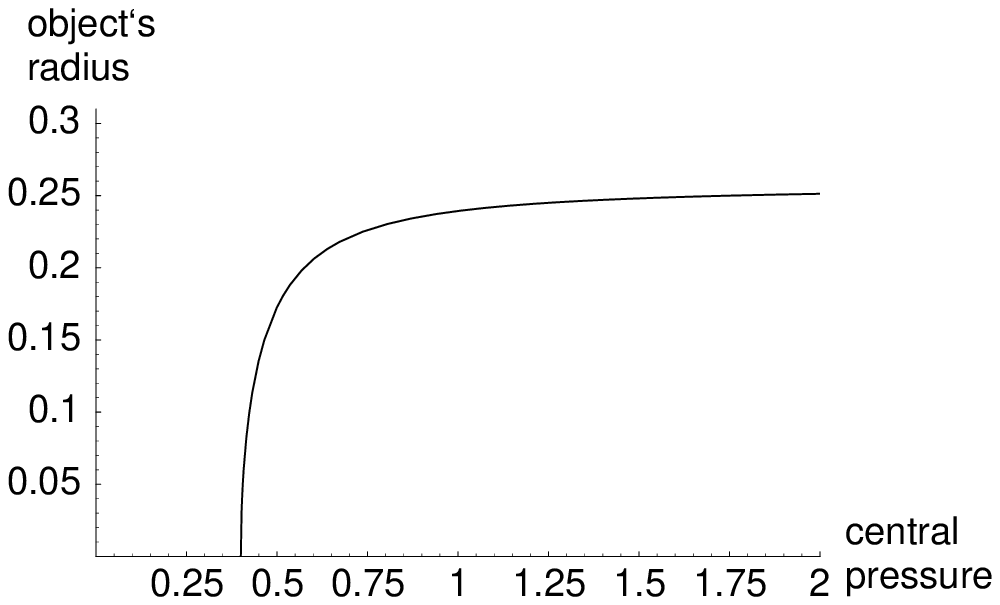,width=\linewidth} 
\caption{Radius as a function of central pressure $R=R(P_{c})$}
\label{fig:r_pc_lvanish}
\end{minipage}\hfill
\begin{minipage}[h]{.46\linewidth}
\centering\epsfig{figure=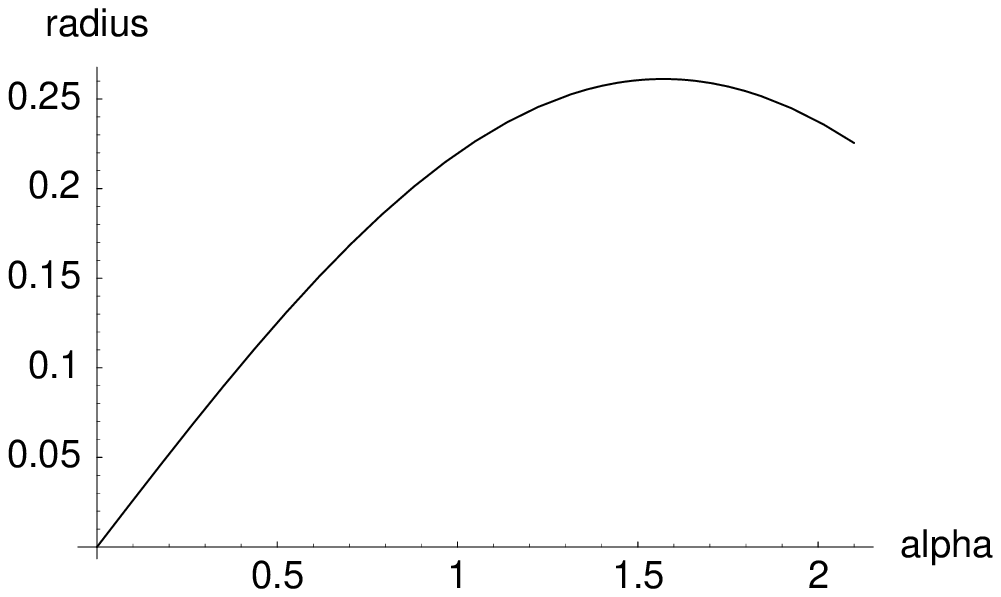,width=\linewidth} 
\caption{Radial coordinate \mbox{$r=r(\alpha)$}}
\label{fig:radius_lvanish}
\end{minipage}
\end{figure}

Figure~\ref{fig:press_lvanish} shows pressure (\ref{eqn_press_lvanish}).
Figures~\ref{fig:r_pc_lvanish} and~\ref{fig:radius_lvanish} show the radius of the
object as a function of the central pressure and the radius as a function of
the new variable $\alpha$, respectively. Constant density and central pressure 
are both set to one, $\rho_{0}=1$, $P_{c}=1$.  

\newpage

\subsubsection{Decreasing solutions with two regular centres}
\label{subsec_2centres_a}

\subsubsection*{$\Lambda_{0} \leq \Lambda < \Lambda_{E}$}
As before the pressure is given by  
\begin{align*}
        P(\alpha) &=\rho_{0}\frac{\left(\frac{\Lambda}{4 \pi \rho_{0}} -1\right)+C\cos \alpha}{3-C\cos \alpha}.
\end{align*}
Assume that the pressure is decreasing near the first centre $\alpha=0$.
This gives an upper bound of the cosmological constant
\begin{align}
	\Lambda_{E} := 4 \pi \rho_{0} \left( 3 \, P_{c}/\rho_{0} +1 \right),
\label{lambdaE}
\end{align}
where $\Lambda_{E}$ is the cosmological constant of the Einstein static universe.
These possible values $\Lambda_{0} \leq \Lambda < \Lambda_{E}$ imply:

The pressure is decreasing is near the first centre $\alpha=0$ but remains
positive for all $\alpha$ because $\Lambda \geq \Lambda_{0}$. 
Therefore there exists a second centre at $\alpha=\pi$. At the second
centre of the 3-sphere the pressure becomes
\begin{align}
	P(\alpha=\pi) = \rho_{0} \frac{ \left( \frac{\Lambda}{4 \pi \rho_{0}} -1 \right) -C}{3 + C}.
	\label{eqn_p_sec_centre}
\end{align}
It only vanishes if $\Lambda = \Lambda_{0}$. The solution is inextendible. 
The second centre is also regular. This is easily shown with Gauss coordinates. Recall metric
(\ref{eqn_metric}) written in $\alpha$ 
\begin{align*}
        ds^{2}=-\left(\frac{P_{c}+\rho_{0}}{P(\alpha)+\rho_{0}}\right)^{2} dt^{2}
        + \frac{1}{k} \left( d\alpha^2 + \sin^{2}\alpha d\Omega^{2} \right).
\end{align*}
So one already used nearly Gauss coordinates up to a rescaling. 
With \mbox{$\chi = \alpha/\sqrt{k}$} this becomes
\begin{align}
        ds^{2}=-\left(\frac{P_{c}+\rho_{0}}{P(\chi)+\rho_{0}}\right)^{2} dt^{2}
        + d\chi^2 + \left(\frac{\sin \sqrt{k} \chi}{\sqrt{k}} \right)^{2} d\Omega^{2}.
\label{eqn_metric_gauss}
\end{align}
Thus the radius in terms of the Gauss coordinate is
\begin{align}
	r(\chi) = \frac{1}{\sqrt{k}} \sin(\sqrt{k} \chi),
\end{align}
and both centres are regular because 
\begin{align}
	\frac{d}{d\chi} r(\chi) = \cos(\sqrt{k} \chi) = \pm 1 \mbox{\ for\ } \sqrt{k}\chi=0, \pi.
\end{align} 

Solutions of this kind are generalisations of the Einstein static universe. These 3-spheres
have a homogenous density but do not have constant pressure. They have a given central
pressure $P_{c}$ at the first regular centre which decreases monotonically towards the second 
regular centre. Generalisations of the Einstein static universe have been published earlier
in Ref.~\cite{Ibrahim:1976}.\footnote{I would like to thank Aysel 
Karafistan for bringing this reference to my attention.}

\begin{figure}[h]
\noindent

\begin{minipage}[h]{.46\linewidth}
\centering\epsfig{figure=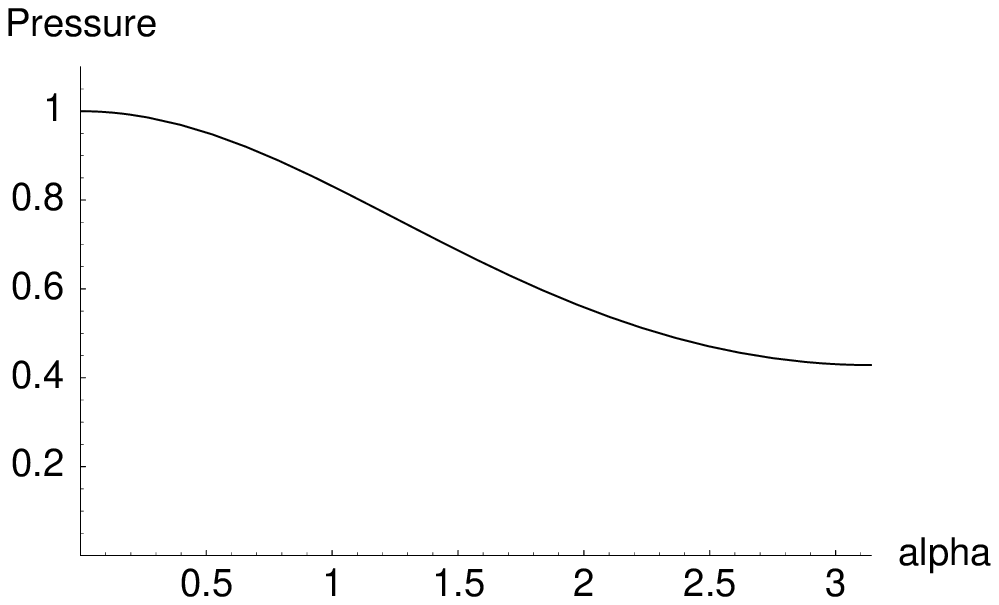,width=\linewidth} 
\caption{Pressure function $P(\alpha)$ with $\rho_{0}=1$, $P_{c}=1$ and
	$\Lambda=12\pi$}
\label{fig:press_2ca}
\end{minipage}\hfill
\begin{minipage}[h]{.46\linewidth}
\centering\epsfig{figure=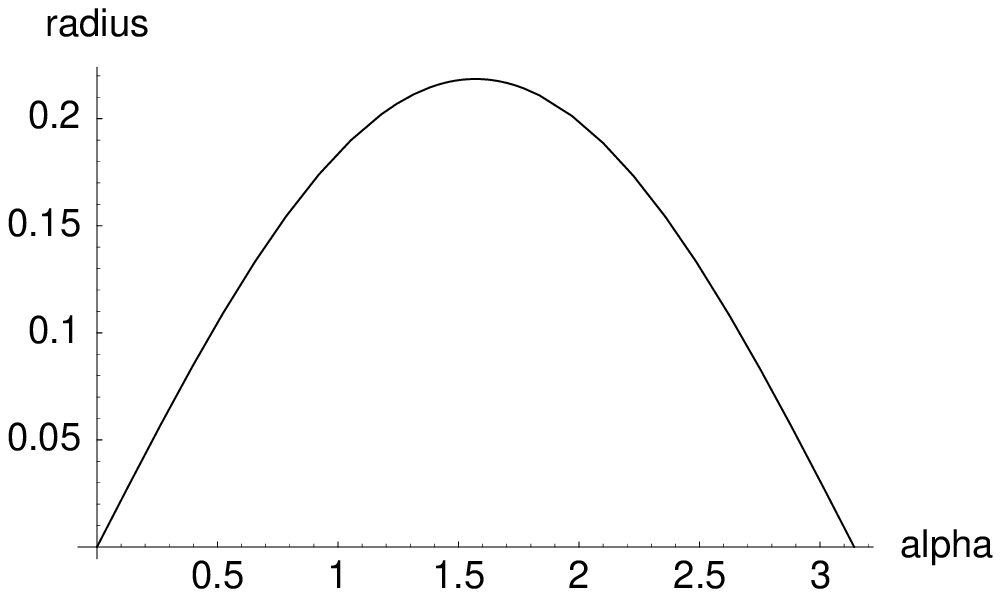,width=\linewidth} 
\caption{Radial coordinate \mbox{$r=r(\alpha)$}}
\label{fig:radius_2ca}
\end{minipage}

\end{figure}

The pressure is plotted in figure~\ref{fig:press_2ca} with $\rho_{0}=1$ and $P_{c}=1$.
Figure~\ref{fig:radius_2ca} shows the radius as a function of $\alpha$.

\newpage

\subsubsection{The Einstein static universe}
\label{subsec_einstein}

\subsubsection*{$\Lambda=\Lambda_{E}$}
Assume a constant pressure function. Then the pressure gradient vanishes and
therefore the right-hand side of (\ref{tov_const}) has to vanish. This gives 
an unique relation between density, central pressure and cosmological constant.
One obtains (\ref{lambdaE})
\begin{align*}
        \Lambda = \Lambda_{E} = 4\pi \left( 3P_{E} + \rho_{0} \right),
\end{align*}
where $P_{E}=P_{c}$ was used to emphasise that the given central pressure corresponds
to the Einstein static universe and is homogenous. 
Its metric is given by
\begin{align*}
        ds^{2}=-dt^{2}+ \mathcal{R}_{E} \left( d\alpha^{2}+ \sin^{2} \alpha d\Omega^{2} \right),
\end{align*}
where 
\begin{align*}
        \mathcal{R}_{E}=\frac{1}{\sqrt{4\pi (P_{E}+\rho_{E})}}.
\end{align*}
$\mathcal{R}_{E}$ is the radius of the three dimensional hyper-sphere $t=\text{const.}$
The time component was rescaled to $1$.

Thus for a given density $\rho_{0}$ there exists for every choice of a central pressure $P_{c}$ a unique cosmological
constant given by (\ref{lambdaE}) such that an Einstein static universe is the solution to (\ref{tov_const}).

\begin{figure}[h]
\noindent

\begin{minipage}[h]{.46\linewidth}
\centering\epsfig{figure=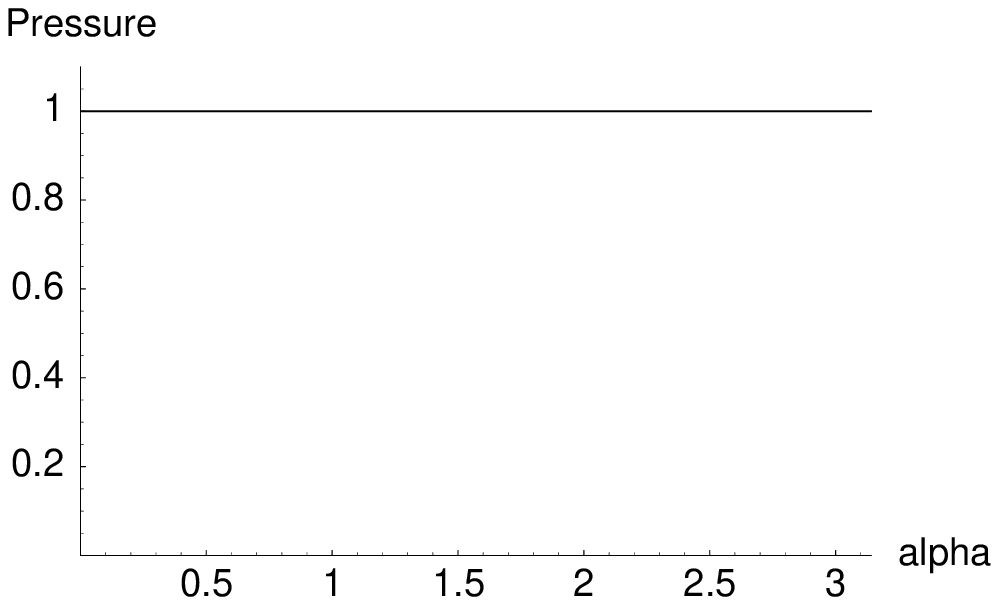,width=\linewidth} 
\caption{Pressure function $P(\alpha)$ with $\rho_{0}=1$, $P_{c}=1$ and
	$\Lambda=16\pi$}
\label{fig:press_einstein}
\end{minipage}\hfill
\begin{minipage}[h]{.46\linewidth}
\centering\epsfig{figure=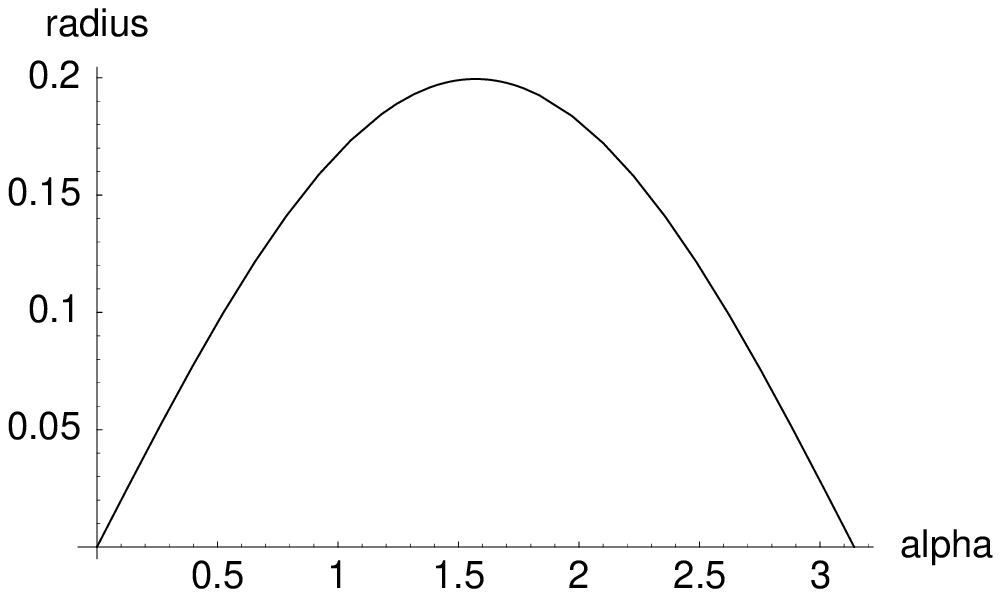,width=\linewidth} 
\caption{Radial coordinate \mbox{$r=r(\alpha)$}}
\label{fig:radius_einstein}
\end{minipage}

\end{figure}

The constant pressure of the Einstein static universe is plotted 
in figure~\ref{fig:press_einstein} with $\rho_{0}=1$ and $P_{c}=1$.
Figure~\ref{fig:radius_einstein} shows the radius as a function of $\alpha$.
Solutions with two centres all have the same radial coordinate $r=r(\alpha)$.

\subsubsection{Increasing solutions with two regular centres}
\label{subsec_2centres_b}

\subsubsection*{$\Lambda_{E} < \Lambda < \Lambda_{S}$}
Integration of (\ref{tov_const}) again leads to
\begin{align}
	P(\alpha) &=\rho_{0}\frac{\left(\frac{\Lambda}{4 \pi \rho_{0}} -1\right)+C\cos \alpha}{3-C\cos \alpha}.
	\label{eqn_p_inc}
\end{align}
Assume that the pressure is finite at the second centre.
This leads to an upper bound of the cosmological constant defined by
\begin{align}
        \Lambda_{S} := 4\pi \rho_{0} \left( 6 \, P_{c}/\rho_{0} + 4 \right).
\end{align}
The possible values of the cosmological constant imply:

The pressure $P(\alpha)$ is increasing near the first regular centre. It
increases monotonically up to $\alpha = \pi$, where one has a second regular centre.
This situation is similar to the case where  
\mbox{$\Lambda_{0} \leq \Lambda < \Lambda_{E}$}.
These solutions are also describing generalisations of the 
Einstein static universe.

More can be concluded. The generalisations are symmetric with
respect to the Einstein static universe. By symmetric one means
the following. Instead of writing the pressure as a function of $\alpha$
depending on the given values $\rho_{0}$, $P_{c}$ and $\Lambda$ 
one can eliminate the cosmological constant with the pressure at 
the second centre $P(\alpha=\pi)$, given by (\ref{eqn_p_sec_centre}). 
Let $P_{c1}$ and $P_{c2}$ denote the pressures of the first and second centre, 
respectively. Then the pressure can be written as
\begin{align}
	P(\alpha)[P_{c1},P_{c2}]=
	\rho_{0} \frac{2P_{c1} P_{c2} \frac{1}{\rho_{0}}+ (P_{c1}+P_{c2}) + (P_{c1}-P_{c2}) \cos \alpha}
	{2\rho_{0} +  (P_{c1}+P_{c2}) - (P_{c1}-P_{c2}) \cos \alpha}.
\end{align}
This implies
\begin{align}
	P(\frac{\pi}{2}+\alpha)[P_{c1},P_{c2}]=P(\frac{\pi}{2}-\alpha)[P_{c2},P_{c1}].
\end{align}
Thus the pressure is symmetric to $\alpha=\pi/2$ if both central pressures
are exchanged and therefore this is the converse situation to the case
where $\Lambda_{0} \leq \Lambda < \Lambda_{E}$.

Another way of looking at it is to consider the pressure in terms of
effective values, mentioned in section~\ref{subsec_tovl}.
Then the pressure reads
\begin{align}
	P^{\mathrm{eff}}(\alpha) = \rho_{0}^{\mathrm{eff}} \frac{-1+C^{\mathrm{eff}} \cos \alpha}{3-C^{\mathrm{eff}} \cos \alpha},
	\label{eqn_peff}
\end{align}
where
\begin{align}
	C^{\mathrm{eff}} = 
	\frac{3 P_{c}^{\mathrm{eff}}+ \rho_{0}^{\mathrm{eff}}}{P^{\mathrm{eff}}_{c} + \rho_{0}^{\mathrm{eff}}}.
\end{align}
The effective pressure at the second centre is given by $P^{\mathrm{eff}}_{c2} = P^{\mathrm{eff}}(\alpha=\pi)$.
Writing $P^{\mathrm{eff}}_{c1}$ for the effective pressure at the fist centre and
use (\ref{eqn_peff}) one can express $\rho_{0}^{\mathrm{eff}}$ by the two effective
central pressures. Since this leads to an equation quadratic in $\rho_{0}^{\mathrm{eff}}$
there exist two effective constant densities for given effective central pressures.
Therefore the decomposition of the effective energy-momentum tensor~(\ref{eqn_teff})
in perfect fluid part and cosmological constant part is not unique. 

If the central pressures are equal the dependence on $\alpha$ 
vanishes and one is left with the Einstein static universe.

\begin{figure}[!h]
\noindent

\begin{minipage}[h]{.46\linewidth}
\centering\epsfig{figure=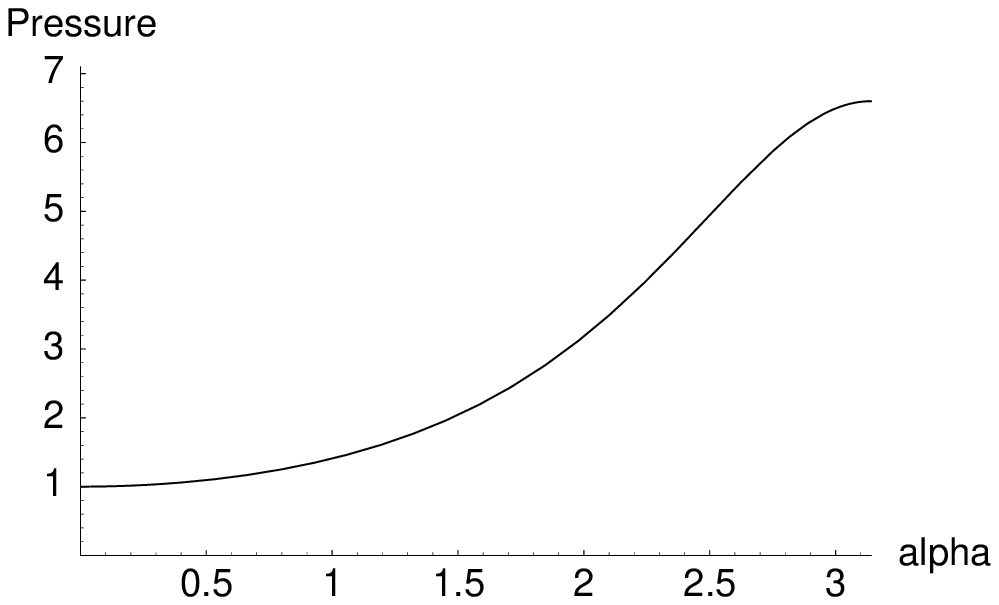,width=\linewidth} 
\caption{Pressure function $P(\alpha)$ with $\rho_{0}=1$, $P_{c}=1$ and
	$\Lambda=30\pi$}
\label{fig:press_2cb}
\end{minipage}\hfill
\begin{minipage}[h]{.46\linewidth}
\centering\epsfig{figure=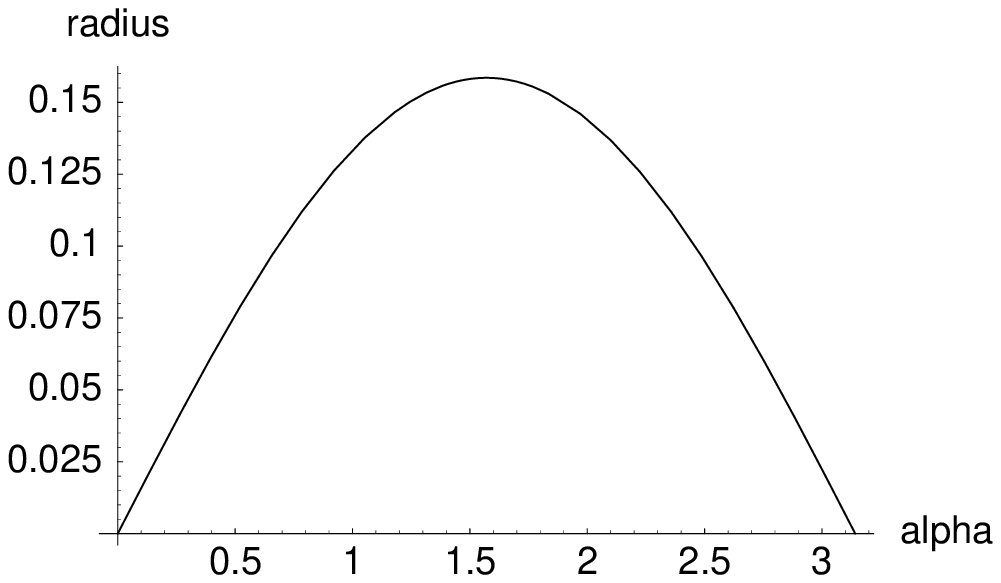,width=\linewidth} 
\caption{Radial coordinate \mbox{$r=r(\alpha)$}}
\label{fig:radius_2cb}
\end{minipage}

\end{figure}

The pressure of solutions with increasing pressure near the first centre
is shown in figure~\ref{fig:press_2cb}. The radial coordinate is plotted
in figure~\ref{fig:radius_2cb}, see the last two cases.

\newpage

\subsubsection{Solutions with geometric singularity}
\label{subsec_sing}

\subsubsection*{$\Lambda \geq \Lambda_{S} $}
In this case it is assumed that $\Lambda$ exceeds the upper limit $\Lambda_{S}$.
Then (\ref{eqn_p_inc}) implies that the pressure is increasing near the centre and
diverges before $\alpha = \pi$ is reached. In appendix~B it is
shown that the divergence of the pressure implies divergence of the squared
Riemann tensor (\ref{Riemann_const}). Thus these solutions have a geometric singularity with unphysical
properties. Therefore they are of no further interest.  

\begin{figure}[!ht]
\noindent
\centering\epsfig{figure=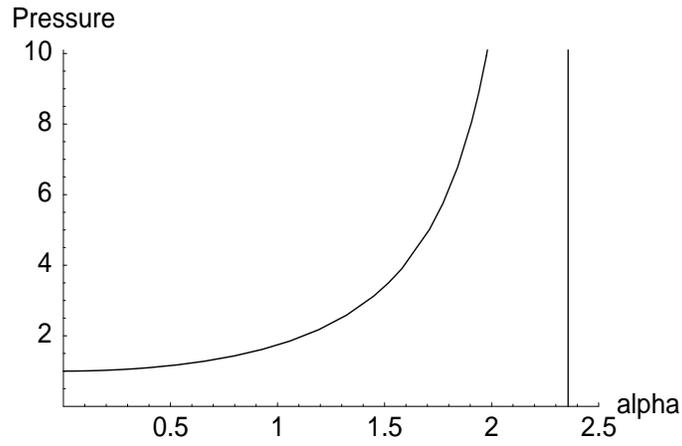,width=9cm,height=6cm} 
\caption{Pressure function $P(\alpha)$ with $\rho_{0}=1$, $P_{c}=1$ and
	$\Lambda=50\pi$}
\label{fig:press_sing}
\end{figure}

Figure~\ref{fig:press_sing} shows the divergent pressure.

\newpage
\subsection{Overview of constant density solutions}
\begin{tabular}[t]{|c|c|p{.55\linewidth}|} \hline
cosmological & spatial & short description of the solution\\ 
constant & geometry & \mbox{} \\ \hline
$\Lambda<-8\pi \rho_{0}$ 		& hyperboloid 	& stellar model with exterior Schwarzschild-anti-de Sitter solution \\
\mbox{} 				& \mbox{} 	& \mbox{} \\
$\Lambda=-8\pi \rho_{0}$		& Euclidean	& stellar model with exterior Schwarzschild-anti-de Sitter solution \\
\mbox{} 				& \mbox{} 	& \mbox{} \\
$-8\pi \rho_{0}<\Lambda<0$		& 3--sphere	& stellar model with exterior Schwarzschild-anti-de Sitter solution  \\ 
\mbox{} 				& \mbox{} 	& \mbox{} \\ \hline
$\Lambda =0 $			& 3--sphere	& stellar model with exterior Schwarzschild solution  \\ 
\mbox{} 				& \mbox{} 	& \mbox{} \\ \hline
$0<\Lambda<4\pi \rho_{0}$		& 3--sphere	& stellar model with exterior Schwarzschild-de Sitter solution  \\
\mbox{} 				& \mbox{} 	& \mbox{} \\
$\Lambda=4\pi \rho_{0}$		& 3--sphere	& stellar model with exterior Nariai solution  \\
\mbox{} 				& \mbox{} 	& \mbox{} \\
$4\pi \rho_{0}<\Lambda<\Lambda_{0} $	& 3--sphere	& decreasing pressure model with exterior Schwarzschild-de Sitter solution; the group orbits are decreasing at the boundary			\\
\mbox{} 				& \mbox{} 	& \mbox{} \\
$\Lambda_{0}\leq\Lambda<\Lambda_{E}$	& 3--sphere	& solution with two centres, pressure decreasing near the first; generalisation of the Einstein static universe 			\\	
\mbox{} 				& \mbox{} 	& \mbox{} \\
$\Lambda=\Lambda_{E}$		& 3--sphere	& Einstein static universe			\\
\mbox{} 				& \mbox{} 	& \mbox{} \\
$\Lambda_{E}<\Lambda<\Lambda_{S}$	& 3--sphere	& solution with two centres, solution increasing near the first; generalisation of the Einstein static universe 	\\
\mbox{} 				& \mbox{} 	& \mbox{} \\
$\Lambda \geq \Lambda_{S}$		& 3--sphere	& increasing pressure solution with geometric singularity \\ \hline
\end{tabular}

%% file: existence-and.tex
\section{Solutions with given equation of state}
\label{exis}

This chapter  analyses the system of differential equations 
for a given monotonic equation of state. The choice of central pressure 
and central density together with the cosmological constant uniquely 
determines the pressure. The chapter is based on \cite{RendallSchmidt}, 
where the following analysis was done without cosmological term.
Geometrised units where $c^{2}=1/\lambda=1$ are used.

The existence of a global solution for cosmological constants 
satisfying $\Lambda < 4\pi \rho_{b}$ is shown, where $\rho_{b}$
denotes the boundary density.

It turns out that solutions with negative cosmological constant
are similar to those with vanishing cosmological constant.
For positive cosmological constants new properties arise.

In section~\ref{sub_buchdahl} Buchdahl variables are introduced. These 
new variables will help to proof the existence of global solutions and
of stellar models. They are also helpful to derive analogous Buchdahl inequalities.    

Section~\ref{sub_existence}  shows the existence of a regular
solution at the centre. After remarks on extending the solution
and a possible coordinate singularity the existence of global solutions
is shown in section~\ref{sub_global}.

For stellar models an analogous Buchdahl inequality is derived. 
Solutions without singularities are constructed. In particular
for positive cosmological constants this is interesting because
a second object has to be put in the spacetime to get
a singularity free solution.

The chapter ends with some remarks on finiteness of the radius.

\subsection{Buchdahl variables}
\label{sub_buchdahl}
In chapter~\ref{costov} metric function $\nu(r)$ was eliminated in (\ref{b}) 
and (\ref{c}) to give the TOV-$\Lambda$ equation.  Alternatively 
the pressure can be eliminated in the same equations. To do 
this Buchdahl \cite{Buchdahl:1959} originally introduced the 
following variables:
\begin{align*}
	y^{2} & = e^{-a(r)}=1-2 w(r)r^{2} \\
	\zeta & = e^{\nu/2} \\
	    x & =r^{2}.
\end{align*}
It seems to be natural to do exactly the same, but using the new 
expression for $e^{-a(r)}$ which involves the cosmological constant. 
Alternatively it can be seen as $e^{-a(r)}$ written in effective values.
The new Buchdahl variables are
\begin{align}
	y^{2} & = e^{-a(r)}=1-2w(r)r^{2}-\frac{\Lambda}{3}r^{2} \\
	\zeta & = e^{\nu/2} \label{zeta}	\\
	    x & =r^{2}.     	
\end{align}
The derivation of the equation one looks for is unpleasant. Anyway 
it is a straight forward calculation only involving basic algebra. 
Nonetheless the derivation is shown in detail because of the interesting 
result: The contribution of the cosmological constant is fully absorbed by the
new Buchdahl variable $y$, no other terms involving $\Lambda$ 
appear in the equation.

Start with the second field equation (\ref{b}). Putting in the new variables leads to
\begin{align*}
	y^{2}+y^{2}\sqrt{x} \, 2 \sqrt{x} \, \nu_{,x}-1 = \left(8\pi P -\Lambda \right)x. 
\end{align*}
With
\begin{align*}
	\frac{y^{2}-1}{x}& =-2w - \frac{\Lambda}{3},
\end{align*}
this gives
\begin{align}
	8\pi P-\frac{2}{3}\Lambda=4y^{2}\frac{\zeta_{,x}}{\zeta}-2w,
	\label{PP}
\end{align}
where $ \frac{d}{dr}=2\sqrt{x}\frac{d}{dx} $ was used. The term $\frac{\zeta_{,x}}{\zeta}$ comes into
the equation because $\nu_{,x}=2\frac{\zeta_{,x}}{\zeta}$ which follows from its definition (\ref{zeta}).
Differentiating with respect to $x$ implies
\begin{align}
	8\pi P_{,x}=4\left(2y y_{,x}\frac{\zeta_{,x}}{\zeta}-y^{2}
			\frac{\zeta_{,xx}\zeta-\zeta_{,x}^{2}}{\zeta^{2}}\right)-2w_{,x}.
	\label{PX}
\end{align}
Next the conservation equation (\ref{c}) is rewritten. This then gives
\begin{align}
	P_{,x}=-\frac{\zeta_{,x}}{\zeta}\left(P+\rho \right)=-\frac{\zeta_{,x}}{\zeta}\left(P+\frac{1}{4\pi}
	(3w+2x w_{,x})\right),
	\label{PPX}
\end{align}
where (\ref{mdd}) was used to eliminate $\rho$.
Putting equations (\ref{PP}) and (\ref{PX}) in (\ref{PPX}) eliminates the pressure. 
The equation in full details is
\begin{multline*}
	\frac{1}{2\pi}\left( 2y y_{,x} \frac{\zeta_{,x}}{\zeta} - y^{2} \frac{\zeta_{,xx}\zeta-
	\zeta_{,x}^{2}}{\zeta^{2}} \right) - \frac{1}{4\pi}w_{,x} = \\
	-\frac{\zeta_{,x}}{\zeta} \left( 
	\frac{1}{8\pi} \left(4y^{2} \frac{\zeta_{,x}}{\zeta}+ \frac{2}{3}\Lambda
	\right)	-\frac{1}{4\pi}w+\frac{1}{4\pi} \left(2x w_{,x}+3w\right) 
	\right).
\end{multline*}
It is obvious that the terms with $y^{2}\zeta_{,x}^{2}$ drop out. Multiplying with $\zeta$ and using that
\begin{align}
	2y y_{,x}=-2w-2x w_{,x}-\frac{\Lambda}{3},
	\label{use}
\end{align}
the following equation is obtained
\begin{align}
	y^{2}\zeta_{,xx}-\left(w+x w_{,x}+\frac{\Lambda}{6} \right)
			\zeta_{,x}-\frac{1}{2}w_{,x}\zeta = 0.
\end{align}
Using again (\ref{use}) one finds that
\begin{align}
	y^{2}\zeta_{,xx}+y y_{,x}\zeta_{,x}-\frac{1}{2}w_{,x}\zeta = 0,
	\label{be}
\end{align}
or rewritten using the product rule
\begin{align}
	\left(y\zeta_{,x}\right)_{,x}-\frac{1}{2}\frac{w_{,x}\zeta}{y} = 0.
	\label{yzeta}
\end{align}
The last two  differential equations are identical to these obtained without a cosmological 
constant and as mentioned above,
$\Lambda$ is fully absorbed by the new variable $y$. The equation is linear in $\zeta$ and $w$.

Derivation of (\ref{yzeta}) can also be done by rewriting the equation with vanishing
cosmological constant in effective values for pressure and density.
This leads to the Buchdahl variable $y$ with $w_{\mathrm{eff}}$.

\subsection{Existence of a unique regular solution at the centre}
\label{sub_existence}
It is shown that there exists a unique regular solution in a neighbourhood 
of the centre $r=0$ for each central density and given equation of state. 
To do this the following theorem is needed, see \cite{RendallSchmidt} for a proof.

\begin{theorem} 
\label{th_existence_centre}
Let $V$ be a finite dimensional real vector space, $ N:V \rightarrow V$ a 
linear mapping, $G:V \times I \rightarrow V$ a $C^{\infty}$ mapping and 
$g:I \rightarrow V$ a smooth mapping, where $I$ is an
open interval in $R$ containing zero. Consider the equation
\begin{align}
	s\frac{d f}{ds} + N f = s G(s,f(s))=g(s)
	\label{diff}
\end{align}
for a function $f$ defined on a neighbourhood of $0$ and $I$ and taking 
values in $V$. Suppose that each eigenvalue of $N$ has a positive real part. 
Then there exists an open Interval $J$ with $0 \in J \subset I$ and a 
unique bounded $C^{1}$ function $f$ on $J \setminus {0}$ satisfying (\ref{diff}).
Moreover $f$ extends to a $C^{\infty}$ solution of (\ref{diff}) on $J$ if $N,G$ 
and $g$ depend smoothly on a parameter $z$ and the eigenvalues of $N$ are 
distinct then the solution also depends smoothly on $z$.
\end{theorem}

For a given equation of state equations (\ref{tov}) and (\ref{mdd}) are forming a
system of differential equations in $P(r)$ and $w(r)$. The aim is to apply
theorem~\ref{th_existence_centre}.

Recall (\ref{mdd}) and (\ref{tov}) with $x=r^{2}$ and $dr=\frac{dx}{2\sqrt{x}}$. These substitutions give
\begin{align}
x\frac{d w}{d x}+\frac{3}{2}w & = 2\pi \rho \label{rhs1}\\
\frac{d\rho}{d x} & = -\left(\frac{d P}{d\rho}\right)^{-1}\frac{1}{2}\frac{\left( 4\pi P
			+ w -\frac{\Lambda}{3} \right)
		       \left( P + \rho \right)}
		      {1-2w x-\frac{\Lambda}{3}x} .
		\label{rhs2}
\end{align}
Looking for the mathematical structure of the equations one has that the first one is singular at the centre
$x=0$ whereas the second one is regular there. Both equations have different properties, but using the
substitution
\begin{align*}
	\rho =\rho_{c} + x\rho_{1},
\end{align*}
where $\rho_{c}$ is the value of the central density,
the second equation becomes singular at the centre, too. The transformed system is
\begin{align}
x\frac{d w}{d x}+\frac{3}{2}w & =2\pi \rho_{c}+2\pi x\rho_{1} \\
x\frac{d\rho_{1}}{d x}+\rho_{1} & =-\left(\frac{d P}{d\rho}\right)^{-1}\frac{1}{2}\frac{\left(4\pi P
			+ w -\frac{\Lambda}{3} \right)
		       \left( P + \rho_{c} + x\rho_{1} \right)}
		      {1-2w x-\frac{\Lambda}{3}x}.
\end{align}
The system will be put in the required form to
apply theorem~\ref{th_existence_centre}. After substituting 
in \mbox{$\rho =\rho_{c} + x\rho_{1}$} and the corresponding
pressure relation \mbox{$P=P_{c}+xP_{1}(\rho_{1})$}, only one 
more algebraic relation is needed. It is easily verified that
\begin{align*}
	\left(1-2w x-\frac{\Lambda}{3}x\right)^{-1} =
	1+\left(2w x+\frac{\Lambda}{3}x\right)
	\left(1-2w x-\frac{\Lambda}{3}x\right)^{-1}.
\end{align*}
Then the matrix $N$ has the form
\begin{align}
	\left(	\begin{array}{cc}
		3/2	&	0 \\
	 \frac{(P_{c}+\rho_{c})}{2d P/d\rho (\rho_{c})} 	&  1
		\end{array} \right).
\end{align}
It is easy to obtain that $N$ does not change if it is written in terms of effective values.

Now theorem~\ref{th_existence_centre} can be applied. Thus the system has a unique 
bounded solution in the neighbourhood of the centre, which is $C^{\infty}$. 
Therefore one has a unique smooth solution to (\ref{tov}) and (\ref{mdd}) near the centre. 

This disproves Collins \cite{Collins} statement that for given equation of state, central pressure and cosmological constant
the solution is not uniquely determined.

Standard theorems for differential equations imply that the solution can be extended as long as 
the right-hand sides of (\ref{tov}) or (\ref{rhs2}) are well defined.
If the pressure satisfies $P < \infty$ and the denominator of (\ref{rhs2}) does not
vanish, this means $y=1-2wx-(\Lambda/3)x > 0$, then the right-hand sides are
well defined. The second term involve the cosmological constant and therefore 
some new features will arise.

Uniqueness of the solution at the centre implies the following theorem.
\begin{theorem}
Suppose $\rho(P)$, $P_{c}$ and the cosmological constant $\Lambda$ 
are given such that 
\begin{align}
	4\pi P_{c} + \frac{4\pi}{3}\rho(P_{c}) -\frac{\Lambda}{3} = 0,
\end{align}
note that $(4\pi/3)\rho(P_{c})=w_{c}$.
Then the solution is the Einstein static universe with $\Lambda_{E}=\Lambda$.
\end{theorem}

\subsection{Extension of the solution}
\label{extension}
\begin{theorem}
Assume the pressure is decreasing near the centre, this means
\begin{align}
	4\pi P_{c} + \frac{4\pi}{3}\rho(P_{c}) - \frac{\Lambda}{3} > 0.
	\label{cond}
\end{align}
Then the solution is extendible and the pressure is 
monotonically decreasing if
\begin{align}
	4\pi P+w-\Lambda/3 > 0.
\end{align}
\end{theorem}
\emph{Proof.} Suppose that $\rho=\rho(P)$, $P_{c}$ and $\Lambda$ are 
given such that $P$ is decreasing
near the centre, then $w(x)_{,x} \leq 0$.
Using $y > 0$ because of the metric's signature, (\ref{yzeta}) implies
\begin{align}
	\left(y\zeta_{,x}\right)_{,x} \leq 0.
	\label{z>0}
\end{align}
Rewriting (\ref{PP}) gives
\begin{align}
	y\zeta_{,x}=\frac{\zeta}{2y}\left(4\pi P+w-\frac{\Lambda}{3}\right),
	\label{z.ex}
\end{align}
next using the implication of (\ref{z>0}) leads to
\begin{align*}
	y\zeta_{,x} \leq \left(y\zeta_{,x}\right)(0).
\end{align*}
Together with the explicit expression of $y \zeta_{,x}$ in (\ref{z.ex}) this finally shows
\begin{align}
	y \geq \frac{4\pi P+w-\frac{\Lambda}{3}}{4\pi P_{c}+w_{c}
		-\frac{\Lambda}{3}}.
	\label{y>P.}
\end{align}
Therefore the new Buchdahl variable $y$ cannot vanish before the numerator does.
Thus the right-hand sides of (\ref{tov}) or (\ref{rhs2}) are well defined and one
can extend the solution if $4\pi P+w-\Lambda/3 > 0$.

Since $y \geq 0$ and $4\pi P+w-\Lambda/3 > 0$ the sign of the 
right-hand side of (\ref{rhs2}) is strictly negative. Therefore the density and by the 
equation of state the pressure are monotonically decreasing functions.

\subsection{A possible coordinate singularity $y=0$}
For some cosmological constants the new Buchdahl variable $y$ may vanish before 
the pressure does. In the constant density case the vanishing of $y$ corresponded 
to a coordinate singularity and one could extend the solution.
Therefore one would like to show that this is also true for a 
prescribed equation of state. 

The following theorem only indicates that
the vanishing of $y$ corresponds to a coordinate singularity.
The square of the Riemann tensor in Buchdahl variables (\ref{Riemann_buchdahl}) 
is given in appendix~B.
\begin{theorem}
The squared Riemann tensor (\ref{Riemann_buchdahl}) does not diverge 
as $y \rightarrow 0$ if the pressure is finite.
\end{theorem}
\emph{Proof.} This is seen by first showing that $y \rightarrow 0$ implies  
$4\pi P+w-\Lambda/3 \rightarrow 0$, which is the numerator 
of the TOV-$\Lambda$ equation (\ref{tov}) and of (\ref{y>P.}). 
  
Assume that $\rho(P)$, $P_{c}$ and $\Lambda$ are given such that the pressure is decreasing near 
the centre. Assume further that $\Lambda$ is given such that $y$ vanishes before the pressure. 
Then there exists $r_{s}>0$ such that \mbox{$4 \pi P_{s} + w_{s} - \Lambda/3 =0$}. 
Show that $r_{s}=\hat{r}$, where $\hat{r}$ denotes the radius where the  TOV-$\Lambda$ equation 
becomes singular. This is proved indirectly: 

Assume $r_{s} \neq \hat{r}$, i.e. $y_{s} \neq 0$. Then $4 \pi P_{s} + w_{s} - \Lambda/3 =0$ 
implies $P'_{s}=0$ and equation (\ref{z.ex}) gives
\begin{align}
	\zeta_{,x}(x_{s})=0. 
	\label{zeta,x=0}
\end{align}
Evaluating (\ref{be}) at $x_{s}$ leads to
\begin{align}
	y^{2}_{s} \zeta_{,xx}(x_{s}) - \frac{1}{2} \zeta(x_{s}) w_{,x}(x_{s}) = 0.
	\label{be_xs}
\end{align} 
The pressure is assumed to be decreasing near the centre, thus $w_{,x}(x_{s}) \leq 0$. 
Since $y_{s} \neq 0$ (\ref{be_xs}) implies
\begin{align}
	 \zeta_{,xx}(x_{s}) \leq 0.
\end{align} 
Using (\ref{zeta}), the definition of $\zeta$, implies that  $P''(r_{s}) \geq 0$.
On the other hand the above assumptions on the pressure are such 
that $P''(r)$ is smaller than zero near the centre. 
Therefore $P''(r) \leq 0$ on $0 < r \leq r_{s}$. And both inequalities 
together imply $P''(r_{s}) = 0$, which using (\ref{zeta}) combines to
\begin{align}
	\zeta_{,xx}(x_{s}) = 0.
	\label{zeta,xx=0}
\end{align} 
Evaluating (\ref{be}) at $x_{s}$ using (\ref{zeta,x=0}) and (\ref{zeta,xx=0}) finally gives
\begin{align}
	\zeta(x_{s}) w_{,x}(x_{s}) = 0.
\end{align}
Since $\zeta$ cannot vanish  $w_{,x}(x_{s}) = 0$.
It was assumed that $r_{s},x_{s} > 0$ thus equation (\ref{rhs1}) reads
\begin{align}
	0 = x_{s} w_{,x}(x_{s}) = \frac{3}{2} \left( \frac{4\pi}{3} \rho_{s} - w_{s} \right),
\end{align} 
which gives $w_{s} = (4\pi /3)\rho_{s}$. Therefore $\rho(r)=\rho_{s}=\mbox{const.}$ on $0<r \leq r_{s}$.
Conversely $P(r)=\mbox{const.}$, which contradicts the above assumption that the pressure
is decreasing near the centre. Thus $P'(r_{s}) \neq 0$, therefore $y$ has to vanish and $r_{s} = \hat{r}$.
So the situation is similar to the constant density case. Therefore the expression
\begin{align}
	y\frac{\zeta_{,x}}{\zeta} = -\frac{1}{2} \left( \frac{4\pi P + w - \frac{\Lambda}{3}}{y} \right)
	\label{bound}
\end{align} 
does not diverge as $y \rightarrow 0$ 
and shows finiteness of the squared Riemann tensor.

The first term of (\ref{Riemann_buchdahl}) gives $4/x^{2}$, the second and the third
term vanish because (\ref{bound}) implies $y^{2} \zeta_{,x}/\zeta = 0$.
By the above the last term is bounded if the pressure is finite.\\
\mbox{} \\
This indicates that $y=0$ corresponds to a coordinate singularity.

At the possible coordinate singularity the pressure has a well defined value given by 
\begin{align}
	P(\hat{r})=\frac{1}{4 \pi} \left( \frac{\Lambda}{3} - w(\hat{r}) \right).
	\label{phatr}
\end{align}

\subsection{Existence of global solutions with $\Lambda < 4 \pi \rho_{b}$}
\label{sub_global}
The following theorem shows the existence of global solutions.
Two possibilities occur. Either the matter occupies the whole space
or the pressure vanishes at a finite radius, in which case a vacuum
solution is joined. 
\begin{theorem}
Let an equation of state be given such that $\rho$ is defined for $p \geq 0$,
non-negative and continuous for $p \geq 0$, $C^{\infty}$ for $p > 0$ and 
suppose that $d\rho/dp >0$ for $p > 0$. Further let the cosmological constant
be given such that $\Lambda < 4\pi \rho_{b}$.

Then the pressure is decreasing near the centre
and there exists for any positive value of central pressure $P_{c}$ 
a unique inextendible static, spherically 
symmetric solution of Einstein's field equations with cosmological 
constant with a perfect fluid source and equation of state $\rho(P)$.  

If $\Lambda \leq 0$ then the matter either occupies the whole space
with $\rho$ tending to $\rho_{\infty}$ as $r$ tends to infinity or
the matter has finite extend. In the second case a unique Schwarzschild-anti-de Sitter 
solution is joined on as an exterior field.

If $0<\Lambda<4\pi\rho_{b}$ the matter always has finite extend
and a unique Schwarzschild-de Sitter solution is joined on as
an exterior field. 
\end{theorem}
\emph{Proof.} If the cosmological constant is given such that $\Lambda<4\pi\rho_{b}$ then
\begin{align*}
	0	&< \frac{4\pi}{3} \rho_{b} -\frac{\Lambda}{3} \\
		&< 4\pi P_{c} + \frac{4\pi}{3} \rho(P_{c}) - \frac{\Lambda}{3},
\end{align*}
and the pressure is decreasing near the centre, see (\ref{cond}).

Since the pressure is decreasing near the centre the 
denominator of (\ref{y>P.}) is a positive number $\mathcal{D}$. 
Then the numerator of (\ref{y>P.}) can be estimated by
\begin{align}
	y  	&\geq \frac{4\pi P+w-\frac{\Lambda}{3}}{\mathcal{D}} \nonumber \\
		&\geq \frac{w_{b}-\frac{\Lambda}{3}}{\mathcal{D}} 
		\label{eqn_estimate_wb} \\
		&\geq \frac{\frac{4\pi}{3}\rho_{b}-\frac{\Lambda}{3}}{\mathcal{D}}. 
		\label{eqn_estimate_rhob}
\end{align}
Thus if 
\begin{align}
	\Lambda < 4\pi \rho_{b},
	\label{upper_limit}
\end{align}
then\footnote{The argument of \cite{Mars} is
	applicable if the cosmological constant satisfies this condition.}  
the new Buchdahl variable cannot vanish before the pressure, $\rho_{b}$ is
given by the equation of state because $\rho_{b} = \rho(P=0)$.
The coordinate $x_{b}$ where the pressure vanishes will be taken as the 
definition of the stellar object's radius $r_{b}$.

\subsubsection*{$\Lambda \leq 0$}
If $\Lambda \leq 0$ the matter can occupy the whole space because (\ref{y>P.}) 
implies positivity of $y$. 

Suppose $P(x_{b})=0$. At the corresponding radius $r_{b}$ the Schwarzschild-anti-de Sitter solution, (\ref{sds}) with
negative cosmological constant, is joined uniquely by the condition $M=m(r_{b})$. 
In this manner the metric is $C^{0}$ only, because the 
density at the boundary may be non-zero. The metric is $C^{1}$ at $P(r_{b})=0$ if Gauss coordinates relative
to the hypersurface $P(r_{b})=0$ are used. These are given by $\chi(r)=\int_{0}^{r} e^{a(s)/2} ds$. If the boundary 
density does not vanish the Ricci tensor has a discontinuity. Thus the metric is at most $C^{1}$. Without further
assumptions on the boundary density this cannot be improved.    

Assume now that $P(x) > 0$ for all $x>0$. $P(x)$ is monotonically decreasing, therefore 
$\lim_{x \rightarrow \infty} P(x) = P_{\infty}$ exists. This implies that the pressure 
gradient tends to zero as $x \rightarrow \infty$. Because of $y^{-1} \rightarrow 0$ as $x \rightarrow \infty$
equation (\ref{tov}) does not imply that $P_{\infty}=0$, which it does if $\Lambda=0$. 
Thus the equation of state only gives $\rho \rightarrow \rho_{\infty}=\rho(P_{\infty})$ as $x \rightarrow \infty$.

\subsubsection*{$0< \Lambda < 4\pi \rho_{b}$}
If $0< \Lambda < 4\pi \rho_{b}$ then one can estimate the 
pressure~(\ref{phatr}) at the possible coordinate singularity $y=0$.
Note that 
\begin{align}
	\frac{\Lambda}{3} < \frac{4\pi}{3} \rho_{b} \leq w_{b} \leq w(r),
	\label{eqn_lambda3<wb}
\end{align}
which holds for all $r$. Therefore
\begin{align}
	P(\hat{r}) = \frac{1}{4\pi} \left( \frac{\Lambda}{3} - w(\hat{r}) \right) < 0.
\end{align}
Hence there exists $r_{b}$ such that $P(r_{b})=0$. Since the
pressure is decreasing and $P(\hat{r})<0$ it follows that $r_{b} < \hat{r}$.  

Thus if the cosmological constant is positive and \mbox{$\Lambda < 4\pi \rho_{b}$} 
then the pressure always vanishes at some $x_{b}$. At the corresponding $r_{b}$ the 
Schwarzschild-de Sitter solution is joined uniquely by the same 
condition $M=m(r_{b})$. The metric is at most $C^{1}$ because the 
boundary density is larger than zero, this cannot be improved. 

\subsection{Generalised Buchdahl inequality}
For stellar models an analogous Buchdahl inequality is derived.
It turns out that the new inequality nearly coincides with the former (\ref{eqn_Buch}), 
which was derived for constant density. 

The proof of the following theorem solves the field equations 
with constant density written in Buchdahl variables. 
This solution is compared with a decreasing solution and leads
to an analogous Buchdahl inequality.

The solution with constant density is denoted with a tilde above.

\begin{theorem}
\label{theorem_buch}
Let the cosmological constant be given such that $\Lambda < 4\pi \rho_{b}$.
Then for stellar models there holds
\begin{align}
	\sqrt{1-2w_{b} r_{b}^{2}-\frac{\Lambda}{3}r_{b}^{2}}  
	\geq \frac{1}{3} - \frac{\Lambda}{9 w_{b}}.
\end{align}
\end{theorem}
\emph{Proof.} Assume that $\rho(P)$, $P_{c}$ and 
$\Lambda$ are given such that the pressure is decreasing
near the centre. In section~\ref{extension} it was shown that 
this means that pressure and mean density are decreasing functions.

Equation (\ref{yzeta}) implies 
\begin{align}
	(\tilde{y}\tilde{\zeta}_{,x})_{,x}& =0 	\nonumber \\
	\tilde{y}\tilde{\zeta}_{,x}& =D,
	\label{zetatilde}
\end{align}
where $D$ is a constant of integration. 

If the density is constant $\zeta$ is the only unknown function in
Einstein's field equations.

The function $\zeta$ can be normalised to give $\zeta_{b}=y_{b}$.
The constant $D$ is obtained from (\ref{PP}) evaluated at the boundary, therefore
\begin{align}
	D=\frac{1}{2}\tilde{w}-\frac{\Lambda}{6},
\end{align}
which can be used to integrate (\ref{zetatilde}). 
Notice that (\ref{use}) reads
\begin{align*}
	-2\tilde{y}\tilde{y}_{,x} = 2\tilde{w}+\frac{\Lambda}{3}.
\end{align*}
The right-hand side of (\ref{zetatilde}) can be written as
\begin{align*}
	\tilde{y}\tilde{\zeta}_{,x}=
	\left( 2\tilde{w}+\frac{\Lambda}{3} \right)
	\left( \frac{1}{2}\tilde{w}-\frac{\Lambda}{6} \right)
	\left( 2\tilde{w}+\frac{\Lambda}{3} \right)^{-1}.
\end{align*}
Substitute in $-2\tilde{y}\tilde{y}_{,x}$ because of the first factor and divide by $\tilde{y}$.
Integration then leads to 
\begin{align*}
	\tilde{\zeta}(x)=
	-\frac{\tilde{w}-\frac{\Lambda}{3}}{2\tilde{w}+\frac{\Lambda}{3}}\tilde{y}(x)
	+ \mbox{const.}
\end{align*}
The constant of integration is
\begin{align*}
	\mbox{const.} = \tilde{\zeta}(0)+\frac{\tilde{w}-
	\frac{\Lambda}{3}}{2\tilde{w}+\frac{\Lambda}{3}}.
\end{align*}
So the solution of differential equation (\ref{zetatilde}) is obtained.
Equation (\ref{z>0}) implies
\begin{align*}
	\tilde{y}\tilde{\zeta}_{,x} = \left(y\zeta_{,x}\right)_{b} < y\zeta_{,x}.
\end{align*}
Using that $\tilde{y} > y$ one finds
\begin{align*}
	\zeta(x) \geq \tilde{\zeta}(x) = -\frac{\tilde{w}-\frac{\Lambda}{3}}{2\tilde{w}+\frac{\Lambda}{3}}
	\tilde{y}(x)+\tilde{\zeta}(0)+\frac{\tilde{w}-\frac{\Lambda}{3}}{2\tilde{w}+\frac{\Lambda}{3}}.
\end{align*}
Evaluating this at the boundary and using the normalisation condition it is found that
\begin{align*}
	y_{b}  \geq  -\frac{\tilde{w}-\frac{\Lambda}{3}}{2\tilde{w}+\frac{\Lambda}{3}}
	y_{b} +\tilde{\zeta}(0)+ \frac{\tilde{w}-\frac{\Lambda}{3}}{2\tilde{w}+\frac{\Lambda}{3}}.
\end{align*}
Since $\tilde{\zeta}(0)$ is positive some algebra
leads to the analogous Buchdahl inequality 
\begin{align}
	y_{b} \geq \frac{1}{3} - \frac{\Lambda}{9 \tilde{w}}.
	\label{buchdahlw}
\end{align}
One compares solutions with decreasing mean density and solution
with constant density. The compared constant density corresponds 
to the boundary mean density. Therefore (\ref{buchdahlw}) reads
\begin{align}
	y_{b} \geq \frac{1}{3} - \frac{\Lambda}{9 w_{b}},
	\label{buchdahlrho}
\end{align}
which nearly equals the analogous Buchdahl inequality (\ref{eqn_Buch}) 
if the expression for $y_{b}$ is used. The only difference is that the equation
contains the boundary mean density $w_{b}$ rather than the boundary density $\rho_{b}$.
Explicitly written out leads to
\begin{align}
	\sqrt{1-2w_{b} r_{b}^{2}-\frac{\Lambda}{3}r_{b}^{2}}  
	\geq \frac{1}{3} - \frac{\Lambda}{9 w_{b}},
	\label{eqn_buch_sqr}
\end{align}
and holds for all monotonically decreasing densities and
proves theorem~\ref{theorem_buch}. \\ 
\mbox{} \\
Equation~(\ref{eqn_lambda3<wb}) implies that the right-hand sides 
of the last two equations~(\ref{eqn_buch_sqr}) and~(\ref{buchdahlrho})
are positive. This improves~(\ref{y>P.}), which only gave $y>0$. 
Solving inequality~(\ref{eqn_buch_sqr}) for $r_{b}$ gives an 
analogue of (\ref{eqn_buchdahl<>})
\begin{align}
        	r_{b}^{2} \leq \frac{\frac{1}{9} \left(4-\frac{\Lambda}{3 w_{b}} \right)}{w_{b}}.
	\label{eqn_gen_buchdahl}
\end{align}
As before the only difference is that the boundary mean density is used 
instead of the boundary density itself. Multiply (\ref{eqn_gen_buchdahl}) 
by $w_{b}$ and use that $w_{b}=m(r_{b})/r_{b}^{3}$ gives
\begin{align*}
        	\frac{m(r_{b})}{r_{b}} \leq \frac{1}{9} \left(4-\frac{\Lambda r_{b}^{3}}{3 m(r_{b})} \right).
\end{align*}
Taking all terms on one side and multiplying by $9m(r_{b})r_{b}$ leads to
\begin{align*}
	9m(r_{b})^{2} - 4m(r_{b})r_{b} + \frac{\Lambda}{3} r_{b}^{4} \leq 0.
\end{align*}
Rewriting this with 
\begin{align*}
	9m(r_{b})^{2} - 4m(r_{b})r_{b} = \left(3m(r_{b})-\frac{2}{3}r_{b}\right)^{2} - \frac{4}{9}r_{b}^{2},
\end{align*}
finally gives
\begin{align}
	\left(3m(r_{b})-\frac{2}{3}r_{b}\right)^{2} - r_{b}^{2} \left(\frac{4}{9} -\frac{\Lambda}{3}r_{b}^{2} \right) \leq 0.
	\label{eqn_helpbuch}
\end{align}
Another way of writing the above~(\ref{eqn_helpbuch}) is  
\begin{align}
	\left(3m(r_{b})-\frac{2}{3}r_{b}-r_{b}\sqrt{\frac{4}{9}-\frac{\Lambda}{3}r_{b}^{2}} \right)
	\left(3m(r_{b})-\frac{2}{3}r_{b}+r_{b}\sqrt{\frac{4}{9}-\frac{\Lambda}{3}r_{b}^{2}} \right) \leq 0,
	\label{product}
\end{align}
where $a^{2}-b^{2}=(a-b)(a+b)$ was used.
Putting $\Lambda=0$ shows that the second term has to be positive 
because of positivity of the mass. Then the first term is negative
and implies the Buchdahl inequality (\ref{eqn_buchdahl}).
For negative cosmological constants the square root terms
are always well defined. For positive cosmological constants
they are well defined if
\begin{align}
	r_{b} \leq \sqrt{\frac{4}{3}} \frac{1}{\sqrt{\Lambda}}.
\end{align}
Then both expressions of the product~(\ref{product}) can be combined to
\begin{align}
	3m(r_{b}) \leq \frac{2}{3}r_{b} + r_{b}\sqrt{\frac{4}{9}-\frac{\Lambda}{3} r_{b}^{2}}.
	\label{eqn_buch_mrb}
\end{align}
With $\Lambda=0$ equation~(\ref{eqn_buch_mrb}) directly
implies Buchdahl's inequality~(\ref{eqn_buchdahl3m}). It
holds for arbitrary static fluid balls in which the density
does not increase outwards.

\subsection{Solutions without singularities}
The last two sections showed the existence of stellar models for
cosmological constants $\Lambda < 4\pi \rho_{b}$. 
Equation~(\ref{y>P.}) implies that the boundary of the stellar object 
has a lower bound given by the black-hole event horizon 
and an upper bound given by the cosmological event horizon. 
The upper bound only occurs if the cosmological constant is 
positive, as already said.

Stellar models with $\Lambda \leq 0$ have a lower bound
given by the black-hole event horizon. At the boundary
the Schwarzschild-anti-de Sitter solution is joined on as
an exterior field. Figure~\ref{fig: Penrose_anti1} shows that
this joined exterior field has no singularities. Therefore
stellar models with $\Lambda \leq 0$ have no singularities.
Solutions with cosmological constant satisfying $\Lambda \leq 0$  
are globally static.

For positive cosmological constants the situation is different. But 
one can construct solutions without singularities.

At the boundary $r=r_{b}$ the Schwarzschild-de Sitter solution
is joined $C^{1}$ by the usual procedure introducing Gauss coordinates.
This leads to Penrose-Carter figure~\ref{fig: Penrose1a}. This spacetime
has an infinite sequence of singularities $r=0$ and space-like infinities $r=\infty$.

The surface $r=r_{b}$ can also be found in the vacuum region where
the time-like Killing vector is past directed. This means that a second
stellar object is put in the spacetime. Therefore Penrose-Carter 
figure~\ref{fig: Penrose2} gives a general picture and is not 
restricted to constant density solutions.

Note that solutions with positive cosmological constant are not
globally static because of the dynamical parts of the exterior
Schwarzschild-de Sitter solution.

Solutions without singularities and without horizons were recently 
described by~\cite{Mazur}. They considered an interior de Sitter region
and an exterior Schwarzschild solution separated by a small shell of
matter with equation of state $\rho(P)=P$. The shell replaces both
the de Sitter and the Schwarzschild horizon. The new solution
has no singularities.

\subsection{Remarks on finiteness of the radius}
So far it has been shown that given an equation of state, a central pressure and a cosmological
constant there exist a unique model with finite or infinite extend. This depends on the given
equation of state and on the cosmological constant. This section gives criteria to distinguish these 
two cases, see \cite{RendallSchmidt}.

\subsubsection*{$\Lambda > 0$}
If $\Lambda > 0$ solutions are always finite. Thus given an equation of state $\rho=\rho(P)$ 
and cosmological constant such that  
\begin{align}
	\Lambda < 4\pi \rho(P=0) = 4\pi \rho_{b},
\end{align}
then there always exist a radius $r_{b}$ where the pressure vanishes.

\subsubsection*{$\Lambda \leq 0$}
If $\Lambda \leq 0$ either the pressure vanishes for some finite radius or the density is 
always positive and tends to $\rho_{\infty}$ as $r$ tends to infinity. 

First a necessary condition for finiteness of the radius is derived.
Recall (\ref{tov}) written in the variable~$x$
\begin{align}
	P_{,x}=-\frac{1}{2}\frac{\left( 4\pi P + w -\frac{\Lambda}{3} \right) 
		       \left( P+ \rho \right)}
		      {1-2w x-\frac{\Lambda}{3}x}.
	\label{tovx}
\end{align}
Assume that the stellar model is bounded by $x_{b}$ or the corresponding $r_{b}$.
On the closed interval $[0, x_{b}]$ the function $y$ has a positive minimum $K$. 
The maximal contribution due to the first factor is given by its value at the centre. Therefore
\begin{align*}
	P_{,x} > -L \left( P + \rho \right),
\end{align*}
where the constant $L$ is given by
\begin{align}
	L = \frac{1}{2K} \left(4\pi P_{c} + w_{c} - \frac{\Lambda}{3} \right).
	\label{L}
\end{align}
Compare the above inequality with
\begin{align*}
	P^{*}_{,x} = L \left(P^{*} + \rho^{*} \right),
\end{align*}
which can be integrated to give 
\begin{align*}
	\int_{0}^{P_{c}^{*}} \frac{d P}{P + \rho} = L x_{b} = L r_{b}^{2}.
\end{align*}
Thus we obtain a bound for the integral. Hence if the star has finite boundary then
\begin{align}
	\int_{0}^{P_{c}} \frac{d P}{P + \rho} < \infty.
	\label{necessary}
\end{align}
This gives a necessary condition for the equation of state to give fluid balls of finite extend.
Then equation (\ref{int}) implies that $\nu(r)$ is finite.

Now a sufficient condition is derived.
Since the density distribution is decreasing $m \geq \frac{4\pi}{3} \rho r^{3}$, $w \geq \frac{4\pi}{3} \rho$ 
for all $r > 0$.

Thus (\ref{tovx}) gives
\begin{align}
	P_{,x} < -\frac{1}{2}w\rho < -\frac{2\pi}{3}\rho^{2},
	\label{px,suff}
\end{align}
where one needed that
\begin{align*}
	4\pi P - \frac{\Lambda}{3} > 0,
\end{align*}
which is always achieved as long as $\Lambda \leq 0$.  Then comparing with a $P^{*}$ again and using that
there has to exist a point $x_{b}$ where the pressure vanishes one may integrate (\ref{px,suff}). 
Therefore if
\begin{align}
	\int_{0}^{P_{c}}\frac{d P}{\rho^{2}} < \infty,
	\label{sufficient}
\end{align}
the stellar object has finite radius.
Both criteria (\ref{necessary}), (\ref{sufficient}) only involve the low pressure behaviour of the equation of state.

%% file: appendix.tex
\section*{Appendices}
\addcontentsline{toc}{section}{Appendices}

\section*{A Einstein tensor and energy-momentum conservation}
\addcontentsline{toc}{subsection}{\numberline {A}Einstein tensor and energy-momentum conservation}
\addtocounter{section}{1}
\setcounter{equation}{0}
In the following a static, spherically symmetric metric is considered.
In the used notation it may be written as
\begin{align}
	ds^{2}=-\frac{1}{\lambda} e^{\lambda \nu (r)} dt^{2}+ e^{a(r)} dr^2+
	r^{2} (d\theta ^{2} +\sin^{2}\negmedspace \theta \ d\phi^{2}).
\end{align}
The non vanishing Christoffel symbols are:
\begin{align*}
&\Gamma^{t}_{tr}=\frac{1}{2}\lambda \nu'(r) \\
&\Gamma^{r}_{tt}=\frac{1}{2}\nu'(r) e^{\lambda \nu(r)-a(r)} &
&\Gamma^{r}_{rr}=\frac{1}{2}a'(r) \\
&\Gamma^{r}_{\theta \theta}=-r e^{-a(r)} &
&\Gamma^{r}_{\phi \phi}=-r\sin^{2}\theta e^{-a(r)}  \\
&\Gamma^{\theta}_{\theta r}=\frac{1}{r} &
&\Gamma^{\theta}_{\phi \phi}=-\cos\theta \sin\theta \\
&\Gamma^{\phi}_{\phi r}=\frac{1}{r} &
&\Gamma^{\phi}_{\phi \theta}=\cot\theta.
\end{align*}
The trace terms are:
\begin{align*}
&\Gamma^{\sigma}_{t \sigma}=0 &         
&\Gamma^{\sigma}_{r \sigma}=\frac{2}{r}+\frac{1}{2}(\lambda \nu'(r)+a'(r)) \\
&\Gamma^{\sigma}_{\theta \sigma}=\cot\theta & 
&\Gamma^{\sigma}_{\phi \sigma}=0.
\end{align*}
The Ricci tensor is defined by
\begin{align}
	R_{\mu \nu}=\Gamma^{\rho}_{\mu \nu ,\rho} - \Gamma^{\rho}_{\mu \rho ,\nu} 
	+ \Gamma^{\sigma}_{\mu \nu}\Gamma^{\rho}_{\sigma \rho} 
	- \Gamma^{\sigma}_{\mu \rho}\Gamma^{\rho}_{\sigma \nu}.
\end{align}
Its components are:
\begin{align*}
R_{tt}&=e^{\lambda \nu(r)-a(r)} \left( \frac{1}{2}\nu''(r)+\frac{1}{4}\lambda \nu'(r)^{2}+\frac{1}{r}\nu'(r)-\frac{1}{4}a'(r)\nu'(r) \right) \\
R_{rr}&= -\frac{1}{2}\lambda \nu''(r)-\frac{1}{4}\lambda^{2} \nu'(r)^{2}+\frac{1}{4}a'(r) \lambda \nu'(r)+\frac{1}{r}a'(r) \\
R_{\theta \theta}&= 1-e^{-a(r)}+\frac{1}{2}r a'(r)e^{-a(r)}-\frac{1}{2}r\lambda\nu'(r)e^{-a(r)} \\
R_{\phi \phi}&= \sin^{2}\theta R_{\theta \theta} \\
R_{\mu \nu}&= 0 \mbox{\ \ if $\mu \neq \nu$}.
\end{align*}
The Ricci scalar is
\begin{align*}
	R=g^{\mu \nu}R_{\mu \nu} & = -\lambda \nu''(r)e^{-a(r)} - \frac{1}{2}\lambda^{2} \nu'(r)^{2}e^{-a(r)} + 
	\frac{1}{2}a'(r)\lambda\nu'(r)e^{-a(r)} \\
	& +\frac{2}{r^{2}}-\frac{2e^{-a(r)}}{r^{2}}+\frac{2}{r}a'(r)e^{-a(r)}-\frac{2}{r}\lambda\nu'(r)e^{-a(r)}.
\end{align*}
The Einstein tensor is defined by
\begin{align}
	G_{\mu \nu}=R_{\mu \nu}-\frac{1}{2}R g_{\mu \nu}.
\end{align}
Only its four diagonal components remain. They are given by:
\begin{align}
G_{tt} & =\frac{1}{\lambda r^{2}}e^{\lambda \nu (r)}\frac{d}{dr}\left( r-re^{-a(r)}\right) 
\label{G00} \\
G_{rr} & =\frac{1}{r^{2}}\left(1+r\lambda \nu'(r)-e^{a(r)}\right) 
\label{G11} \\
G_{\theta \theta} & =r^{2}e^{-a(r)}\frac{1}{2}(\lambda \nu''(r)-\frac{1}{r}a'(r)+\frac{1}{r}\lambda \nu'(r)+\frac{1}{2}\lambda^{2} \nu'(r)^{2}-\frac{1}{2}a'(r)\lambda \nu'(r)) 
\label{G22} \\
G_{\phi \phi} & =\sin^{2}\theta G_{\theta \theta}. \label{G33}
\end{align}
The Einstein tensor has vanishing covariant derivative and therefore implies 
energy-momentum conservation,
\begin{equation}
	\nabla_{\nu}T^{\mu \nu}=0.
\end{equation}
For a perfect fluid in a static, spherically symmetric spacetime the 
energy-momentum tensor has the form
\begin{equation}
	T_{\mu \nu}=(\rho(r) +\lambda P(r) )U_{\mu}U_{\nu} + P(r)g_{\mu \nu},
	\label{tmunu}
\end{equation}
where 4-velocity $U_{t}=(-1/\lambda)e^{\lambda\nu(r) /2}$ is given for a fluid at rest. 
Conservation of this quantity gives four equations of motion. Because of symmetries, only the radial 
component $\mu = r$ does not  equal zero. By using the Christoffel symbols derived above, it is found that
\begin{align*}
	0=\left(\rho(r)+\lambda P(r) \right)\frac{\nu'(r)}{2}e^{-a(r)} + P'(r) e^{-a(r)},
\end{align*}
which gives
\begin{align}
	2P'(r) = -( \lambda P(r) + \rho(r) ) \nu'(r).
	\label{emc}
\end{align} 

\newpage
\section*{B Geometric invariants}
\addcontentsline{toc}{subsection}{\numberline {B}Geometric invariants}
\addtocounter{section}{1}
\setcounter{equation}{0}

\subsection*{Constant density solutions}
Metric (\ref{eqn_metric}) of the constant density solutions 
for $\Lambda > -8 \pi \rho_{0}$ can be written
\begin{align}
	ds^{2}=-\left(\frac{P_{c}+\rho_{0}}{P(\alpha)+\rho_{0}}\right)^{2}dt^{2}
        	+\frac{1}{k}(d\alpha^{2}+\sin^{2} \negmedspace \alpha \ d\Omega^{2}).
\end{align}
Integration of (\ref{tov_const}) gives
\begin{align}
	P(\alpha)=\rho_{0}\frac{\left(\frac{\Lambda}{4 \pi \rho_{0}} -1\right)+C\cos \alpha}{3-C\cos \alpha}.
	\label{Pressure}
\end{align}
From this one can find the square of the Riemann tensor. It reads
\begin{multline}
	R_{\mu \nu \lambda \kappa}R^{\mu \nu \lambda \kappa} =
	12\left( \frac{8\pi}{3}\rho_{0} +\frac{\Lambda}{3} \right)^{2} \\ \times
	\frac{81-C^{2}\cos^{2}\alpha \left( 9-6C\cos\alpha-2C^{2}\cos^{2}\alpha \right)}
	{\left(3+C\cos\alpha \right)^{2} \left(3-C\cos\alpha \right)^{2} },
	\label{Riemann_const}
\end{multline}
and gives two important implications. 

First it shows that the radial coordinate $r$ behaves badly
as $r \rightarrow \hat{r}$, which corresponds to $\alpha \rightarrow \pi /2$. Thus
 \begin{align}
	R_{\mu \nu \lambda \kappa}R^{\mu \nu \lambda \kappa}(\alpha=\pi /2) =
	12\left( \frac{8\pi}{3}\rho_{0}+\frac{\Lambda}{3} \right)^{2},
\end{align}
and one clearly has a coordinate singularity because the square of Riemann tensor is finite.
It was already pointed out that this is only a coordinate singularity when the geometry of the
interior solutions was discussed. 

Secondly, the square of the Riemann tensor diverges as the pressure diverges. Compare the 
denominators of (\ref{Pressure}) and (\ref{Riemann_const}). Thus there is a geometric singularity
if the pressure is divergent. 

\subsection*{Buchdahl variables}
In Buchdahl variables 
\begin{align*}
	y^{2} & = e^{-a(r)} = 1-2w(r)r^{2}-\frac{\Lambda}{3}r^{2}, \\
	\zeta & = e^{\nu/2},\ x =r^{2},
\end{align*}
the static spherically symmetric metric can be written
\begin{align}
	ds^{2}=-\zeta(x)^{2} dt^{2} + \frac{dx^2}{4x y(x)^{2}} + x(d\theta ^{2} +\sin^{2}d\phi^{2}).
\end{align}
In these coordinates the square of the Riemann tensor becomes
\begin{multline}
	R_{\mu \nu \lambda \kappa}R^{\mu \nu \lambda \kappa} =
	\left[ 2 \, \frac{1-y^{2}}{x} \right]^{2} + 2 \left[ 2 \left( y^{2} \right)_{,x} \right]^{2} \\ +
	2 \left[ y^{2} \frac{\zeta_{,x}}{\zeta}\right]^{2} +
	\left[ \frac{\left(\zeta_{,x}^{\, 2}  \, x y^{2} \right)_{,x} } {\zeta \zeta_{,x}} \right]^{2}.	
	\label{Riemann_buchdahl}
\end{multline}
The square of the Weyl tensor is given by
\begin{align}
	C_{\mu \nu \lambda \kappa}C^{\mu \nu \lambda \kappa} =
	\frac{64}{3} \frac{y^{2}}{\zeta^{2}} 
	\left( \left(y\zeta_{,x}\right)_{,x}+\frac{1}{2}\frac{w_{,x}\zeta}{y} \right).
	\label{weyl}
\end{align}
\subsection*{Conformal flatness of constant density solutions}
In chapter~\ref{exis} Einstein's field equation with cosmological constant for a perfect fluid 
were rewritten in terms of Buchdahl variables. This gave equation (\ref{yzeta}), stated again
\begin{align}
	\left(y\zeta_{,x}\right)_{,x}-\frac{1}{2}\frac{w_{,x}\zeta}{y} = 0.
	\label{yzeta_app}
\end{align}
If the density is assumed to be constant then $w_{,x}=0$. Therefore equation (\ref{yzeta_app})
implies that the square of the Weyl tensor (\ref{weyl}) vanishes. In spherical symmetry this then
implies that the Weyl tensor vanishes. Thus constant density solutions
are conformally flat. Furthermore it follows that this is not true if $w_{,x} \neq 0$.

\subsection*{Invariants for other metrics}
To complete this appendix the invariants of the Schwarzschild-de Sitter (\ref{sds}) 
and Nariai (\ref{eqn_nariai_r}) solution are given.
For Schwarzschild-de Sitter they are given by
\begin{align}
	R_{\mu \nu \lambda \kappa}R^{\mu \nu \lambda \kappa} &= 48 \frac{M^{2}}{r^{6}} + \frac{8}{3} \Lambda^{2} \\
	C_{\mu \nu \lambda \kappa}C^{\mu \nu \lambda \kappa} &= 48 \frac{M^{2}}{r^{6}}.
\end{align}
The Nariai metric gives
\begin{align}
	R_{\mu \nu \lambda \kappa}R^{\mu \nu \lambda \kappa} &= 8 \Lambda^{2} \\
	C_{\mu \nu \lambda \kappa}C^{\mu \nu \lambda \kappa} &= \frac{16}{3} \Lambda^{2}.
\end{align}
 
\newpage
\section*{C Penrose-Carter diagrams}
\addcontentsline{toc}{subsection}{\numberline {C}Penrose-Carter diagrams}
\addtocounter{section}{1}
\setcounter{equation}{0}

\begin{figure}[ht!]
\begin{center}
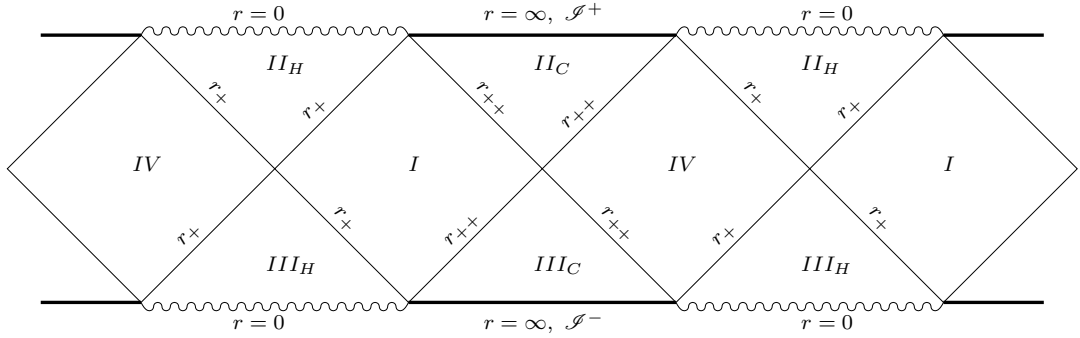
\end{center}
\caption{Penrose-Carter diagram for Schwarzschild-de Sitter space. The Killing
	vector $\partial/\partial t$ is time-like and future-directed in regions $I$ and
	time-like and past-directed in regions $IV$. In the others regions it is space-like.
	The surfaces $r=r_{+}$ and $r=r_{++}$ are black-hole and cosmological event
	horizons, respectively. $\scri^{+}$ and $\scri^{-}$ are the space-like infinities.}
\label{fig: Penrose}
\end{figure}

\begin{figure}[ht!]
\begin{center}
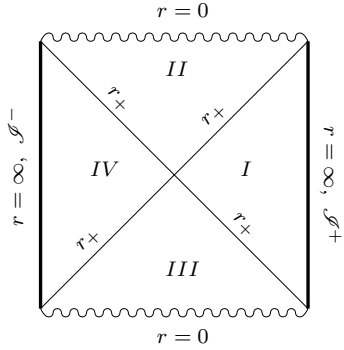
\end{center}
\caption{Penrose-Carter diagram for Schwarzschild-anti-de Sitter space. 
	The surface $r=r_{+}$ is the black-hole event horizons.
	$\scri^{+}$ and $\scri^{-}$ are the time-like infinities.}
\label{fig: Penrose_anti1}
\end{figure}

\begin{figure}[ht!]
\begin{center}
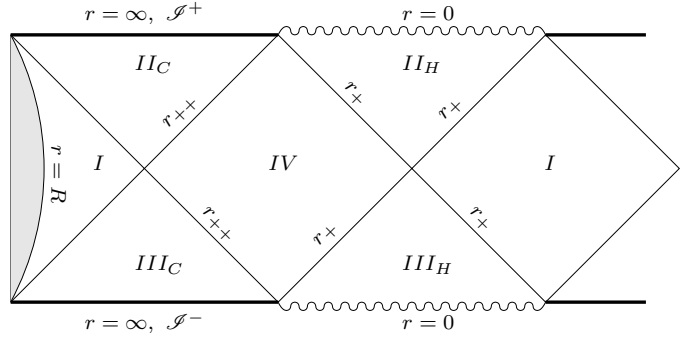
\end{center}
\caption{Penrose-Carter diagram with one stellar object having a radius $R$ which lies between
	the two horizons. The group orbits are increasing at the boundary. 
	The shaded region is the matter solution with regular centre. 
	There is still an infinite sequence of  singularities $r=0$ and 
	space-like infinities $r=\infty$. }
\label{fig: Penrose1a}
\end{figure}

\begin{figure}[ht!]
\begin{center}
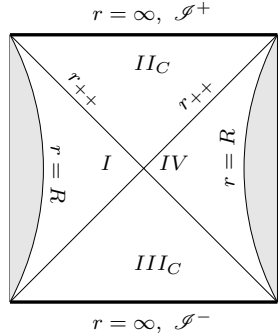
\end{center}
\caption{Penrose-Carter diagram with two stellar objects having radii $R$ which lie between
	the two horizons. Since the group orbits are increasing up to $R$ the vacuum 
	part contains the cosmological event horizon $r_{++}$. 
	This solution with two objects has no singularities. Because of regions $II_{C}$ 
	and $III_{C}$ this spacetime is not globally static.}
\label{fig: Penrose2}
\end{figure}

\begin{figure}[ht!]
\begin{center}
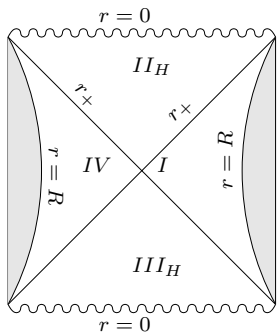
\end{center}
\caption{Penrose-Carter diagram with two stellar objects having radii $R$ which lie between
	the two horizons. The group orbits of the interior solutions are decreasing where
	the vacuum solution is joined. Thus the $r=0$ singularity of the vacuum part is present.}
\label{fig: Penrose3}
\end{figure}

%% file: penrose.pstex_t
\begin{picture}(0,0)%
\includegraphics{penrose}%
\end{picture}%
\setlength{\unitlength}{2763sp}%
\begingroup\makeatletter\ifx\SetFigFont\undefined%
\gdef\SetFigFont#1#2#3#4#5{%
  \reset@font\fontsize{#1}{#2pt}%
  \fontfamily{#3}\fontseries{#4}\fontshape{#5}%
  \selectfont}%
\fi\endgroup%
\begin{picture}(9624,2955)(289,-5431)
\put(1876,-4636){\rotatebox{45.0}{\makebox(0,0)[lb]{\smash{\SetFigFont{8}{9.6}{\familydefault}{\mddefault}{\updefault}{\color[rgb]{0,0,0}$r_{+}$}%
}}}}
\put(3001,-3511){\rotatebox{45.0}{\makebox(0,0)[lb]{\smash{\SetFigFont{8}{9.6}{\familydefault}{\mddefault}{\updefault}{\color[rgb]{0,0,0}$r_{+}$}%
}}}}
\put(6676,-4636){\rotatebox{45.0}{\makebox(0,0)[lb]{\smash{\SetFigFont{8}{9.6}{\familydefault}{\mddefault}{\updefault}{\color[rgb]{0,0,0}$r_{+}$}%
}}}}
\put(7801,-3511){\rotatebox{45.0}{\makebox(0,0)[lb]{\smash{\SetFigFont{8}{9.6}{\familydefault}{\mddefault}{\updefault}{\color[rgb]{0,0,0}$r_{+}$}%
}}}}
\put(3226,-4336){\rotatebox{315.0}{\makebox(0,0)[lb]{\smash{\SetFigFont{8}{9.6}{\familydefault}{\mddefault}{\updefault}{\color[rgb]{0,0,0}$r_{+}$}%
}}}}
\put(8026,-4336){\rotatebox{315.0}{\makebox(0,0)[lb]{\smash{\SetFigFont{8}{9.6}{\familydefault}{\mddefault}{\updefault}{\color[rgb]{0,0,0}$r_{+}$}%
}}}}
\put(4276,-4636){\rotatebox{45.0}{\makebox(0,0)[lb]{\smash{\SetFigFont{8}{9.6}{\familydefault}{\mddefault}{\updefault}{\color[rgb]{0,0,0}$r_{++}$}%
}}}}
\put(5326,-3586){\rotatebox{45.0}{\makebox(0,0)[lb]{\smash{\SetFigFont{8}{9.6}{\familydefault}{\mddefault}{\updefault}{\color[rgb]{0,0,0}$r_{++}$}%
}}}}
\put(4576,-5386){\makebox(0,0)[lb]{\smash{\SetFigFont{8}{9.6}{\familydefault}{\mddefault}{\updefault}{\color[rgb]{0,0,0}$r=\infty, \mbox{\ }\scri^{-}$}%
}}}
\put(5626,-4336){\rotatebox{315.0}{\makebox(0,0)[lb]{\smash{\SetFigFont{8}{9.6}{\familydefault}{\mddefault}{\updefault}{\color[rgb]{0,0,0}$r_{++}$}%
}}}}
\put(2326,-2611){\makebox(0,0)[lb]{\smash{\SetFigFont{8}{9.6}{\familydefault}{\mddefault}{\updefault}{\color[rgb]{0,0,0}$r=0$}%
}}}
\put(2326,-5386){\makebox(0,0)[lb]{\smash{\SetFigFont{8}{9.6}{\familydefault}{\mddefault}{\updefault}{\color[rgb]{0,0,0}$r=0$}%
}}}
\put(7426,-2611){\makebox(0,0)[lb]{\smash{\SetFigFont{8}{9.6}{\familydefault}{\mddefault}{\updefault}{\color[rgb]{0,0,0}$r=0$}%
}}}
\put(7426,-5386){\makebox(0,0)[lb]{\smash{\SetFigFont{8}{9.6}{\familydefault}{\mddefault}{\updefault}{\color[rgb]{0,0,0}$r=0$}%
}}}
\put(2101,-3211){\rotatebox{315.0}{\makebox(0,0)[lb]{\smash{\SetFigFont{8}{9.6}{\familydefault}{\mddefault}{\updefault}{\color[rgb]{0,0,0}$r_{+}$}%
}}}}
\put(6901,-3211){\rotatebox{315.0}{\makebox(0,0)[lb]{\smash{\SetFigFont{8}{9.6}{\familydefault}{\mddefault}{\updefault}{\color[rgb]{0,0,0}$r_{+}$}%
}}}}
\put(4501,-3211){\rotatebox{315.0}{\makebox(0,0)[lb]{\smash{\SetFigFont{8}{9.6}{\familydefault}{\mddefault}{\updefault}{\color[rgb]{0,0,0}$r_{++}$}%
}}}}
\put(1426,-3961){\makebox(0,0)[lb]{\smash{\SetFigFont{8}{9.6}{\familydefault}{\mddefault}{\updefault}{\color[rgb]{0,0,0}$IV$}%
}}}
\put(6226,-3961){\makebox(0,0)[lb]{\smash{\SetFigFont{8}{9.6}{\familydefault}{\mddefault}{\updefault}{\color[rgb]{0,0,0}$IV$}%
}}}
\put(8701,-3961){\makebox(0,0)[lb]{\smash{\SetFigFont{8}{9.6}{\familydefault}{\mddefault}{\updefault}{\color[rgb]{0,0,0}$I$}%
}}}
\put(3901,-3961){\makebox(0,0)[lb]{\smash{\SetFigFont{8}{9.6}{\familydefault}{\mddefault}{\updefault}{\color[rgb]{0,0,0}$I$}%
}}}
\put(2626,-3061){\makebox(0,0)[lb]{\smash{\SetFigFont{8}{9.6}{\familydefault}{\mddefault}{\updefault}{\color[rgb]{0,0,0}$II_{H}$}%
}}}
\put(2626,-4861){\makebox(0,0)[lb]{\smash{\SetFigFont{8}{9.6}{\familydefault}{\mddefault}{\updefault}{\color[rgb]{0,0,0}$III_{H}$}%
}}}
\put(5026,-3061){\makebox(0,0)[lb]{\smash{\SetFigFont{8}{9.6}{\familydefault}{\mddefault}{\updefault}{\color[rgb]{0,0,0}$II_{C}$}%
}}}
\put(7426,-3061){\makebox(0,0)[lb]{\smash{\SetFigFont{8}{9.6}{\familydefault}{\mddefault}{\updefault}{\color[rgb]{0,0,0}$II_{H}$}%
}}}
\put(5026,-4861){\makebox(0,0)[lb]{\smash{\SetFigFont{8}{9.6}{\familydefault}{\mddefault}{\updefault}{\color[rgb]{0,0,0}$III_{C}$}%
}}}
\put(7426,-4861){\makebox(0,0)[lb]{\smash{\SetFigFont{8}{9.6}{\familydefault}{\mddefault}{\updefault}{\color[rgb]{0,0,0}$III_{H}$}%
}}}
\put(4576,-2611){\makebox(0,0)[lb]{\smash{\SetFigFont{8}{9.6}{\familydefault}{\mddefault}{\updefault}{\color[rgb]{0,0,0}$r=\infty, \mbox{\ }\scri^{+}$}%
}}}
\end{picture}

%% file: penrose_anti1.pstex_t
\begin{picture}(0,0)%
\includegraphics{penrose_anti1}%
\end{picture}%
\setlength{\unitlength}{2763sp}%
\begingroup\makeatletter\ifx\SetFigFont\undefined%
\gdef\SetFigFont#1#2#3#4#5{%
  \reset@font\fontsize{#1}{#2pt}%
  \fontfamily{#3}\fontseries{#4}\fontshape{#5}%
  \selectfont}%
\fi\endgroup%
\begin{picture}(2939,3060)(6031,-5461)
\put(6676,-4636){\rotatebox{45.0}{\makebox(0,0)[lb]{\smash{\SetFigFont{8}{9.6}{\familydefault}{\mddefault}{\updefault}{\color[rgb]{0,0,0}$r_{+}$}%
}}}}
\put(8026,-4336){\rotatebox{315.0}{\makebox(0,0)[lb]{\smash{\SetFigFont{8}{9.6}{\familydefault}{\mddefault}{\updefault}{\color[rgb]{0,0,0}$r_{+}$}%
}}}}
\put(6901,-3211){\rotatebox{315.0}{\makebox(0,0)[lb]{\smash{\SetFigFont{8}{9.6}{\familydefault}{\mddefault}{\updefault}{\color[rgb]{0,0,0}$r_{+}$}%
}}}}
\put(7426,-3061){\makebox(0,0)[lb]{\smash{\SetFigFont{8}{9.6}{\familydefault}{\mddefault}{\updefault}{\color[rgb]{0,0,0}$II$}%
}}}
\put(7801,-3511){\rotatebox{45.0}{\makebox(0,0)[lb]{\smash{\SetFigFont{8}{9.6}{\familydefault}{\mddefault}{\updefault}{\color[rgb]{0,0,0}$r_{+}$}%
}}}}
\put(7426,-4861){\makebox(0,0)[lb]{\smash{\SetFigFont{8}{9.6}{\familydefault}{\mddefault}{\updefault}{\color[rgb]{0,0,0}$III$}%
}}}
\put(6751,-3961){\makebox(0,0)[lb]{\smash{\SetFigFont{8}{9.6}{\familydefault}{\mddefault}{\updefault}{\color[rgb]{0,0,0}$IV$}%
}}}
\put(6151,-4411){\rotatebox{90.0}{\makebox(0,0)[lb]{\smash{\SetFigFont{8}{9.6}{\familydefault}{\mddefault}{\updefault}{\color[rgb]{0,0,0}$r=\infty, \mbox{\ }\scri^{-}$}%
}}}}
\put(8851,-3511){\rotatebox{270.0}{\makebox(0,0)[lb]{\smash{\SetFigFont{8}{9.6}{\familydefault}{\mddefault}{\updefault}{\color[rgb]{0,0,0}$r=\infty, \mbox{\ }\scri^{+}$}%
}}}}
\put(7351,-5461){\makebox(0,0)[lb]{\smash{\SetFigFont{8}{9.6}{\familydefault}{\mddefault}{\updefault}{\color[rgb]{0,0,0}$r=0$}%
}}}
\put(7351,-2536){\makebox(0,0)[lb]{\smash{\SetFigFont{8}{9.6}{\familydefault}{\mddefault}{\updefault}{\color[rgb]{0,0,0}$r=0$}%
}}}
\put(8101,-3961){\makebox(0,0)[lb]{\smash{\SetFigFont{8}{9.6}{\familydefault}{\mddefault}{\updefault}{\color[rgb]{0,0,0}$I$}%
}}}
\end{picture}

%% file: penrose1a.pstex_t
\begin{picture}(0,0)%
\includegraphics{penrose1a}%
\end{picture}%
\setlength{\unitlength}{2763sp}%
\begingroup\makeatletter\ifx\SetFigFont\undefined%
\gdef\SetFigFont#1#2#3#4#5{%
  \reset@font\fontsize{#1}{#2pt}%
  \fontfamily{#3}\fontseries{#4}\fontshape{#5}%
  \selectfont}%
\fi\endgroup%
\begin{picture}(6045,2955)(3868,-5431)
\put(6676,-4636){\rotatebox{45.0}{\makebox(0,0)[lb]{\smash{\SetFigFont{8}{9.6}{\familydefault}{\mddefault}{\updefault}{\color[rgb]{0,0,0}$r_{+}$}%
}}}}
\put(7801,-3511){\rotatebox{45.0}{\makebox(0,0)[lb]{\smash{\SetFigFont{8}{9.6}{\familydefault}{\mddefault}{\updefault}{\color[rgb]{0,0,0}$r_{+}$}%
}}}}
\put(8026,-4336){\rotatebox{315.0}{\makebox(0,0)[lb]{\smash{\SetFigFont{8}{9.6}{\familydefault}{\mddefault}{\updefault}{\color[rgb]{0,0,0}$r_{+}$}%
}}}}
\put(5326,-3586){\rotatebox{45.0}{\makebox(0,0)[lb]{\smash{\SetFigFont{8}{9.6}{\familydefault}{\mddefault}{\updefault}{\color[rgb]{0,0,0}$r_{++}$}%
}}}}
\put(4276,-3736){\rotatebox{270.0}{\makebox(0,0)[lb]{\smash{\SetFigFont{8}{9.6}{\familydefault}{\mddefault}{\updefault}{\color[rgb]{0,0,0}$r=R$}%
}}}}
\put(5626,-4336){\rotatebox{315.0}{\makebox(0,0)[lb]{\smash{\SetFigFont{8}{9.6}{\familydefault}{\mddefault}{\updefault}{\color[rgb]{0,0,0}$r_{++}$}%
}}}}
\put(7426,-2611){\makebox(0,0)[lb]{\smash{\SetFigFont{8}{9.6}{\familydefault}{\mddefault}{\updefault}{\color[rgb]{0,0,0}$r=0$}%
}}}
\put(7426,-5386){\makebox(0,0)[lb]{\smash{\SetFigFont{8}{9.6}{\familydefault}{\mddefault}{\updefault}{\color[rgb]{0,0,0}$r=0$}%
}}}
\put(6901,-3211){\rotatebox{315.0}{\makebox(0,0)[lb]{\smash{\SetFigFont{8}{9.6}{\familydefault}{\mddefault}{\updefault}{\color[rgb]{0,0,0}$r_{+}$}%
}}}}
\put(6226,-3961){\makebox(0,0)[lb]{\smash{\SetFigFont{8}{9.6}{\familydefault}{\mddefault}{\updefault}{\color[rgb]{0,0,0}$IV$}%
}}}
\put(8701,-3961){\makebox(0,0)[lb]{\smash{\SetFigFont{8}{9.6}{\familydefault}{\mddefault}{\updefault}{\color[rgb]{0,0,0}$I$}%
}}}
\put(5026,-3061){\makebox(0,0)[lb]{\smash{\SetFigFont{8}{9.6}{\familydefault}{\mddefault}{\updefault}{\color[rgb]{0,0,0}$II_{C}$}%
}}}
\put(7426,-3061){\makebox(0,0)[lb]{\smash{\SetFigFont{8}{9.6}{\familydefault}{\mddefault}{\updefault}{\color[rgb]{0,0,0}$II_{H}$}%
}}}
\put(5026,-4861){\makebox(0,0)[lb]{\smash{\SetFigFont{8}{9.6}{\familydefault}{\mddefault}{\updefault}{\color[rgb]{0,0,0}$III_{C}$}%
}}}
\put(7426,-4861){\makebox(0,0)[lb]{\smash{\SetFigFont{8}{9.6}{\familydefault}{\mddefault}{\updefault}{\color[rgb]{0,0,0}$III_{H}$}%
}}}
\put(4576,-2611){\makebox(0,0)[lb]{\smash{\SetFigFont{8}{9.6}{\familydefault}{\mddefault}{\updefault}{\color[rgb]{0,0,0}$r=\infty, \mbox{\ }\scri^{+}$}%
}}}
\put(4576,-5386){\makebox(0,0)[lb]{\smash{\SetFigFont{8}{9.6}{\familydefault}{\mddefault}{\updefault}{\color[rgb]{0,0,0}$r=\infty, \mbox{\ }\scri^{-}$}%
}}}
\put(4651,-3961){\makebox(0,0)[lb]{\smash{\SetFigFont{8}{9.6}{\familydefault}{\mddefault}{\updefault}{\color[rgb]{0,0,0}$I$}%
}}}
\end{picture}

%% file: penrose2.pstex_t
\begin{picture}(0,0)%
\includegraphics{penrose2}%
\end{picture}%
\setlength{\unitlength}{2763sp}%
\begingroup\makeatletter\ifx\SetFigFont\undefined%
\gdef\SetFigFont#1#2#3#4#5{%
  \reset@font\fontsize{#1}{#2pt}%
  \fontfamily{#3}\fontseries{#4}\fontshape{#5}%
  \selectfont}%
\fi\endgroup%
\begin{picture}(2466,2940)(3868,-5431)
\put(4651,-5386){\makebox(0,0)[lb]{\smash{\SetFigFont{8}{9.6}{\familydefault}{\mddefault}{\updefault}{\color[rgb]{0,0,0}$r=\infty, \mbox{\ }\scri^{-}$}%
}}}
\put(5476,-3436){\rotatebox{45.0}{\makebox(0,0)[lb]{\smash{\SetFigFont{8}{9.6}{\familydefault}{\mddefault}{\updefault}{\color[rgb]{0,0,0}$r_{++}$}%
}}}}
\put(4726,-3961){\makebox(0,0)[lb]{\smash{\SetFigFont{8}{9.6}{\familydefault}{\mddefault}{\updefault}{\color[rgb]{0,0,0}$I$}%
}}}
\put(4426,-3136){\rotatebox{315.0}{\makebox(0,0)[lb]{\smash{\SetFigFont{8}{9.6}{\familydefault}{\mddefault}{\updefault}{\color[rgb]{0,0,0}$r_{++}$}%
}}}}
\put(5026,-3061){\makebox(0,0)[lb]{\smash{\SetFigFont{8}{9.6}{\familydefault}{\mddefault}{\updefault}{\color[rgb]{0,0,0}$II_{C}$}%
}}}
\put(5026,-4861){\makebox(0,0)[lb]{\smash{\SetFigFont{8}{9.6}{\familydefault}{\mddefault}{\updefault}{\color[rgb]{0,0,0}$III_{C}$}%
}}}
\put(4651,-2611){\makebox(0,0)[lb]{\smash{\SetFigFont{8}{9.6}{\familydefault}{\mddefault}{\updefault}{\color[rgb]{0,0,0}$r=\infty, \mbox{\ }\scri^{+}$}%
}}}
\put(4276,-3736){\rotatebox{270.0}{\makebox(0,0)[lb]{\smash{\SetFigFont{8}{9.6}{\familydefault}{\mddefault}{\updefault}{\color[rgb]{0,0,0}$r=R$}%
}}}}
\put(5251,-3961){\makebox(0,0)[lb]{\smash{\SetFigFont{8}{9.6}{\familydefault}{\mddefault}{\updefault}{\color[rgb]{0,0,0}$IV$}%
}}}
\put(5926,-4111){\rotatebox{90.0}{\makebox(0,0)[lb]{\smash{\SetFigFont{8}{9.6}{\familydefault}{\mddefault}{\updefault}{\color[rgb]{0,0,0}$r=R$}%
}}}}
\end{picture}

%% file: penrose3.pstex_t
\begin{picture}(0,0)%
\includegraphics{penrose3}%
\end{picture}%
\setlength{\unitlength}{2763sp}%
\begingroup\makeatletter\ifx\SetFigFont\undefined%
\gdef\SetFigFont#1#2#3#4#5{%
  \reset@font\fontsize{#1}{#2pt}%
  \fontfamily{#3}\fontseries{#4}\fontshape{#5}%
  \selectfont}%
\fi\endgroup%
\begin{picture}(2424,2910)(1489,-5386)
\put(3001,-3511){\rotatebox{45.0}{\makebox(0,0)[lb]{\smash{\SetFigFont{8}{9.6}{\familydefault}{\mddefault}{\updefault}{\color[rgb]{0,0,0}$r_{+}$}%
}}}}
\put(2326,-2611){\makebox(0,0)[lb]{\smash{\SetFigFont{8}{9.6}{\familydefault}{\mddefault}{\updefault}{\color[rgb]{0,0,0}$r=0$}%
}}}
\put(1876,-3736){\rotatebox{270.0}{\makebox(0,0)[lb]{\smash{\SetFigFont{8}{9.6}{\familydefault}{\mddefault}{\updefault}{\color[rgb]{0,0,0}$r=R$}%
}}}}
\put(2326,-5386){\makebox(0,0)[lb]{\smash{\SetFigFont{8}{9.6}{\familydefault}{\mddefault}{\updefault}{\color[rgb]{0,0,0}$r=0$}%
}}}
\put(2626,-3061){\makebox(0,0)[lb]{\smash{\SetFigFont{8}{9.6}{\familydefault}{\mddefault}{\updefault}{\color[rgb]{0,0,0}$II_{H}$}%
}}}
\put(2626,-4861){\makebox(0,0)[lb]{\smash{\SetFigFont{8}{9.6}{\familydefault}{\mddefault}{\updefault}{\color[rgb]{0,0,0}$III_{H}$}%
}}}
\put(2851,-3961){\makebox(0,0)[lb]{\smash{\SetFigFont{8}{9.6}{\familydefault}{\mddefault}{\updefault}{\color[rgb]{0,0,0}$I$}%
}}}
\put(2176,-3961){\makebox(0,0)[lb]{\smash{\SetFigFont{8}{9.6}{\familydefault}{\mddefault}{\updefault}{\color[rgb]{0,0,0}$IV$}%
}}}
\put(2101,-3211){\rotatebox{315.0}{\makebox(0,0)[lb]{\smash{\SetFigFont{8}{9.6}{\familydefault}{\mddefault}{\updefault}{\color[rgb]{0,0,0}$r_{+}$}%
}}}}
\put(3526,-4111){\rotatebox{90.0}{\makebox(0,0)[lb]{\smash{\SetFigFont{8}{9.6}{\familydefault}{\mddefault}{\updefault}{\color[rgb]{0,0,0}$r=R$}%
}}}}
\end{picture}

%% file: deutsch_ent.tex
\section*{Deutsche Zusammenfassung}
\addcontentsline{toc}{section}{Deutsche Zusammenfassung} 

Die vorliegende Arbeit untersucht statische, kugelsymmetrische 
L"osungen der Einsteinschen Feldgleichungen mit kosmologischer 
Konstante. F"ur die Materie wird eine ideale Fl"ussigkeit mit 
vorgegebener monotoner Zustandsgleichung angenommen. 
Es werden neue L"osungstypen beschrieben.

Bemerkenswert ist, da"s die kosmologische Konstante
innerhalb dieser Fragestellung lange vernachl"assigt wurde. 

Das erste Kapitel ist eine Einf"uhrung. Sie ordnet die Thematik 
innerhalb der Gravitationstheorie ein und geht auf Teile vorhandener 
Literatur ein.

Obwohl schon Weyl~\cite{Weyl} diese L"osungen untersuchte und
wesentliche Ergebnisse darstellte, wurden sie kaum weiter beachtet.
Erst wieder Kriele~\cite{Kriele}, Stuchl\'{\i}k~\cite{Stuchlik} und Winter~\cite{Winter}
untersuchten solche L"osungen. 

Im zweiten Teil werden die relevanten Gleichungen hergeleitet 
und deren mathematische Struktur beschrieben. 

Wird f"ur die Materie eine ideale Fl"ussigkeit angenommen, so 
erh"alt man eine Verallgemeinerung der Tolman-Oppenheimer-Volkoff 
Gleichung mit kosmologischer Konstante (\ref{tov}), kurz 
TOV-$\Lambda$ Gleichung. Sei nun eine Zustandsgleichung
$\rho = \rho(P)$ vorgegeben. Dann bilden die beiden Gleichungen
(\ref{tov}) und (\ref{mdd}) ein System von gew"ohnlichen Differentialgleichungen 
in den Funktionen $P(r)$ und $w(r)$.

F"ur das Vakuum lassen sich die Feldgleichungen einfach l"osen. 
Man erh"alt die bekannte Schwarzschild-de Sitter L"osung (\ref{sds}) f"ur positive  
kosmologische Konstante bzw. die Schwarzschild-anti-de Sitter 
L"osung f"ur das negative Vorzeichen. Dies ist die einzige statische, 
kugelsymmetrische L"osung der Vakuumfeldgleichungen
mit nichtkonstanten Gruppenbahnen.

Penrose-Carter Diagramm~\ref{fig: Penrose} zeigt die
Schwarzschild-de Sitter Raum-Zeit mit dem Ereignishorizont
des Schwarzen Loches und dem kosmologischen Ereignishorizont.

Das Penrose-Carter Diagramm~\ref{fig: Penrose_anti1} zeigt die 
Schwarzschild-anti-de Sitter Raum-Zeit. In ihr gibt es
nur den Ereignishorizont des Schwarzen Loches.

Mit kosmologischer Konstante gibt es eine weitere statische,
kugelsymmetrische L"osung der Feldgleichungen f"ur das Vakuum,
die Nariai L"osung \cite{Nariai1,Nariai2}. Das Volumen der
Gruppenbahnen ist hierbei konstant. Einer der neuen L"osungstypen
erfordert diese L"osung f"ur den Vakuumteil.

Nach dieser Beschreibung der Gleichungen wird kurz auf deren 
Newtonschen Grenzwert eingegangen. Wie zu erwarten, ergibt sich die
Newtonsche Theorie mit einem zus"atzlichen $\Lambda$-Term (\ref{newton}). 
Das Newtonsche Analogon der TOV-$\Lambda$ Gleichung ist 
die Euler Gleichung der Hydrodynamik im hydrodynamischen 
Gleichgewicht, wieder mit korrigierten Potential, (\ref{euler}). 

Der dritte Abschnitt befa"st sich mit den speziellen konstante
Dichte L"osungen. Die TOV-$\Lambda$ Gleichung wird f"ur den 
einfachen Fall einer Zustandsgleichung $\rho=\rho(P) \rightarrow \rho=\rho_{0}=\mbox{const.}$, 
n"amlich homogener Dichte, integriert. Der zus"atzlich 
vorhandene $\Lambda$-Term er"offnet neue L"osungen. Die 
$t=\mbox{const.}$ Hyperfl"achen sind bei verschwindender 
kosmologischer Konstante 3-Sph"aren. $\Lambda$ kann nun 
auch so gew"ahlt werden, da"s diese Hyperfl"achen euklidisch 
oder hyperbolisch werden. Die L"osungen mit konstanter Dichte 
sind konform flach, unabh"angig von der kosmologischen
Konstanten. Mit den im vierten Kapitel eingef"uhrten neuen 
Buchdahlvariablen kann leicht gezeigt werden, da"s dies 
direkt aus den Feldgleichungen folgt. 

Konstante Dichte L"osungen wurden nur bis zu einer oberen Grenze 
$\Lambda < 4 \pi \rho_{0}$ untersucht \cite{Stuchlik}. 
Bis zu dieser Grenze beschreiben die L"osungen ausschlie"slich Modelle
von Sternen. Es existiert immer ein Radius, an dem der Druck verschwindet
und an dem eine Vakuuml"osung angeschlossen wird.
F"ur diese Modelle l"a"st sich eine analoge Buchdahl Ungleichung
herleiten. Sie besagt, da"s der Radius stets zwischen dem Ereignishorizonts
des entsprechenden Schwarzen Loches und dem kosmologischen 
Ereignishorizont liegt. Der kosmologische Ereignishorizont existiert
nur im Falle positiver kosmologischer Konstante. Daher haben
Modelle mit $\Lambda \leq 0$ im allgemeinen keine obere Schranke 
f"ur den Radius.

Diese L"osungen k"onnen frei von Singularit"aten konstruiert werden. 
Falls $\Lambda \leq 0$ ist dies offensichtlich, da die Au"senrauml"osungen
frei von Singularit"aten ist. F"ur $\Lambda > 0$ 
mu"s ein zweites Objekt in die Raum-Zeit gesetzt werden, damit diese 
L"osungen frei von Singularit"aten werden, siehe Penrose-Carter 
Diagramm~\ref{fig: Penrose2}.

Wird eine gr"o"sere kosmologische Konstante vorgegeben, so wird die 
Koordinatensingularit"at erreicht, noch bevor der Druck verschwindet. 
An ihr divergiert der Druckgradient, der Druck jedoch bleibt endlich. 
Da es sich lediglich um eine Koordinatensingularit"at handelt, 
k"onnen entsprechende L"osungen fortgesetzt werden. In diesem dritten 
Kapitel werden die verschiedenen L"osungen durch die 
kosmologische Konstante parametrisiert. 

Entspricht $\Lambda$ der Grenze, so verschwindet der
Druck bei der Koordinatensingularit"at. Nur in diesem Fall erfordert die
Innenrauml"osung, da"s die Nariai Metrik angeschlossen wird. In den
anderen F"allen ist es immer die Schwarzschild-de Sitter L"osung, die
den Vakuumteil beschreibt. Wird der kosmologische Term nun gr"o"ser
gew"ahlt, dann verschwindet der Druck nach der Koordinatensingularit"at. 
Dort ist das Volumen der Gruppenbahnen fallend und man hat den Teil der
Schwarzschild-de Sitter Metrik anzuschlie"sen, der die $r=0$ Singularit"at
enth"alt, siehe Penrose-Carter Diagramm~\ref{fig: Penrose3}. 
Eine weitere Vergr"o"serung f"uhrt dazu, da"s der Druck nicht
mehr verschwinden kann, er ist strikt positiv. Es ergeben sich 
L"osungen mit zwei Zentren, die in Fall konstanter Dichte beide regul"ar sind.
Dies sind Verallgemeinerungen des statischen Einstein Universums.
Sie haben homogene Dichte aber inhomogenen Druck. Der
im ersten Zentrum vorgegebene Zentraldruck f"allt monoton bis zum
zweiten Zentrum. Bei weiterer Vergr"o"serung ergibt sich der 
Einstein Kosmos selbst, um dann deren Verallgemeinerungen 
mit ansteigendem Druck bis zum zweiten Zentrum zu beschreiben. 
In Abschnitt~\ref{subsec_2centres_b} wurde gezeigt, da"s sich die
Verallgemeinerungen des statischen Einstein Universums mit 
ansteigendem Druck im ersten Zentrum und diese mit fallendem Druck
im ersten Zentrum ineinander "uberf"uhren lassen. 

Als letztes, bei noch gr"o"serer kosmologischer Konstante ist der Druck 
monoton ansteigend und divergiert. Dort hat die Raum-Zeit eine geometrischen 
Singularit"at. Diese L"osungen sind unphysikalisch und werden nicht 
weiter diskutiert. Am Ende dieses Kapitels werden alle konstante
Dichte L"osungen in einer "Ubersicht dargestellt.

Das vierte Kapitel besch"aftigt sich zun"achst mit der Existenz von globalen
L"osungen. Mit neuen Buchdahlvariablen kann diese bis zu einer oberen 
Grenze der kosmologischen Konstante (\ref{upper_limit}) bewiesen werden. 
Die entsprechenden L"osungen 
werden beschrieben und mit denen konstanter Dichte verglichen.
Die obere Grenze, bis zu der die Existenz einer globalen L"osung gezeigt werden
kann, schlie"st aber die neuen L"osungstypen aus.

F"ur endliche L"osungen wird die verallgemeinerte Buchdahlungleichung (\ref{eqn_buch_sqr})
hergeleitet. Sie gilt f"ur alle statischen, kugelsymmetrischen Fl"ussigkeitskugeln
deren Dichte nach au"sen nicht ansteigt. Damit wird gezeigt, da"s L"osungen
ohne Singularit"aten nicht nur im Fall konstanter Dichte konstruiert werden 
k"onnen, sondern da"s diese L"osungen f"ur beliebige Zustandsgleichungen 
diese Eigenschaft besitzen.

Den Abschlu"s des Kapitels bilden Bemerkungen "uber die Existenz eines endlichen
Radius f"ur eine vorgegebene Zustandsgleichung. Es werden eine notwendige und
eine hinreichende Bedingung hergeleitet.

Im Anhang dieser Arbeit werden Ricci und Einstein Tensor sowie einige 
geometrische Invarianten berechnet. Dar"uber hinaus sind dort die
Penrose-Carter Diagramme dargestellt.